\documentclass[final,1p,times]{elsarticle}

\usepackage{amssymb}
\usepackage{amsthm}
\usepackage{amsmath}
\usepackage{multirow}
\usepackage{gensymb}
\usepackage{booktabs}
\usepackage{subcaption}
\usepackage[table]{xcolor}
\usepackage{diagbox}

\theoremstyle{assumption}
\newtheorem{assumption}{Assumption}[section]
\theoremstyle{definition}
\newtheorem{definition}{Definition}[section]
\theoremstyle{proposition}
\newtheorem{proposition}{Proposition}[section]
\theoremstyle{lemma}
\newtheorem{lemma}{Lemma}[section]
\theoremstyle{corollary}
\newtheorem{corollary}{Corollary}[section]
\DeclareMathOperator*{\argmax}{argmax}

\journal{Artificial Intelligence}

\begin{document}

\begin{frontmatter}



\title{Grounded Predictions of Teamwork as a One-Shot Game: A Multiagent Multi-Armed Bandits Approach}


\affiliation[inst1]{organization={Artificial Intelligence Research Institute, Barcelona, IIIA-CSIC},
            addressline={Campus UAB Carrer de Can Planas, Zona 2}, 
            city={Barcelona},
            postcode={08193}, 
            country={Spain}}
\author[inst1]{Alejandra López de Aberasturi Gómez\corref{corresponding author}}
\ead{alejandra@iiia.csic.es}

\author[inst1]{Carles Sierra}
\ead{sierra@iiia.csic.es}

\author[inst1]{Jordi Sabater-Mir}
\ead{jsabater@iiia.csic.es}

\begin{abstract}
Humans possess innate collaborative capacities. However, effective teamwork often remains challenging. This study delves into the feasibility of collaboration within teams of rational, self-interested agents who engage in teamwork without the obligation to contribute. Drawing from psychological and game theoretical frameworks, we formalise teamwork as a one-shot aggregative game, integrating insights from Steiner's theory of group productivity. We characterise this novel game's Nash equilibria and propose a multiagent multi-armed bandit system that learns to converge to approximations of such equilibria. Our research contributes value to the areas of game theory and multiagent systems, paving the way for a better understanding of voluntary collaborative dynamics. We examine how team heterogeneity, task typology, and assessment difficulty influence agents' strategies and resulting teamwork outcomes. Finally, we empirically study the behaviour of work teams under incentive systems that defy analytical treatment. Our agents demonstrate human-like behaviour patterns, corroborating findings from social psychology research. 
\end{abstract}



\end{frontmatter}

\section{Introduction}\label{sec:introduction}
Cooperation and alliances have long been integral to human prosperity, and remain vital in today's world \citep{apicella2019evolution}. While cooperation is not unique to humans, our species stands out for its ability to apply teamwork skills across diverse domains and to conceptualise teamwork itself \citep{teamworkinMAS2010}. Here, we understand teamwork as defined by The American Heritage Dictionary \citep{teamwork}: \textit{Cooperative effort by the members of a group or team to achieve a common goal.} Human cooperation extends beyond small groups to encompass large numbers of unrelated individuals engaging in various forms of collective action, including activities that involve personal costs for the benefit of others, such as contributing to public goods \citep{bowles2003}. This kind of collective action is essential for addressing modern challenges outlined in sustainable development objectives, as stated by The United Nations General Assembly \citep{unitednations}. However, while collaboration comes naturally to us, achieving effective teamwork is still challenging.

From a mathematical perspective, teamwork has been extensively modelled within cooperative game theory, which assumes that all team members share a common reward and there is no conflict among them. Solution concepts like the Shapley Value aim to fairly divide the total utility created by the team. This approach works well in closely-knit teams where cooperation is clearly beneficial. However, in typical human teamwork scenarios, securing binding collaboration agreements is often unattainable, leaving room for defection and calling for the application of non-cooperative game theory.

On the other hand, computer science, particularly in the field of Multiagent Systems (MAS), has extensively studied teamwork and how to design multiagent systems that cooperate effectively \citep{teamworkinMAS2010,tambe1997towards,wooldridge2009introduction}. Specifically, considerable attention has been directed towards team composition\footnote{\citet{andrejczuk2016concise} define team composition as the process of deciding which agents should be part of a team.} and team formation\footnote{Conversely, \citet{andrejczuk2016concise} define team formation as the process undertaken by agents to learn to work together in a team.} within groups of cooperative, self-interested agents \citep{andrejczuk2016concise}. Although the objectives necessitating cooperation vary among algorithms (\textit{e.g.}, minimising overall costs, maximising social utility, or enhancing outcome quality), the prevalent approach assumes that agents need cooperation to succeed individually, making intentionally defective strategies unviable. This approach primarily seeks to determine how to achieve maximum performance from the agent team, whether in terms of maximising benefits or minimising costs. In this sense, much of the focus has been on prescriptive approaches: \textit{How can we design each agent to ensure effective cooperation? How should these agents work together to solve problems?} \citep{wooldridge2009introduction}.
In short, the focus of most work on multiagent systems has been addressing issues associated with \textit{self-interested} agents who require cooperation to achieve individual success, and resolving the resulting conflicts of interest to enable this cooperation. Surprisingly, the problem of predicting the performance of human teams, or more broadly, performance in teams without enforced cooperation, has been largely overlooked. When addressing this challenge, a descriptive approach, rather than a prescriptive one, would be more suitable: \textit{What social effects arise when each (self-interested) agent in a society adopts a particular, not necessarily cooperative, policy?}. The scarce efforts in computer science to tackle this issue have often lacked an integrative, cross-disciplinary understanding of progress made in fields like economics, game theory, and social psychology. For instance, \citet{andrejczuk2018composition} highlighted that in MAS models of human team formation and performance, there's often an unrealistic assumption that agents have complete knowledge of their environment and teammates. Plus, factors that are crucial according to psychology and economics, such as time constraints, task type and team heterogeneity, are often overlooked in MAS. 

In this paper, we address this gap in the literature by modelling human teamwork through a descriptive approach. We move away from the assumptions of complete agent knowledge and enforced cooperation, instead integrating insights from psychology and game theory to answer the question of how to learn theoretically grounded predictions of team performance.

This research question has two main components: Firstly, our goal is to identify the Nash Equilibria (NE) in a non-cooperative game that models teamwork. Secondly, we aim to propose a learning framework where agents converge to the teamwork outcomes predicted by the NE of such a game. 

As per the second aspect, the topic of learning in games studies how less than fully rational players grope for optimality over time, as well as understand possible barriers to reaching equilibrium. Specifically, reinforcement learning (RL) agents using smooth-best responses can apply value-based learning techniques to approximate the NE of any normal-form game, as demonstrated by \citet{leslie2005individual}. The architecture of RL agents naturally lends itself to representing beliefs, preferences, intentions, and actions \citep{jara2019theory}. When combined in a multiagent system, the interactive learning of RL agents facilitates the emergence of complex behaviours \citep{baker2020emergent, ndousse2020learning, marl-book}. Thus, Multiagent Reinforcement Learning (MARL) provides a suitable framework to explore the emergence of social dynamics during teamwork through a descriptive approach. Traditionally, teams in MARL have been fully cooperative, with agents sharing a common reward \citep{zhang2021multi}. However, the field is evolving towards more realistic multiagent teams, including mixed-motive scenarios and social dilemmas beyond zero-sum games \cite{leibo2017multi}, more detailed modelling of agents' motivations \citep{ndousse2020learning, hughes2018inequity, jaques2019social, mckee2020social, rodriguez2021multi, madhushani2023heterogeneous}, and innovative multiagent negotiation techniques \cite{kramar2022negotiation}.

These efforts have contributed to the rise of a new research stream called ``Cooperative AI'' \citep{dafoe2020open}. The goal of Cooperative AI is to promote interdisciplinary dialogue with natural, social, and behavioural sciences. Cooperative AI leverages recent advances in artificial intelligence, including reinforcement learning, to study various aspects of cooperation and to solve cooperation problems arising in systems of (human and/or machine) agents. However, despite the development of more realistic multiagent teams in MARL, a Cooperative AI approach has not yet been used to model teamwork without presuming enforced cooperation. We see an opportunity to improve the modelling of human teamwork by harnessing MARL frameworks to represent selfish, interactive learning agents who are not compelled to cooperate, drawing on insights from psychology.

\subsection{Our contribution}

Our research addresses a pivotal question: How can we effectively represent \textbf{(1) a teamwork setting without enforced cooperation} and \textbf{(2) a team of reward-maximising agents} to learn theoretically grounded predictions of team performance? To answer this, we model teamwork as a mixed-motive game, drawing inspiration from social psychology research on human behaviour in teams. We characterise the game's equilibria, which predict teamwork outcomes. Building on previous research in learning in games, we propose a Multiagent Multi-Armed Bandit (MA-MAB) system to learn approximations of the Nash Equilibria (NE) of the game. Our methodology is grounded in aggregative games as the foundational framework. As will be explained shortly, aggregative games are a subset of pure-strategy non-cooperative games that express payoffs for each player as a function of their strategy and an aggregate representing all interactions in the game. This condensed payoff representation overcomes the complexities associated with an increasing number of players \cite{jensen2018aggregative}. 
Although previous research has used bandits to play games such as Cournot oligopoly one-shot games \citep{taywade2022}, zero-sum games \citep{leslie2005individual}, and repeated Stackelberg games \citep{qian2016restless}, to the best of our knowledge, this is the first instance where aggregative games have been explicitly combined with Multiagent Multi-Armed Bandit Models. 
This work aims to contribute to Cooperative AI \citep{dafoe2020open}, both theoretically and practically:

\begin{itemize}

    \item We introduce \textit{teamwork games}, a novel aggregative game that explicitly models teamwork and considers the evaluation of teamwork outcomes as part of the players' utility function. Our work builds on the most popular research stream in interdependence theory \citep{lewin1948resolving} on teamwork: Steiner's \citep{steiner1972group} \emph{theory of group productivity}. We identify the branch of game theory that allows us to mathematically formulate the main ingredients of Steiner's theory: \emph{general public good games} \citep{cornes2007weak, cornes2007aggregative}. Extending this framework, we incorporate variables that have been previously overlooked, including the weights of team members' opinions within the team, task complexity, individual expertise, preferences, and time constraints.

    \item We characterise the NE of this novel game, and find that they exhibit variations of the equilibria observed in general public good games \citep{cornes2007weak, cornes2007aggregative}. These variations are influenced by elements unique to our model, such as the teamwork evaluation function and player expertise.

    \item We propose a multiagent framework in which agents learn strategic behaviour that approximates the NE of the game. Building on the results in learning in games by \citet{leslie2005individual}, we conceptualise the allocation of individual work contributions to the team task as a multi-armed bandit problem with Boltzmann action selection.
    
    \item We validate the proposed MA-MAB system by empirically proving the convergence of the learned strategies towards the approximated NE of the game. In over $140$ experiments, the model achieved a near-perfect fit ($\xi^2 = 0.992$) between the equilibrium value of team productivity and the actual productivity to which the MA-MAB model converged. Additionally, we analyse the impact of isolated variables on team productivity when the bandits face teamwork scenarios:
    
    \begin{itemize}
        \item When examining how agent productivity varies with different types of tasks, our findings align with \citeauthor{steiner1972group}'s hypotheses on team composition and productivity. Specifically, tasks where the weakest team members significantly impact the outcome are better performed by homogeneous teams with similar expertise levels. In additive tasks, where each member's contribution adds up to the outcome, team heterogeneity does not affect productivity. Conversely, tasks where the strongest members have the most influence are better handled by teams with diverse expertise levels.

        \item When we analyse the effects of increasing the difficulty of the teamwork evaluation function, we find that productivity rises with task difficulty, especially when passing to the next evaluation level is achievable. However, this trend halts once the task requirements exceed the team's capacity. At this point, agents either refrain from increasing their contribution or reduce it. 

        \item We find that agents with lower expertise are more influenced by changes in task type, while highly skilled agents are more responsive to changes in the teamwork evaluation function than to the task's nature.               
    \end{itemize}

    \item After validating the MA-MAB framework, we expand our analysis beyond analytically solvable games and empirically study the policies of our agents after learning with a discontinuous evaluation function, resembling a binary pass/fail assessment.

\end{itemize}

The remainder of the article is structured as follows. Section \ref{sec:preliminaries} explores the theoretical frameworks related to teamwork and decision-making from the perspectives of game theory, social psychology, and multiagent systems. Section \ref{sec:background} introduces key concepts from aggregative games, forming the foundation of our mathematical framework for investigating teamwork. Section \ref{sec:model} builds on this framework to establish a model of human teamwork. Section \ref{sec:equilibria} examines the equilibria in teamwork games, highlighting the effects of different parameter regimes on the team outcome. Section \ref{sec:MA-MAB_learning} conceptualises the decision-making process of team members as a multiagent multi-armed bandit problem, detailing the learning dynamics involved. Section \ref{sec:experiments} presents our experimental setup and discusses the results, focusing on team productivity, individual strategies, and simulations beyond traditional aggregative game theory. Finally, Section \ref{sec:conclusions} draws conclusions, discusses limitations, and outlines potential future work.

\section{Preliminaries}\label{sec:preliminaries}

In this section, we explore the theoretical frameworks related to teamwork or decision-making from the perspectives of the three main areas integrated in this paper: game theory, social psychology, and multiagent systems.

\subsection{Aggregative games and the public goods dilemma}\label{subsec:aggregative_games}

From a mathematical perspective, numerous efforts have been made to model teamwork, particularly within cooperative game theory.  Cooperative game theory approaches assume that all team members share a common reward, and thus there is no potential for conflict between them. Solution concepts in cooperative game theory, such as the Shapley Value, have been developed to be ``fair'' divisions of the total utility created by the team or coalition \citep{zhang2021multi, hernandez2010rankings, hiller2019structure, yan2020evaluating}. 

While this approach may be effective for tightly-knit teams where the benefits of cooperation (or the drawbacks of non-cooperation) are evident, it often falls short in typical human teamwork settings. In these scenarios, it's not feasible to establish binding agreements that ensure the cooperation of all team members. Consequently, it becomes necessary to acknowledge the possibility that team members may shirk their collective responsibilities and seek to maximise their own utility by free-riding on the efforts of others. This perspective on teamwork is further explored through mixed games, notably in the context of the voluntary provision of public goods game\footnote{In this basic version of the game, participants privately decide how many tokens to contribute to a communal pool. These tokens are aggregated, resulting in a collective outcome known as the ``public good''. The benefits from this public good are evenly distributed among all players, while each participant retains the tokens they choose not to contribute.} \cite{myatt2009evolution, CARPENTER2009221, banos2019monitoring}. Indeed, teamwork can be viewed as a public good, presenting members with a dilemma between contributing to the collective task or exploiting others' efforts (social loafing). Despite the potential benefits of cooperation, individuals may have varying preferences regarding the provision of the common good, and practical constraints often fail to prevent free-rider behaviours \citep{kameda1992social}. In this paper, we adopt this perspective of teamwork as a mixed-motive game.

Aggregative games, a subset of pure-strategy non-cooperative games, express payoffs for each player as a function of their strategy and an aggregate representing all interactions in the game. This approach simplifies analysis as it avoids the complexities associated with an increasing number of players \cite{jensen2018aggregative}. By aggregating team interactions, analysts can simplify calculations, eliminating the need to consider individual strategies to compute the best response. Notably, some longstanding economic models exhibit this aggregative structure, including the voluntary provision of public goods game \cite{corchon2021aggregative}, which is the focus of this paper.

In the literature on aggregative games, the emphasis lies in computing Nash Equilibria (NE) and understanding how NE are influenced by changes in exogenous parameters (an analysis called comparative statics), along with addressing their existence \cite{jensen2018aggregative, acemoglu2013aggregate}. The antecedents of using aggregative games as an analytical tool for the voluntary contribution to public goods are presented by \citet{cornes2007aggregative}. Their work adapted the canonical model of public goods proposed by \citet{bergstrom1992uniqueness} ---which considers public goods as the unweighted sum of all individual contributions--- to the aggregative games formalism.

However, as early as 1983, \citet{hirshleifer1983weakest} had already suggested two types of public goods ---best-shot (where the best individual performance determines team productivity) and weakest-link (where the poorest individual performance determines team productivity)--- that are not captured by summation technology but hold empirical significance. \citet{cornes1993dyke} and \citet{cornes2007weak} address these observations by exploring games where the total public good level is generated by individual contributions based on a more general, not necessarily summative, production technology. For further insights, interested readers may consult works such as \cite{Cornes2016} and \cite{buchholz2017theory}.

Using aggregative games, \citet{jensen2022diversity} examines the effect of diverse incentive systems on team performance within heterogeneous teams characterised by varying skill sets and expertise levels. Although Jensen's approach shares similarities with ours, it differs in that his analysis posits that contributions from \textit{cognitively} similar members in a team are additive. The non-linearities in his model arise from the complexity of coordinating efforts among members of different cognitive categories\footnote{\citet{jensen2022diversity} defines a cognitive category as a specific combination of approaches, perspectives, and abilities.} within diverse teams, rather than variations in task type.

\subsection{Social psychological theories of group productivity}\label{subsec:social_psychology}

In the field of psychology, social dilemmas, especially the public goods dilemma, have been thoroughly examined through interdependence theory \citep{de1999trust, balliet2013trust}. The theory of interdependence, developed by \citeauthor{lewin1948resolving}, views groups as dynamic entities shaped by the interdependence among their members \citep{lewin1948resolving}. 
Understanding teamwork as an instance of the public goods dilemma, we elaborate on a renowned theory of team effectiveness aligned with interdependence theory: \citeauthor{steiner1972group}'s model \citep{steiner1972group}. While we acknowledge that team effectiveness encompasses more than mere productivity \cite{mcgrath1964social}, in this paper, we focus on productivity as the metric to evaluate teamwork \cite{campion1993relations, kerr2004group}. Plus, we focus our attention on the internal variables within a group that affect and are best predictors of group productivity (\textit{i.e.} closed group paradigm \citep{casal2011psicología}). \citeauthor{steiner1972group}'s model identifies four key elements within a group that influence productivity:

\begin{table}[h]
    \centering    
    \caption{Steiner's task taxonomy based on individual contributions and group outcome \citep{steiner1972group}.}
    \label{tab:steiner_task_taxonomy}
    \begin{tabular}{|c|p{5cm}|p{5cm}|}
        \hline
        \rowcolor[HTML]{E3E3E3} \textbf{Task Type} & \textbf{Characteristics} & \textbf{Example} \\
        \hline
        Additive & Group success results from the sum of individual efforts, with each member's contribution adding to the collective outcome. & A group project where each member researches and writes a specific section, contributing to a comprehensive report. \\
        \hline
        Conjunctive & Group performance is determined by the contribution of the group member who performs most poorly. & In times of disaster, community survival depends on each person fulfilling their duty. Heroism emerges when any individual can become the weak link in dire circumstances \cite{hirshleifer1983weakest, caparros2020corona}. \\
        \hline
        Disjunctive & Group performance is determined by the contribution of the member who performs the most. & Imagine multiple anti-missile batteries defending a city against a single incoming nuclear-armed ballistic missile. Success hinges on whether the best shot can destroy the incoming threat \cite{hirshleifer1983weakest}. \\
        \hline
        Discretionary & Members' contributions are combined through weighted averaging, often in tasks involving judgment. & When tasked with estimating the temperature of a room, group members engage in discussion, and their final consensus is usually an implicit weighted average of their opinions, where the weights signify the importance of each member's opinion within the team.\\
        \hline        
    \end{tabular}

\end{table}
\begin{enumerate}
\item \textbf{Individuals' profiles:} Individuals contribute various resources to the task, ranging from skills and expertise levels to effort and personality traits.

\item \textbf{Task Type:} Steiner categorised tasks based on how individual contributions combine to shape the group's performance (see Table \ref{tab:steiner_task_taxonomy}); which parallels \citeauthor{cornes2007weak}'s observation that the nature of the public good dictates how the team resources should be combined to achieve a favourable collective outcome. For instance, in activities like tug-of-war, each participant's physical effort directly contributes to the total force exerted. However, dynamics are different in scenarios like marching soldiers or an orchestra, where the slowest or least proficient member can determine the overall performance. Steiner's task taxonomy is still relevant, explaining how the nature of a collective task determines the integration of individual contributions into the final outcome. Additionally, he distinguished between \textbf{unitary} tasks (tasks that cannot be subdivided into smaller subtasks and require a single skill for completion) and \textbf{divisible} tasks (tasks that are readily divided into subtasks, each performable by a different individual, usually involving multiple skills). In this paper, we focus solely on unitary tasks.

\item \textbf{Group Interaction Processes:} Team members' preferences and intentions drive their action choices. Introducing an evaluation system that assesses group production will most possibly impact those preferences, reshaping the interdependencies among members and thereby affecting productivity. \citet{steiner1972group} identified three main types of payoff systems: \emph{promotive} or cooperative (each team member receives a high payoff when their behaviour is highly beneficial to their partners and to themselves), \emph{contrient} or competitive (team members exert contrary effects upon one another's payoff), and \emph{independent} (where each team member's payoff is unaffected by the rest of the team's actions). In this paper, we focus on the first paradigm, studying the impact of introducing incentives for cooperation during group evaluation. 

\item \textbf{Group Outcome:} Individual decisions are influenced by the foreseen group's outcome, particularly whether the group's outcome will satisfy the individuals' preferences or not.
\end{enumerate}

Steiner clearly distinguished between potential performance, representing a group's capabilities, and actual performance, reflecting its accomplishments. If somehow one knew a team's potential performance, its actual performance would be obtained by subtracting some ``lost'' processes from its potential performance. These processes refer to any factor hindering the group from reaching its full potential. For instance, Steiner labelled as \emph{motivation loss} the situation when one or several members exert less effort than their potential in a collective task. One factor contributing to motivation loss is \textit{social loafing}, a free-rider behaviour characterised by decreased individual contributions when working in a group compared to working alone, assuming others are also working \citep{kerr1983motivation, karau1993social, cornes1996theory, fehr1999theory}. Conversely, \textit{social compensation} entails individuals putting in extra effort when anticipating lower performance from group members. This phenomenon arises from the desire to compensate for perceived skill or motivation gaps among colleagues to complete a task successfully. The literature on social compensation suggests that individuals are more likely to work harder in a group setting if \textbf{1)} at least one group member anticipates insufficient effort from others for success, and \textbf{2)} the task holds significant importance to those individuals \cite{williams1991social}.

\subsection{Multiagent multi-armed bandits (MA-MAB) and decision-making}\label{subsec:MA-MAB} 

Human behaviour, particularly decision-making, can be likened to a search problem of actions that lead to favourable outcomes. Decision-makers must navigate the balance between exploring new options and exploiting known ones. Researchers often employ multi-armed bandit (MAB) models to depict human decision-making within exploration-exploitation tasks. These models represent decision-makers faced with multiple options, each associated with uncertain rewards \citep{wu2018generalization, reverdy2014modeling}. The classical bandit problem simplifies decision-making into a one-move one-person game with random payoffs for each action. In a $k$-armed bandit problem some action payoffs remain unknown, and observing the payoff from one action provides no information about the distributions governing the payoffs of other choices \citep{sutton2018reinforcement}. Specifically, each action (arm) $a$ is associated with an expected reward if selected, denoted by $q(a)$, which represents the action's value.

Each agent maintains an estimate of the value of action $a$ at time $t$, denoted as $Q_t(a)$. The value estimates are computed incrementally. Let $Q_t$ denote the estimate after $t-1$ previous selections for a particular action. Given this estimate and the $t^{\text{th}}$ reward $R_t$, the value estimate is updated using the following rule:

\begin{equation}\label{eq:update_rule}
    Q_{t+1} = Q_t + l_r^t \cdot (R_t - Q_t)
\end{equation}

In this equation, $l_r \in \left(0,1\right]$ is the learning rate parameter and modulates the influence of new rewards on the estimated value. Higher learning rates imply greater adaptability to new information, while lower rates prioritise stability in prior estimates. We will pick a learning rate with a decay schedule that ensures asymptotic stochastic convergence, meeting the conditions:

\[
\begin{cases}
    \sum \limits_{t = 0}^{\infty}l_r^t = \infty \\
    \sum \limits_{t = 0}^{\infty}(l_r^t)^2 < \infty
\end{cases}
\]

\citet{leslie2005individual} elaborate on how employing action selection methods based on smooth best responses can steer a multiagent multi-armed bandit system toward a set of strategies approaching a Nash equilibrium. Specifically, the soft-max action selection method \citep{sutton2018reinforcement} facilitates this process, calculating the probability of selecting an action based on its estimated value and a temperature parameter $\tau > 0$:
\[
\mathbb{P}(a) = \frac{\mathrm{e}^{Q_t(a)/(Q_t^\text{max}\cdot \tau)}}{\sum \limits_a \mathrm{e}^{Q_t(a)/(Q_t^\text{max}\cdot \tau)}}
\]

At time $t$, $Q_t^\text{max}$ is the value of the best-valued action(s). This action selection method implements smooth best responses. The use of smooth best responses implies that Nash Equilibria (NE) are no longer fixed points in the strategy space. Instead, the algorithm will converge to Nash distributions. Nash distributions are joint strategies where each agent plays smooth best responses to the rewards arising during interaction with others. According to \citet{govindan2003short}, NE of a game are approximated by Nash distributions for sufficiently low temperatures. Therefore, if $\tau$ is sufficiently low and the agents interact for a sufficiently long period, the learning algorithm can be blindly applied in any game. If convergence occurs, a Nash distribution must have been reached \citep{leslie2005individual}.

\section{Background: generalised aggregative games}\label{sec:background}

In this section, we introduce key concepts from aggregative games that underpin one of our main contributions developing a theoretical framework to investigate teamwork formally. In this introductory section, we use intentionally broad and standard notation to ensure compatibility with most work in the field. However, as we delve into the formulation of our specific problem (Section \ref{sec:model}), certain adaptations to the notation will be necessary to accommodate the nuances of our model. For a comprehensive review of aggregative games, the interested reader may refer to \citet{jensen2018aggregative} and \citet{corchon2021aggregative}.

Throughout, we will only consider games with a finite number of players, and a set of $n$ players will be denoted by $\mathcal{I}=\{1,\ldots,n\}$. Unless otherwise stated, we adhere to the following notation: each player $i$ possesses an action or strategy $d$-dimensional set $\mathcal{A}_i \subseteq \mathbb{R}^d$, with $a_i \in\mathcal{A}_i $ denoting a typical element. A joint action set is denoted $\mathcal{A}\equiv \prod_{i=1}^{n}\mathcal{A}_i$, with a joint action written as $\textbf{a} = \left(a_1, \ldots, a_n\right) \in \mathcal{A}$. For any fixed player $i$, the vector of opponents' joint actions is $\textbf{a}_{-i}= \left(a_1, \ldots,a_{i-1}, a_{i+1},\ldots, a_n\right) \in \mathcal{A}_{-i}$, with $ \mathcal{A}_{-i}\equiv \prod_{j\neq i}\mathcal{A}_j$. We will consider that player $i$'s payoff function depends on their individual action and the vector of their opponent's joint actions: $u_i:\mathcal{A}_i \times \mathcal{A}_{-i}\rightarrow \mathbb{R}$. A (pure strategy Nash) equilibrium is a joint strategy $\hat{\textbf{a}} = (\hat{a}_i,\hat{\textbf{a}}_{-i})$ such that:

\begin{equation}\label{eq:Nash_Equilibrium}
    u_i(\hat{a}_i,\hat{\textbf{a}}_{-i}) \geq u_i(a_i, \hat{\textbf{a}}_{-i}) \; \forall a_i \in \mathcal{A}_i \; \text{, } i \in \mathcal{I}
\end{equation}

Aggregative games are a subset of pure strategy non-cooperative games \citep{jensen2018aggregative}. Their distinctive feature is that, given a player $i$, their utility function $u_i$ can be simplified into a \textit{reduced} utility function of $i$'s action $a_i$ and of some finite aggregate of the set of player's joint actions, $G(\textbf{a})$:

\begin{equation}
    \widetilde{u_i}(a_i, G(\textbf{a}))) \equiv u_i(a_i, \textbf{a}_{-i}) 
\end{equation}
Depending on how the aggregate $G(\textbf{a})$ is computed, aggregative games can be linearly aggregative, generalised aggregative or quasi-aggregative. We will focus on the second type:
\begin{definition}[Generalised Aggregative Game \citep{jensen2018aggregative}]\label{def:generalized_aggregative_game}
A non-cooperative game $\Gamma = \left(\mathcal{A}_i, \widetilde{u_i}\right)_{i\in \mathcal{I}}$ where $\mathcal{A}_i \subseteq \mathbb{R}^{d} \: \forall i$ is termed generalised aggregative if:
\begin{enumerate}
    \item There exists an aggregator or social composition function $G: \mathcal{A}\rightarrow \mathbb{R}$ (the \textit{aggregator}) and a reduced utility function $\widetilde{u_i}:\mathcal{A}_i \times \mathbb{R}$.
    
    \item The aggregator $G$ is an additively separable function: $G(\textbf{a}) = H(\sum_{i=1}^n h_i(a_i))$ where $H:\mathbb{R}\rightarrow \mathbb{R}$ and $h_i:\mathcal{A}_i\rightarrow \mathbb{R}$ are strictly increasing functions. 
    
    \item Each player's utility function $u_i$ can be re-written using their reduced utility function $\widetilde{u_i}$:
    \[
        u_i(\textbf{a}) =\widetilde{u_i}(a_i, G(\textbf{a}))    
    \]
     
    \item The values in the range of $G$ are the set of possible aggregates:
    \[
    \mathcal{G} \equiv \{G(\textbf{a}):\textbf{a}\in \mathcal{A}\}
    \]
    An aggregate is considered an equilibrium aggregate if $\hat{\textbf{a}}$ represents a pure strategy NE for the game.
\end{enumerate}
\end{definition}

During the computation of equilibria in generalised aggregative games, a crucial tool emerges: the \emph{replacement correspondence}, closely linked to the concept of the \emph{best-response correspondence}. Simply put, the best-response of a player $i$ represents their optimal strategy in response to their opponents' strategies:

\begin{definition}[Best-response correspondence \citep{jensen2018aggregative}]\label{def:best-response_correspondence}
    In a generalised aggregative game $\Gamma = \left(a_i, \widetilde{u_i}\right)_{i\in \mathcal{I}}$, player $i$'s best-response correspondence, denoted $b_i:\mathcal{A}_{-i}\rightarrow 2^{\mathcal{A}_i} \cup \varnothing$, is given by:
    \begin{equation}\label{eq:best-response_correspondence}
        b_i(\textbf{a}_{-i}) =  \argmax_{a_i \in \mathcal{A}_i} \widetilde{u_i}\left(a_i, H\left(\sum_{j=1}^n h_j(a_j)\right)\right)
    \end{equation} 
\end{definition}

By writing $\sum_{j\neq i}h_j(a_j)$ as the sum of all players' strategies except for player $i$, we can define player $i$'s reduced best-response as a function of this sum:
    \begin{equation}\label{eq:reduced_best_response}
        \widetilde{b_i}\left(\sum_{j\neq i}h_j(a_j)\right) = \argmax_{a_i \in \mathcal{A}_i} \widetilde{u_i}\left(a_i, H\left(h_i(a_i) + \sum_{j\neq i}h_j(a_j)\right)\right)
    \end{equation}       

By construction, $b_i(\textbf{a}_{-i}) = \widetilde{b_i}(\sum_{j\neq i}h_j(a_j)) \: \forall \textbf{a}_{-i}\in \mathcal{A}_{-i}$. In other words, a player’s best-reply is a function of the aggregate of the other players' actions.

The best-response function tells us the optimal action for a player considering their opponents' actions. Conversely, the replacement correspondence asks: given an outcome aggregate $G$, what actions could player $i$ have taken that are consistent with $G$ being the aggregate? In essence, it helps us identify the feasible actions for player $i$ when we know the value of the outcome aggregate.

\begin{definition}[Replacement Correspondence \citep{jensen2018aggregative, acemoglu2013aggregate}]\label{def:replacement_correspondence}
Let $\Gamma = \left(\mathcal{A}_i, \widetilde{u_i}\right)_{i\in \mathcal{I}}$ be a generalised aggregative game. The replacement correspondence for player $i$, denoted as $r_i:\mathcal{G} \rightarrow 2^{\mathcal{A}_i} \cup \varnothing$, is defined as:
    \begin{equation}\label{eq:replacement_correspondence}
      r_i(G) \equiv \{a_i \in \mathcal{A}_i: a_i \in \widetilde{b_i}(H^{-1}(G)-h_i(a_i))\}
    \end{equation}
\end{definition}
Where $G$ is some aggregate value and we have used that $\sum_{j\neq i}h_j(a_j) = H^{-1}(G) - h_i(a_i)$.

The aggregate replacement correspondence, denoted as $R$, expands on this concept. It computes the set of possible outcome aggregates resulting from each player's replacement correspondence over a given aggregate value. This concept proves instrumental in finding a Nash Equilibrium (NE) of an aggregative game, particularly as the number of players increases:

\begin{definition}[Aggregate Replacement Correspondence \citep{acemoglu2013aggregate}]\label{def:aggregate_replacement_correspondence} Let $\Gamma = \left(\mathcal{A}_i, \widetilde{u_i}\right)_{i\in \mathcal{I}}$ be a generalised aggregative game. The aggregate replacement correspondence, denoted as $R:\mathcal{G} \rightarrow 2^{\mathcal{G}}\cup \varnothing$, is defined as:
\begin{equation}\label{eq:aggregate_replacemente_correspondence}
    R(G) \equiv \{G(\textbf{a}) \in \mathcal{G}: a_i \in r_i(G) \: \forall i \in \mathcal{I}\}
\end{equation}
\end{definition}

\citet{jensen2018aggregative} uses the aggregate replacement correspondence to define an equilibrium aggregate through the following proposition:

\begin{proposition}\label{prop:eq_aggregate}
    An aggregate $\hat{G}$ is an equilibrium aggregate if and only if it is a fixed point of the aggregate replacement correspondence, expressed as:

    \begin{equation}\label{eq:fix_point}
        \hat{G} \in R(\hat{G})
    \end{equation}
\end{proposition}

Thus, to investigate the equilibria of a generalised aggregative game, we can proceed as follows:
    \begin{enumerate}
        \item Compute $R(G)$ as in Eq. (\ref{eq:aggregate_replacemente_correspondence}).
        \item Obtain the equilibrium values $\hat{G}$ as the fixed points of $R(G)$: $\hat{G} \in R(\hat{G})$
        \item By Proposition \ref{prop:eq_aggregate}, any fixed point $\hat{G}$ determines a set $\{\hat{\textbf{a}}\}$ of (possibly unique) pure strategy Nash equilibria, which we compute by applying the set of replacement correspondences $\{r_i\}_{i \in \mathcal{I}}$ over $\hat{G}$:
        \begin{equation}
            \{\hat{\textbf{a}}\} = \{(\hat{a}_1, \ldots, \hat{a}_n) \in r_1(\hat{G}) \times r_2(\hat{G}) \times \ldots \times r_n(\hat{G})\}
        \end{equation}
    \end{enumerate}

We will see that in some contexts, it is convenient to compute equilibria by means of the \textit{share} correspondence:
\begin{definition}[Share Correspondence]\label{def:share_function}
    Let $\Gamma = \left(\mathcal{A}_i, \widetilde{u_i}\right)_{i\in \mathcal{I}}$ be a generalised aggregative game where player $i$ has a replacement correspondence $r_i(G)$. Then, for any aggregate $G >0$, the share correspondence $s_i(G): \mathcal{G}\rightarrow 2^{\left[0,1\right]}$ is given by:

    \begin{equation}
        s_i(G) = \left\{\frac{h_i(a_i)}{H^{-1}(G)} : a_i \in r_i(G) \right\}
    \end{equation}
\end{definition}

Rather than indicating the absolute contribution that player $i$ makes to the teamwork outcome (as $r_i$ does), this correspondence provides insights into the \textit{relative} contribution by $i$ concerning the total teamwork outcome. Whenever the replacement correspondence is a function, the share correspondence is also a function. According to \citet{hartley2000share}, $i$'s share function answers the following question:

\textit{Given an aggregate value $G$, is there a proportion, $b \in [0, 1]$, such that, if $b\cdot G$ were taken away from $G$, player $i$'s best response to the remaining quantity, $(1 - b)\cdot G$, would precisely compensate for the proportion $b$?}

Suppose we can define a replacement and share function for every player in a game. Then, a strategy profile $\hat{\textbf{a}} = (\hat{a}_1, \ldots, \hat{a}_n)$ is an equilibrium with equilibrium level $\hat{G}$ if and only if 
\begin{equation}
    S(\hat{G}) = \sum_{j=1}^N s_j(\hat{G}) = 1
\end{equation}

where $ S(\hat{G})$ is the \textit{aggregate share function} \citep{jensen2018aggregative}. 

Finally, it is pertinent to introduce the concept of normal good, given its relevance for this paper:

\begin{definition}[Normal Good]\label{def:normal_goods}
A \emph{normal good} is a type of good\footnote{A good is a commodity or service that can be utilised to satisfy human wants and that has exchange value.} whose demand increases as consumer income rises. Given a good $x$ and a consumer's income $y$, the Income Elasticity of Demand $\xi$ measures the sensitivity of demand for $x$ to changes in $y$:
    \begin{equation}\label{eq:income_elasticity}
        \xi = \frac{\Delta x/x}{\Delta y/y}
    \end{equation}

We say that a good is normal if $0 < \xi < 1$. 

\end{definition}

In essence, normal goods, like food and clothing, demonstrate a positive correlation between demand and income.

\section{Teamwork games}\label{sec:model} 

Having presented generalised aggregative games, we build on this framework to establish a model of human teamwork. Transitioning from the broader scope of generalised aggregative games, we now focus specifically on general public good games, as outlined by \citet{cornes2007weak}. A general public good game is a generalised aggregative game where the aggregator $G$ (also known as the public good or the social composition function) is a Constant Elasticity of Substitution (CES) function of the players' strategies: :
\begin{equation}\label{eq:CES}
    G = \left(\sum_{i=1}^n \beta_i \cdot g_i^\rho\right)^{\frac{1}{\rho}}
\end{equation}

where $\rho \neq 0$ is the substitution parameter, $g_i$ is player $i$'s contribution to the public good and $\beta_i$ represents the weight given to player $i$'s contribution: $\beta_i> 0 \: \forall i$. In a CES function, the substitution parameter $\rho$ remains consistent across all players. We will see that together with the set of parameters $\{\beta_i\}_{i\in \mathcal{I}}$, it characterises the \textbf{task type}: additive, conjunctive, disjunctive or discretionary. 

Notice that in the context of general public good games, the strategies consist of gifts or contributions to the public good and are usually denoted by $g_i$. In line with the definition of a generalised aggregative game (Def. \ref{def:generalized_aggregative_game}), we see that in this case, $G = H(\sum_{i=1}^n h_i(g_i))$ with $H(z) = z^{1/\rho}$ and $h_i(g_i) = \beta_i \cdot g_i^\rho$. Furthermore\footnote{In this paper, we denote by $\mathbb{R}^{+}$ the set of positive real numbers, $\mathbb{R}_0^{+}$ the set of non-negative real numbers including zero, and $\mathbb{R}^{*}$ the set of all real numbers except zero.}, $g_i \in \mathbb{R}_0^{+}$.

This section introduces a teamwork model based on general public good games: teamwork games. The fundamental idea is that the output of teamwork can be seen as a public good to which team members have the option to contribute. When individuals work together, they subliminally face a dilemma: to spend their time on activities that yield individual benefits (\textit{e.g.}, leisure) while delegating the task to others, or to strive for the common good (at the expense of their time and energy). The former choice carries the risk that everyone opts for the same strategy, leading to the task not being completed, with subsequent penalties from a manager. The latter entails the risk of being the sole bearer of the collective enterprise on one's shoulders (sucker's payoff). Players' agency is manifested through the time they choose to allocate to the task. Therefore, in a teamwork game, a player $i$'s strategies or actions are now abstract and dimensionless, representing the percentage of a turn that the player allocates to the task. The contribution or gift $g_i$ is calculated based on this action, but it is not a strategy per se, but a consequence of it. Thus, adjustments to the standard notation presented in \ref{subsec:aggregative_games} are necessary to suit the specifics of our model.
In the remainder of this section, we carry out those adjustments and elucidate the pivotal elements of our model.

\subsection{Model}
We adopt productivity as our metric to evaluate teamwork, drawing from \citeauthor{steiner1972group}'s theoretical foundation \citep{steiner1972group}. Given a unitary task, we assume an objective and quantitative measure of the team's productivity in terms of work units, denoted by $G$ and given by Eq. (\ref{eq:CES}).

\begin{definition}[Unitary Team Task]\label{def:task}
Tasks can be completed through a sequence of rounds or turns. A unitary team task $\mathcal{T}$ is characterised by a scalar $G^{\mathcal{T}} \in \mathbb{R}_0^{+}$ representing the team productivity (in work units) needed to complete the task, a pair $\left[\Delta_t, N\right] \in \mathbb{N}^{2}$ indicating \textbf{1)} the length (in time units) of each turn and \textbf{2)} the duration (in turns) of a task, and a set of parameters $\Theta^\mathcal{T} = \left(\rho, \beta_1, \ldots, \beta_n\right)$ (with $\rho \in \mathbb{R}^{*}$ and $\beta_i \in \mathbb{R}^{+}\: \forall i$) which specify the task's typology ( additive, conjunctive, disjunctive, or discretionary):

\[
\mathcal{T} := (G^{\mathcal{T}},\left[\Delta_t, N\right],\Theta^\mathcal{T})
\]
\end{definition}
Given a vector of contributions to the task $\left(g_1, \ldots, g_n\right)$ resulting in $G$ work units as per Eq. (\ref{eq:CES}), the team task will be deemed accomplished after $t$ turns if the team's productivity units suffice to complete the task within the designated timeframe. Formally, the following conditions must be met simultaneously:
\begin{equation}\label{eq:end_conditions}
    \left\{\begin{matrix}
        G\geq G^{\mathcal{T}}  \\
        t \leq N \\
    \end{matrix}\right.
\end{equation}

Players exhibit varying levels of proficiency in completing tasks. When two players with different ability levels dedicate the same amount of time to a task, their contributions will inevitably differ. More precisely, we define \textit{expertise} as the derivative of productivity with respect to time:

\begin{definition}[Expertise Concerning a Unitary Task]\label{def:expertise}
For a player $i$, their expertise $p_i^{\mathcal{T}} \in \left[0,1\right]$ regarding a unitary task $\mathcal{T}$ quantifies the amount of work $i$ can accomplish per unit of time allocated to that task:

\begin{equation}\label{eq:expertise}
   p_i^{\mathcal{T}} = \left(\frac{\mathrm{d}G }{\mathrm{d} t}\right)_{i} \geq 0
\end{equation}

When a team consists of players with varying levels of expertise, we describe it as \textbf{heterogeneous}. Conversely, if all players have the same level of expertise, the team is described as \textbf{homogeneous}. We will describe a team's joint expertise by the vector of expertise levels: $\textbf{p} = (p_1, \ldots, p_n)$.

\end{definition}

Each turn $t$, players' agency manifests through the percentage $a_i$ of that turn they choose to devote to the task. Therefore, a player $i$'s strategies or actions $\mathcal{A}_i \subseteq \mathbb{R}_0^{+}$ are abstract and dimensionless, representing the percentage of the total turn that a player allocates to the task. A joint action is then a vector of these strategic choices: $\mathbf{a} = (a_1, \ldots, a_n) \in \mathcal{A}$, where $ \mathcal{A}\equiv \prod \limits_{i=1}^{n}\mathcal{A}_i$. Consequently, a player $i$ with expertise $p_i^{\mathcal{T}}$ who chooses an action $a_i$ in a turn of duration $\Delta_t$, makes a contribution or gift of $g_i$ work units:

\begin{equation}\label{eq:individual_contribution}
  g_i = a_i \cdot p_i^{\mathcal{T}}  \cdot \Delta_t
\end{equation}

In a teamwork game, the aggregative public good $G$ represents the teamwork outcome and is given by a Constant Elasticity of Substitution (CES) function just like in a general public good game. The only difference is that now the contributions are computed as in Eq. (\ref{eq:individual_contribution}):

\begin{equation}\label{eq:CES_in_games_of_group_productivity}
    G = \left( \sum \limits_i^n \beta_i \cdot g_i^\rho\right)^{\frac{1}{\rho}} =\left( \sum \limits_i^n \beta_i \cdot \left(a_i \cdot p_i^{\mathcal{T}}  \cdot \Delta_t\right)^\rho\right)^{\frac{1}{\rho}}
\end{equation}

Setting the parameters of this function appropriately, we obtain a teamwork model incorporating the main ingredients of \citeauthor{steiner1972group}'s theory of group productivity:

\begin{enumerate}
    \item $\beta_i$ represents the weight given to the contribution of team member $i$.
    
    \item Together with the set of parameters $\{\beta_i\}_{i\in \mathcal{I}}$, the substitution parameter $\rho$ characterises the \textbf{task type}: additive, conjunctive, disjunctive or discretionary (Section \ref{sec:preliminaries}). It remains consistent across all players. 

    \item \textbf{Additive} tasks are those where $\rho = \beta_i = 1 \: \forall i$. In this case, all contributions to the public good are aggregated additively. From the point of view of game theory, this setting reverts to the canonical public goods model \citep{bergstrom1992uniqueness, cornes2007aggregative}.
    
    \item \textbf{Conjunctive} tasks correspond to $\rho < 1$: the lower a contribution within the set of players, the higher its influence on the team outcome. As $\rho \rightarrow -\infty$ with $\beta_i = 1 \: \forall i$, the task becomes strongly conjunctive, where the poorest individual performance entirely determines the team productivity.  Mathematically, an aggregative public good with $\rho < 1$ is concave in $g_i$ \citep{cornes2007weak}.
    
    \item \textbf{Disjunctive} tasks are with $\rho > 1$: the higher a contribution within the set of players, the higher its influence on the team outcome. As $\rho \rightarrow +\infty$ and $\beta_1 = \beta_2 = \ldots = \beta_n = 1$, the task becomes strongly disjunctive, where the best individual performance entirely determines the team productivity. In other words, the team's result is determined by the player who generates the most work units $g_i$. Mathematically, an aggregative public good with $\rho > 1$ is not concave in $g_i$ \citep{cornes2007weak}. Additive, conjunctive, and disjunctive tasks have been far more extensively studied than discretionary ones in the literature of social psychology, both in theoretical and experimental contexts. Consequently, we will solely focus on these three typologies, leaving the study of discretionary tasks for future work. 
    
    \item Finally, a \textbf{discretionary} task corresponds to a CES function with $\rho = 1$ and $\beta_1, \ldots, \beta_n > 0$, with at least some $i,j \in \mathcal{I} \mid \beta_i \neq \beta_j$. This means that the team outcome is a weighted sum of the individual contributions.

\end{enumerate}
Just as expertise measures the productivity of a player per unit of time, a player's \textit{leisure capacity} $p_i^{L}$ measures a player's ability to enjoy their free time. By leisure, we mean any activity that a player can be involved in while delegating the task to others and that yields individual benefits. We assume that we can quantify leisure in terms of work units $\mathcal{L}$, similar to how we do for work-related tasks. The only distinction is that this work stems from a player investing their spare time in selfish activities.

\begin{definition}[Leisure Capacity]\label{def:leisure_capacity}
Leisure capacity refers to a player's ability to efficiently obtain leisure units during their free time. The leisure capacity value, denoted as $p_i^{L}\in \left[0,1\right]$, is calculated using the equation below:

\begin{equation}\label{eq:leisure_capacity}
    p_i^{L} = \left(\frac{\mathrm{d}\mathcal{L} }{\mathrm{d} t}\right)_{i} \geq 0
\end{equation}

In other words, the formula calculates how much leisure a player can benefit from when investing their free time in private activities. Similar to expertise, leisure capacity is expressed in units of work per unit of time and influences an individual's allocation of free time between team tasks and leisure pursuits.
\end{definition}

With this definition, we can write the private contribution $x_i$ representing leisure units allocated to private activities by $i$ as:

\begin{equation}\label{eq:private_good}
    x_i = (1-a_i) \cdot p_i^L \cdot \Delta_t  = p_i^L \cdot \Delta_t - g_i \frac{p_i^L}{p_i^{\mathcal{T}}} 
\end{equation}
Where in the second equality we have used Eq. (\ref{eq:individual_contribution}). Notice how in economical terms, $\Delta_t \cdot p_i^L$ plays the role of $i$'s exogenous income. This expression of the private contribution allows us to write the following budget constraint for teamwork games:

\begin{assumption}[Individual budget constraints]\label{assum:budget}
    
Suppose $i$'s exogenous income is $\Delta_t \cdot p_i^L$. Then, player $i$’s budget constraint requires that
    \[
        \Delta_t \cdot p_i^L =  x_i + g_i \frac{p_i^L}{p_i^{\mathcal{T}}}
    \]
    
\end{assumption}

Going forward, we denote $\Theta^\mathcal{T}$ as $\Theta$ unless specified otherwise. As already stated in Section \ref{sec:introduction}, in this paper we use the full array of tools in aggregative games to analyse teamwork as a one-shot game. Thus, we will concentrate on one-shot tasks: $\mathcal{T}= (G^{\mathcal{T}},\left[\Delta_t, N=1\right],\Theta)$ . Furthermore, for the sake of simplicity, we adopt the convention that $p_i^L = 1$ for all $i$ (\textit{i.e.}, we assume that all players are capable of fully enjoying their free time):
    \begin{equation}\label{eq:exogenous_income_group_productivity}
         \Delta_t \equiv x_i + \frac{g_i}{p_i^{\mathcal{T}}} 
    \end{equation}
    
Solving for $x_i$ in Eq.(\ref{eq:exogenous_income_group_productivity}) and adding $G$ to both sides yields what is known as player $i$'s full income $M$:

\begin{definition}[Full Income]\label{def:full_income}
A player's full income comprises the sum of their free time and the accessible public good, encompassing all resources available to the player.
    \begin{equation}\label{eq:full_income_group_productivity}
        M \overset{\underset{\mathrm{def}}{}}{=} x_i + G \equiv \Delta_t - \frac{g_i}{p_i^{\mathcal{T}}} + (G_{-i}^\rho + \beta_i g_i^\rho)^\frac{1}{\rho}   
    \end{equation}
where we have used that $G = (G_{-i}^\rho + \beta_i g_i^\rho)^\frac{1}{\rho}$. Notice that $M$ incorporates the contributions of others explicitly as a component of player $i$’s income endowment, which will prove to be useful in forthcoming sections. 

\end{definition}
    
Because teamwork games will be defined as general public good games with further restrictions, it is noteworthy to mention other two assumptions applying to general public good games:

\begin{assumption}[Well-behaved preferences]\label{assum:well-behaved}
    In a general public good game,  the (reduced) utility is a function $\widetilde{u_i}$ of $i$'s private good and of the public good $G$ for all $i$:
        \[\widetilde{u_i}(x_i, G)\]
    This utility is everywhere strictly increasing and strictly quasi-concave in both arguments. It is also continuously differentiable.
\end{assumption}

\begin{assumption}[Normality]\label{assum:normality}
    In a general public good game, both the private good and the public good are normal for every player (Def. \ref{def:normal_goods}).
\end{assumption}

As anticipated in Section \ref{sec:introduction}, one of the contributions of this work is an expansion of \citep{cornes2007weak} that explicitly incorporates diverse teamwork evaluation functions designed to assess team performance. Having presented the key elements of our model, we now formalise the concept of teamwork evaluation. 

\begin{definition}[Evaluation Function]\label{def:assessment_function}
Let $\Gamma = \left(\mathcal{A}_i, \widetilde{u_i}\right)_{i\in \mathcal{I}}$ denote a general public good game, where $\mathcal{A}_i$ represents player $i$'s action set, and $G(\textbf{a}) \in \mathcal{G}$ is the public good (team outcome) that the set of players produces according to Eq. (\ref{eq:CES_in_games_of_group_productivity}). A function $\sigma: \mathcal{G} \rightarrow \left[p,q\right] \subset \mathbb{R}_0^{+}$ is an evaluation function if it satisfies the following properties:

\begin{enumerate}
    \item $\sigma$ is monotonically strictly increasing.
    \item $\sigma$ is differentiable.
    \item $\sigma$ has a second derivative $\sigma^{\prime\prime}$ that satisfies the following inequality for all $G$ in the domain of $\sigma$:
    \[
        \sigma^{\prime}(G)^2 - \sigma^{\prime\prime}(G)\cdot\sigma(G) > 0 
    \] 
\end{enumerate}
The infimum $p$ of $\sigma$ denotes the minimum assessment that teamwork can achieve, while the supremum $q$ represents the maximum attainable assessment.
\end{definition}

The introduction of the evaluation function is crucial in our context because teamwork games differ from general public good games in a significant way: a player's utility depends not only on their contribution but also on how the evaluation function evaluates the team's performance. With all the aforementioned components in place, we can now define a teamwork game as follows:

\begin{definition}[Teamwork Game]\label{def:game_group_productivity}
Consider a set of players $\mathcal{I} = \{1, \ldots, n\}$ engaged in a general public good game $\Gamma = (\mathcal{A}_i, \widetilde{u_i})_{i \in \mathcal{I}}$. Given a one-shot task $\mathcal{T} := (G^{\mathcal{T}},\left[\Delta_t, N=1\right],\Theta)$, an evaluation function  $\sigma: \mathcal{G} \rightarrow \left[p,q\right] \subset \mathbb{R}$ (Definition \ref{def:assessment_function}), a joint expertise $\textbf{p}^{\mathcal{T}}=(p_1^{\mathcal{T}},\ldots, p_n^{\mathcal{T}})$, and a vector $\textbf{p}^{L}=(p_1^L,\ldots, p_n^L)$ representing the set of players' leisure capacities, we define the triplet $(\Gamma, \mathcal{T}, \sigma)$ as a \textit{teamwork game} if it satisfies the following conditions:

\begin{enumerate}

    \item The game is a static (simultaneous), non-cooperative game where strategic choices are represented by the percentages $(a_1,\ldots, a_n) \in \mathbb{R}_0^{+}$ of the total available time $\Delta_t$ allocated by each player to the task.          
    
    \item The strategic choices $(a_1, \ldots, a_n)$ correspond to individual contributions $(g_1,\ldots, g_n)$ through Equation (\ref{eq:individual_contribution}). Thus, $g_i \geq 0 \: \forall i$.
    
    \item The aggregator $G: \mathcal{A}\rightarrow \mathbb{R}$ is a CES function with parameters $\Theta = \{\rho, \beta_1, \ldots, \beta_n\}$:
    \[
        G = \left(\sum_i^n \beta_i \cdot g_i^\rho \right)^{\frac{1}{\rho}}
    \]
    
    \item For a player $i$, their private good $x_i$ is given by Equation (\ref{eq:private_good}).

    \item Player $i$'s reduced utility (or utility for short) is a function of their private good $x_i$ and of the assessment $\sigma(G)$ of the form
    \begin{equation}\label{eq:utility_GoGP}
        \widetilde{u_i}(x_i, \sigma(G))
    \end{equation}

    where $\widetilde{u_i}(x_i, G)$ was the original utility function of the general public good game $\Gamma = (\mathcal{A}_i, \widetilde{u_i})_{i \in \mathcal{I}}$.
\end{enumerate}   
\end{definition}
Building upon Definitions \ref{def:assessment_function} and \ref{def:game_group_productivity}, we introduce the following Lemma to ensure that assumptions \ref{assum:well-behaved} and \ref{assum:normality} about general public good games still apply to teamwork games:

\begin{lemma}\label{lemma:assessment_guarantees}
Consider a teamwork game $\left(\Gamma, \mathcal{T}, \sigma\right)$ as defined in Definition \ref{def:game_group_productivity}. Suppose the individual private contribution $x_i$ and the teamwork outcome $G$ are normal and desirable goods under certain preferences represented by the utility function $\widetilde{u_i}(x_i, G)$ in the general public good game $\Gamma$. Then, the following assertions hold:
\begin{enumerate}
    \item Player $i$'s utility function $\widetilde{u_i}(x_i, \sigma(G))$ in the teamwork game $\left(\Gamma, \mathcal{T}, \sigma\right)$ is differentiable and quasi-concave.
    \item The private contribution $x_i$ and the teamwork outcome $G$ remain normal goods under this utility function.
\end{enumerate}

\end{lemma}

The interested reader can find proof of the above Lemma in \ref{app:assessment_guarantees}. Because of the mapping between $\textbf{a} = (a_1, \ldots, a_n)$ and $\textbf{g}=(g_1, \ldots, g_n)$, a joint action $\textbf{a}$ is fully specified by the latter without ambiguity. Consequently, when computing the Nash Equilibria of a teamwork game in the rest of the paper, we will provide the best responses in terms of gifts $g_i$ instead of percentages of turn $a_i$ unless stated otherwise. Specifically, we will make a slight abuse of notation and assume that the best-response correspondence $b_i(G):\mathcal{A}_{-i} \rightarrow \mathcal{A}_{i}$ and the replacement correspondence (Def. \ref{def:replacement_correspondence}) $r_i(G):\mathcal{G}\rightarrow \mathcal{A}_{i}$ are functions that output individual 
contributions $g_i$ (instead of actions $a_i$). Making this mild departure from notation conventions will provide greater clarity when discussing the impact of introducing an evaluation function into the game.

\section{Equilibria in teamwork games}\label{sec:equilibria}
In this section, we examine the equilibria in teamwork games using the aggregative games toolbox. As stated in Section \ref{sec:preliminaries}, the (reduced) utility function of player $i$ in aggregative games, $\tilde{u}_i$, differs from the usual form of utility in game theory in that its arguments are just two: $i$'s strategy, $a_i$, and some aggregate $G$ of all the other players' strategy profiles. This allows us to compute $i$'s best response as a function (or correspondence) of the aggregate $G$, without needing to distinguish the individual strategic choices of their opponents. 

While this aggregative approach allows computations to scale as the number of players increases, it requires using an alternative to the best-response function, known as the replacement correspondence. The replacement correspondence $r_i(G)$ of player $i$ gives the set of strategies that the player would choose from in an equilibrium where the aggregate of all the contributions (including $i$'s) is $G$. Specifically, if $G = \sum \limits_{i}^n g_i$ and the replacement correspondence is a function, then $g_i = r_i(G)$ satisfies $g_i = b_i(G - g_i)$, where $b_i$ is the traditional best-response function\footnote{In the general case where $G = \left(\sum_i^n \beta_i \cdot g_i^\rho \right)^{\frac{1}{\rho}}$, the corresponding equality would be $r_i(G) = b_i\left(\left[G^{\rho} - \beta_i\cdot g_i^{\rho}\right]^{\frac{1}{\rho}}\right)$}.

We will see that when the task is additive or conjunctive, a set of first-order condition equations can be applied to directly compute the replacement correspondence, and this correspondence will be a function. However, this is not the case with disjunctive tasks: the replacement correspondence will map a set of best strategies, rather than a unique strategy. Furthermore, the first-order conditions will not directly yield the replacement correspondence for disjunctive tasks, necessitating further analysis to find it.

\subsection{Conditions for the replacement correspondence to be defined}\label{subsec:conditions_for_r}

The conditions for the replacement function (or correspondence, where applicable) to be well-defined will vary from one type of task to another. This subsection presents these conditions in an orderly manner: first for additive tasks, then for conjunctive tasks, and finally for disjunctive tasks.
\subsubsection{Additive tasks}\label{subsub:conditions_r_in_additive}

When the task is additive ($\rho = 1$), the teamwork outcome $G$ is a concave function in $g_i$ for all $i$ \citep{cornes2007weak} and our teamwork game has the same aggregate $G$ as a canonical public goods game:

    \begin{equation}\label{eq:BBV_CES}
        G = \sum \limits_{i=1}^n \beta_i \cdot g_i
    \end{equation}

This form of the aggregate together with Lemma \ref{lemma:assessment_guarantees} makes it possible to adopt a result by \citet{cornes2007aggregative} into teamwork games:
Under assumptions [\ref{assum:budget}-\ref{assum:normality}] and with a CES representing an additive task as given by Eq. (\ref{eq:BBV_CES}), the replacement correspondence of player $i$ (see Eq. (\ref{eq:replacement_correspondence})) is a function $r_i: \mathcal{G} \rightarrow \mathcal{A}_i$ with the following properties:

\begin{enumerate}
    \item There exists a finite value, $\overline{G}_i$, at which $r_i(\overline{G}_i) = \overline{G}_i \: \forall i$.
    \item $r_i(G)$ is defined $\forall G \geq \overline{G}_i$.
    \item $r_i(G)$ is continuous.
    \item $r_i(G)$ is everywhere non-increasing in $G$, and is strictly decreasing wherever it is strictly positive.
\end{enumerate}

Where $\overline{G}_i$ represents player $i$’s standalone value, \textit{i.e.}, the level of teamwork outcome that player $i$ would contribute if they were the sole contributor. Put simply: in teamwork games featuring additive tasks, the replacement correspondence used to compute Nash Equilibria becomes a function. The domain of player $i$'s replacement function comprises any value of the teamwork outcome $G$ that equals or exceeds the level they would individually contribute. If $G$ falls below this individual contribution level, $r_i$ will not be defined for player $i$, rendering the Nash Equilibrium uncomputable via the replacement function.
\subsubsection{Conjunctive tasks}\label{subsub:conditions_r_in_conjunctive}
When $\rho < 1$ and $\rho \neq 0$, the task is conjunctive (the lower a contribution within the set of players, the higher its influence in the team outcome) and the teamwork outcome $G$ is a concave function in $g_i$ for all $i$ \citep{cornes2007weak}. We can adapt a central lemma in \citep{cornes2007weak} to determine the circumstances under which a well-defined replacement function exists in a teamwork game where players are involved in a conjunctive task:

\begin{lemma}\label{lemma:existence_weak_link_group_productivity}
    Let $(\Gamma, \mathcal{T}, \sigma)$ be a teamwork game featuring a conjunctive task. For every player, suppose that their private contribution $x_i$ and the team outcome $G$ are normal and desirable goods. Then, the replacement correspondence $r_i(G)$ is a function satisfying the following first-order conditions (FOCs):
    \begin{equation}\label{eq:FOC}
        \left.\frac{\partial \widetilde{u_i}}{\partial g_i}\right|_{\textbf{g}_{-i}} \geq 0
    \end{equation}
            
    with equality whenever $g_i < p_i^{\mathcal{T}}\cdot \Delta_t$. Furthermore, the replacement function is well-defined and has two potential domains, each corresponding to a regime of parameter $\rho$.
    \begin{enumerate}
        \item if $0 < \rho < 1$, $r_i: \left[\overline{G}_i , \infty\right) \rightarrow \mathcal{A}_i$
        \item if $\rho < 0$, $r_i:\left[0, \overline{G}_i\right] \rightarrow \mathcal{A}_i$
    \end{enumerate}

\end{lemma}

In teamwork games featuring conjunctive tasks, the replacement correspondence used to compute Nash Equilibria becomes a function. However, the domain of this function varies based on \textit{how conjunctive} the task is. For quasi-conjunctive tasks where the teamwork outcome depends on the set of the $n$ lowest contributions ($0<\rho <1$), player $i$'s replacement function encompasses any teamwork outcome $G$ equal to or surpassing their standalone level. Conversely, in strongly conjunctive tasks ($\rho <0$), where the teamwork outcome is predominantly determined by the team's lowest contribution, player $i$'s replacement function spans the interval $\left[0, \overline{G}_i\right]$.
The interested reader can find a proof of Lemma \ref{lemma:existence_weak_link_group_productivity} in \ref{app:proof_conjunctive}.

\subsubsection{Disjunctive tasks}\label{subsub:conditions_r_in_disjunctive}
When $\rho > 1$, the task is disjunctive (the higher a contribution within the set of players, the greater its influence in the team outcome) and the team outcome no longer exhibits concavity in $g_i$. This requires an extension of the replacement function to a correspondence \citep{cornes2007weak}.

As warned by \citeauthor{cornes2007weak}, outcomes under non-concave CES functions are notably sensitive to the specific choice of the utility function. In their work, \citeauthor{cornes2007weak} usually concentrate on Cobb-Douglas (CD) preferences, a type of utility function commonly used in economics to describe how consumers allocate their consumption across different goods (in their case, between the private good $x_i$ and the public good $G$):  

    \begin{equation}\label{eq:cobb-douglas_preferences}
        \widetilde{u_i}^{\text{CD}}(x_i, G) = x_i^\alpha \cdot G 
    \end{equation}

with $\alpha > 0$ being the same for all players $i \in \mathcal{I}$ and capturing the relative importance of private consumption compared to the public good. These preferences are particularly suitable for articulating the tension between a player's desire for private good $x_i$ and their desire for the availability of public good $G$: to maximise the utility \(\widetilde{u_i}^{\text{CD}}(x_i, G)\), a balance between \(x_i\) and \(G\) is required. An increase in one of the components will only significantly increase the utility if the other component is also sufficient. A high value of \(\alpha\) indicates that the private good is of relatively greater importance. In contrast, a low value indicates that the public good is more critical to the player's utility.

As expressed in Equation (\ref{eq:utility_GoGP}), teamwork games differ from general public good games in that now a player’s utility hinges on their private contribution $x_i$ and on how the evaluation function appraises the team’s performance, $\widetilde{u_i}(x_i, \sigma(G))$. 
In this section, we narrow our analysis and propose the following utility function, inspired by Cobb-Douglas preferences:

\begin{equation}\label{eq:modified-cobb-douglas-preferences}
    \widetilde{u_i}(x_i, \sigma(G)) = x_i^\alpha \cdot \sigma(G)
\end{equation}
This utility function bears a close resemblance to the Cobb-Douglas function but incorporates the teamwork assessment. It's worth noting that $\alpha >0$ remains uniform across all players, representing the relative significance of leisure to player $i$ compared to the received evaluation. Conversely, $\sigma(G)$ embodies the evaluation function, as outlined in Definition \ref{def:assessment_function}. Decoupling the teamwork outcome $G$ as 
\[
G = \left(\beta_i g_i^\rho + G_{-i}^\rho\right)^{\frac{1}{\rho}}
\]

we can write the best response set for player $i$ in a teamwork game with preferences given by Eq. (\ref{eq:modified-cobb-douglas-preferences}):

    \[
        \mathcal{B}_i(G_{-i}) =  \argmax_{g_i \in [0, \Delta_t p_i^{\mathcal{T}}]} \left\{x_i^\alpha \cdot \sigma\left((\beta_i g_i^\rho + G_{-i}^\rho)^{\frac{1}{\rho}}\right)\right\}
    \]
with $ G_{-i} = \left(\sum \limits_{j\neq i}\beta_j g_j^\rho\right)^{\frac{1}{\rho}}$, and $x_i$ given by Eq. (\ref{eq:private_good}).

Once again, we adapt a lemma by \citet{cornes2007weak} about general public good games to teamwork games to show the existence of three potential behavioural regimes within the set of best responses:

\begin{lemma}\label{lemma:best_response_correspondence_aggregative_games}
    Let $(\Gamma, \mathcal{T}, \sigma)$ be a teamwork game featuring a disjunctive task ($\rho > 1$): 
    
        \[
            G = \left( \sum \limits_i^n \beta_i \cdot g_i^\rho\right)^{\frac{1}{\rho}}\text{with } \rho > 1
        \]
    Suppose that player $i$ has preferences as described in Equation (\ref{eq:modified-cobb-douglas-preferences}), where $\sigma(G)$ denotes an evaluation function (Definition \ref{def:assessment_function}). Then, there exists a threshold value $G_{-i}^{*}>0$ and a positive real-valued function $b_i$ on $\left[0, G_{-i}^{*}\right]$ such that:
    \begin{equation}\label{eq:best_response_set_disjunctive}
        \mathcal{B}_i(G_{-i}) = 
            \begin{cases}
                \left\{0\right\} & \text{if } G_{-i} > G_{-i}^{*} \\
                \left\{0, b_i(G_{-i})\right\} & \text{if } G_{-i} = G_{-i}^* \\
                \left\{b_i(G_{-i})\right\} & \text{if } G_{-i} < G_{-i}^* \\
            \end{cases}
    \end{equation}
\end{lemma}

The interested reader will find a proof of this lemma in \ref{app:proof_lemma}.

Eq.(\ref{eq:best_response_set_disjunctive}) shows that there are three possible behaviour regimes within the set of best responses:

\begin{enumerate}
    \item \textbf{No-contribution regime}: when the provision of the teamwork outcome by others surpasses a certain threshold (\textit{i.e.} exceeds $ G_{-i}^{*}$), player $i$'s best response is not to contribute. 
    \item \textbf{Indifference regime}: if $G_{-i} = G_{-i}^*$, player $i$ becomes indifferent between contributing or abstaining. 
    \item \textbf{Contribution regime:} when the teamwork outcome is suboptimally provided relative to player $i$'s preferences, $i$ will contribute according to their best response. 
\end{enumerate}

\begin{figure}[!h]
    \centering
    \includegraphics[width=0.60\linewidth]{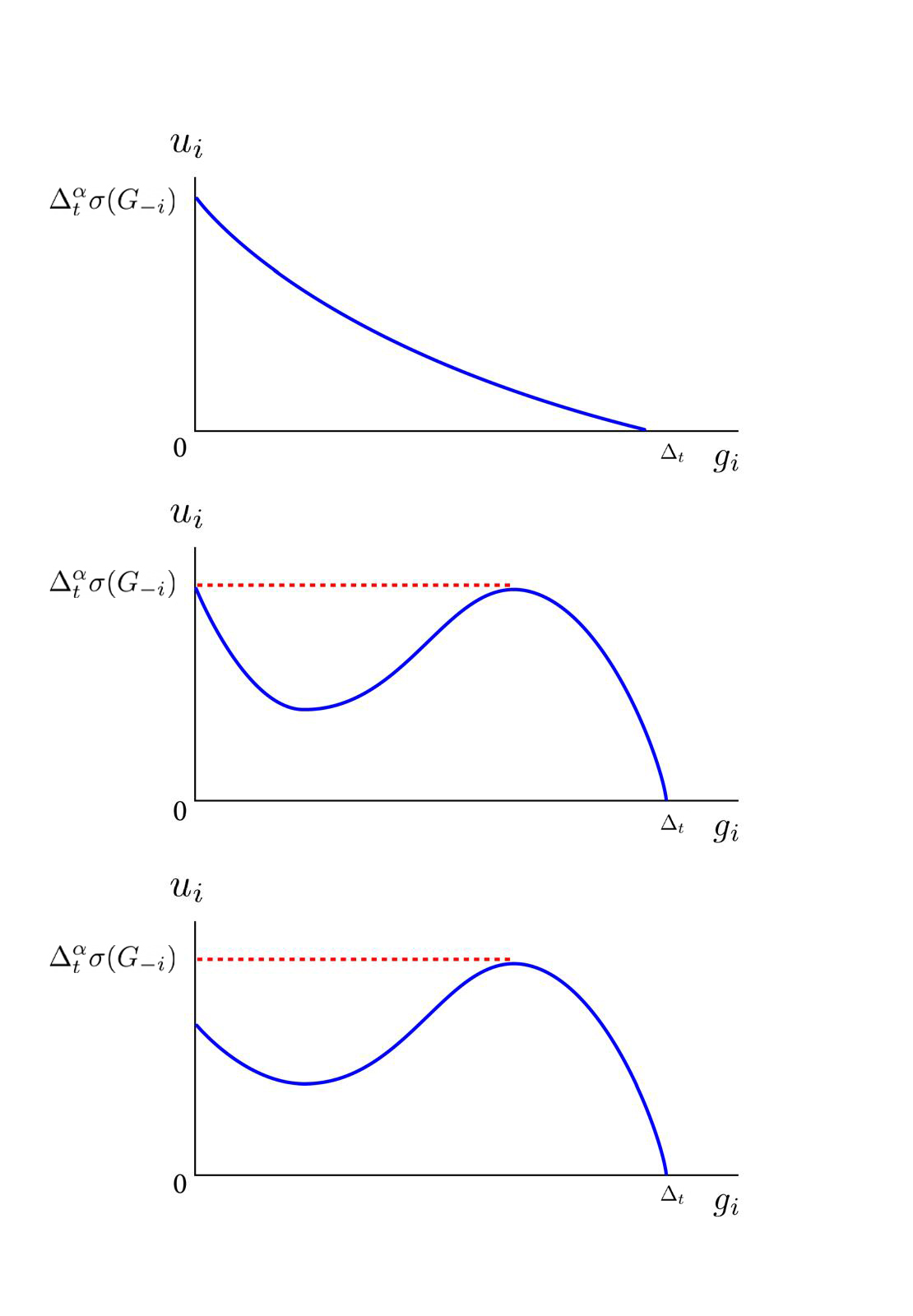}
    \caption{Player $i$'s utility $\widetilde{u_i}(g_i, \sigma(G_{-i}))$, across three regime values of $G_{-i}$. Upper pannel: $G_{-i} > G_{-i}^{*}$. Middle pannel: $G_{-i} = G_{-i}^*$ . Bottom pannel: $G_{-i} < G_{-i}^*$. Adapted from \citep{cornes2007weak}.}
    \label{fig:utility_good_shot}
\end{figure}

We can visually represent player $i$'s utility across these scenarios based on their contribution while holding the provision of the teamwork outcome by others constant (see Fig. \ref{fig:utility_good_shot}). When $G_{-i} > G_{-i}^{*}$ (upper panel), maximum utility occurs at $\widetilde{u_i}(0, G_{-i}) = (\Delta_t\cdot p_i^L)^\alpha \cdot G_{-i}$, where $\Delta_t \cdot p_i^L$ is $i$'s exogenous income and $p_i^L = 1$ for all $i$. At the threshold $G_{-i}^*$, utility maximisation is achieved at two points: $g_i = 0$ and $g_i = b_i(G_{-i}^*)$ (middle pannel). Finally, when $G_{-i} < G_{-i}^*$ (bottom panel), the highest utility is reached for some $b_i(G_{-i}) > 0$.

\subsection{An example}
Before delving into the mathematical formulation of replacement correspondences in teamwork games, we showcase the types of equilibria obtained in teamwork games. We use Eq. (\ref{eq:modified-cobb-douglas-preferences}) to model the players' preferences:
\[
    \widetilde{u_i}(x_i, \sigma(G)) = x_i^\alpha \cdot \sigma(G)
\]

For illustrative purposes, let's assume that enjoying leisure is quadratically more important to the agents than receiving a good evaluation, with $\alpha = 2$ for all players. Furthermore, we take the evaluation function (Def. \ref{def:assessment_function}) to be a logistic function:
\[
    \sigma(G) = \frac{d}{1+ \mathrm{e}^{-\gamma(G-b)}}
\]

The logistic function takes the teamwork outcome \(G\) as input and returns an evaluation within the interval \(\left[0,d\right] \subset \mathbb{R}\). The parameter \(\gamma\) gives a sense of how many units of teamwork outcome are necessary to raise the assessment by one point, and in our case is set to $\gamma = 2$. When \(G=b\), the teamwork outcome receives an evaluation of \(\frac{d}{2}\), making \(b\) the passing point. Table \ref{tab:running_example_NE_teamwork_games} displays the Nash equilibria, calculated numerically, for a heterogenous team with joint expertise $\textbf{p} = (0.3, 0.8)$ across nine teamwork games. These games vary in difficulty, covering easy, medium, and hard levels for additive, conjunctive, and disjunctive tasks.

\begin{table}[!h]
\centering
\caption{Numerically computed Nash equilibria for a heterogeneous team with $\textbf{p} = (0.3, 0.8)$ across nine teamwork games of varying difficulty levels: easy, medium, and hard, for additive, conjunctive, and disjunctive tasks.}
\label{tab:running_example_NE_teamwork_games}
\begin{tabular}{l|c|c|c|}
\cline{2-4}

& \cellcolor[HTML]{EEE0FF}Easy Evaluation & \cellcolor[HTML]{CFC3FF}Medium Evaluation & \cellcolor[HTML]{AB96FF}Hard Evaluation                   \\ \hline
\multicolumn{1}{|l|}{\cellcolor[HTML]{E3E3E3}{Additive} ($\rho = 1$)}    &  (0\%, 45\%)      & (5\%, 64\%)                  &  (33\%, 75\%)             \\ \hline
\multicolumn{1}{|l|}{\cellcolor[HTML]{E3E3E3}{Conjunctive} ($\rho = -500$)}& (56.756\%, 21.351\%) & (60.123\%, 22.620\%)          & (60.198\%, 22.649\%)          \\ \hline
\multicolumn{1}{|l|}{\cellcolor[HTML]{E3E3E3}{Disjunctive} ($\rho = +500$)} &(0\%, 45.1\%)     & (0\%, 65.9\%)              & (0\%, 82.2\%)          \\ \hline
\end{tabular}
\end{table}

A Nash equilibrium is represented by a vector of strategies \(\mathbf{a}\), where \(a_1\) is the strategy of the agent with expertise \(p_1^{\mathcal{T}}\) and \(a_2\) is the strategy of the agent with expertise \(p_2^{\mathcal{T}}\). As detailed in Section \ref{sec:model}, these strategies indicate the percentage of time each player dedicates to the teamwork outcome in equilibrium.

In an additive task, the teamwork outcome results from the sum of individual efforts, with each member's contribution adding to the collective outcome. In a conjunctive task, the outcome is determined by the contribution of the team member who performs the worst. In a disjunctive task, the outcome is determined by the contribution of the member who performs the best. The table shows that increasing the level of difficulty leads to greater time contributions from the players. However, the type of task also affects the equilibria. Although the impact of task type on contributions will be detailed in Section \ref{sec:experiments}, the general trend arises in this table: 
\begin{itemize}
    \item In additive tasks (top row), both members' contributions are equally important, and both increase as the evaluation becomes more difficult.
    
    \item In strongly conjunctive tasks (\(\rho \approx -\infty\)) (middle row), only the smallest contribution determines the outcome. Even if a player wants to contribute generously to the teamwork outcome, their efforts will be wasted if someone else contributes less. Therefore, the rational strategy for players is to match the smallest contribution of the others. Indeed, computing \(a_i \cdot p_i\), one can check that both players in the table are contributing approximately\footnote{We attribute the slight discrepancies in contributions to the use of numerical methods in equilibrium calculations, which may introduce small errors.} the same. 

    \item In strongly disjunctive tasks (\(\rho \approx \infty\)) (bottom row), only the best contribution is added to the outcome. In this case, the teamwork game may have multiple equilibria, with only one player contributing while others free-ride. In our case, the player with more expertise is the only one who contributes in all scenarios, although this is not always the case, and in general there is more than one potential contributor.
\end{itemize}

\subsection{Replacement correspondences and Nash equilibria in teamwork games}\label{subsec:NE}

Once the conditions for $r_i$ to be well-defined have been clarified, we can compute its form. We remind the reader that the output of $r_i$ is the same as $b_i$'s (\textit{i.e} $i$'s best possible strategy), what changes is the function's input ($G$ vs $G_{-i}$). Having the form of the replacement correspondence enables us to derive the aggregate replacement correspondence (see Definition \ref{def:aggregate_replacement_correspondence}). According to Proposition \ref{prop:eq_aggregate}, this allows us to obtain the Nash equilibria of the game through the following procedure:

\begin{enumerate}
        \item Compute $R(G)$ as in Eq. (\ref{eq:fix_point}).
        \item Obtain the equilibrium values $\hat{G}$ as the fixed points of $R(G)$: $\hat{G} \in R(\hat{G})$
        \item By Proposition \ref{prop:eq_aggregate}, any fixed point $\hat{G}$ determines a set $\{\hat{\textbf{a}}\}$ of (possibly unique) pure strategy Nash equilibria, which we compute by applying the set of replacement correspondences $\{r_i\}_{i \in \mathcal{I}}$ over $\hat{G}$:
        \begin{equation}
            \{\hat{\textbf{a}}\} = \{(\hat{a}_1, \ldots, \hat{a}_n) \in r_1(\hat{G}) \times r_2(\hat{G}) \times \ldots \times r_n(\hat{G})\}
        \end{equation}
    \end{enumerate}

This section will now detail the construction of the replacement correspondence for each task type. As outlined previously, the resulting expression can then be directly employed to derive the NE of the game. Whether $r_i$ is a function or a correspondence will depend on the task type (\textit{i.e.} the value of $\rho$). In what follows, we will take the (reduced) utility function $\widetilde{u_i}$ to be that in Eq. (\ref{eq:modified-cobb-douglas-preferences}):

\[
    \widetilde{u_i}(x_i, \sigma(G)) = x_i^\alpha \cdot \sigma(G)
\]
and we will obtain the expression of the function/correspondence $r_i$ for additive, conjunctive and disjunctive tasks. 

\subsubsection{Additive tasks}\label{subsec:r_in_additive}
When tasks are additive $(\rho =1)$, we saw in Section \ref{subsub:conditions_r_in_additive} that if the teamwork's outcome $G$ is greater than or equal to every player's standalone value $\overline{G}_i$, then the replacement correspondence is a function. In those circumstances, the existence and uniqueness of an NE are guaranteed (see \citep{cornes2007aggregative} for a thorough justification). Furthermore, the teamwork outcome $G$ is a concave function in $g_i$ for all $i$ \citep{cornes2007weak}. This implies that this equilibrium can be computed by working through the standard first-order conditions of utility maximisation (Eq. (\ref{eq:FOC})). 

Substituting $\rho = 1$ in Eq. (\ref{eq:CES}), it is easy to check that when an evaluation function over the teamwork outcome $G$ is employed and player $i$'s utility is governed by Equation (\ref{eq:modified-cobb-douglas-preferences}), the Lagrange method yields the following replacement function:

\begin{equation}\label{eq:replacement_function_BBV_w_assessment}
r_i(G) = \max \left\{0, p_i^{\mathcal{T}} \Delta_t-\frac{\alpha}{\beta_i}\cdot \frac{\sigma(G)}{\sigma'(G)} \right\}
\end{equation}

Looking at the non-zero term of Eq.(\ref{eq:replacement_function_BBV_w_assessment}), the minuend represents the potential productivity that player $i$ could achieve if they dedicated an entire turn $\Delta_t$ to contributing to the task. The subtrahend quantifies the extent to which this is not possible: higher values render the subtrahend more negative, indicating a greater impediment to the agent to contribute. 

Thus, when \( p_i^{\mathcal{T}} \Delta_t \) is less than \( \frac{\alpha}{\beta_i} \cdot \frac{\sigma(G)}{\sigma'(G)} \), the result of \( r_i(G) \) is zero and player \( i \) chooses not to contribute. That is, when the expertise is sufficiently low, the best a player can do is free-ride on the efforts of others with greater expertise than themselves. Conversely, if \( p_i^{\mathcal{T}} \Delta_t \) is greater than \( \frac{\alpha}{\beta_i} \cdot \frac{\sigma(G)}{\sigma'(G)} \), then \( r_i(G) = p_i^{\mathcal{T}} \Delta_t - \frac{\alpha}{\beta_i} \cdot \frac{\sigma(G)}{\sigma'(G)} \), indicating that player \( i \) contributes a positive amount determined by this difference. Holding other parameters constant, as \( p_i^{\mathcal{T}} \) increases, it becomes more likely for player \( i \) to contribute more, as \( p_i^{\mathcal{T}} \Delta_t \) becomes larger in comparison to the negative term in the equation. The same applies to the behaviour of \( r_i(G) \) as \( \Delta_t \) increases. In short, the higher a player's expertise, the more likely they are to contribute. Another way to encourage a player to contribute is by providing them with more time $\Delta_t$ (longer turns).

As for $\alpha$, it captures the relative importance that players give to their private time compared to the teamwork outcome. For the case $r_i(G) \neq 0$, it becomes evident that, for a fixed quantity of teamwork outcome $G$ and a fixed weight $\beta_i$, the optimal contribution $g_i$ diminishes as $\alpha > 0$ increases. This is an intuitive outcome, as higher values of $\alpha$ signify a greater emphasis on leisure in the preferences of player $i$, thereby discouraging contributions.

Conversely, as $\beta_i$ increases (with all other factors held constant), the absolute value of the negative term diminishes. Put differently, the greater the importance or weight $\beta_i$ assigned to a player's contributions, the more the player will contribute, assuming other factors remain constant.

Regarding the teamwork evaluation function, higher values of the fraction \(\frac{\sigma(G)}{\sigma'(G)}\) render the subtrahend in Eq. (\ref{eq:replacement_function_BBV_w_assessment}) more negative, indicating a greater impediment to the agent's contribution. This ratio can increase either due to a rise in the numerator or a decrease in the denominator. For simplicity, let us consider for a moment that either the numerator or the denominator changes while the other remains constant. 

If the numerator increases at a point \(G_0\) while \(\sigma'(G)\) remains constant, it means that the evaluation has become more lenient, or in other words, the passing threshold has been lowered, making it easier to pass. Perhaps counterintuitively, this lowering of the passing threshold inhibits player contributions rather than incentivising them, as the same utility is obtained with less effort than before.

If we fix the value of the assessment at some teamwork outcome value, \(\sigma(G_0)\), but decrease the slope of the evaluation function \(\sigma'(G)\), then more teamwork outcome units \(\Delta G\) will be needed to increase the evaluation from \(\sigma(G_0)\) to \(\sigma(G_0) + 1\). In this situation, a player would prefer an evaluation function with a steeper slope, as an increase in their contributions would result in greater utility. In other words, decreasing the slope of the evaluation at a point on the curve \(\sigma(G_0)\) inhibits contributions.

In both cases, the fraction \(\frac{\sigma(G)}{\sigma'(G)}\) increases. In conclusion: an easier pass or a less steep evaluation function inhibits players from contributing more. And vice-versa: an optimal contribution \(g_i\) will increase as the fraction decreases, as the opportunity cost in terms of leisure time (\textit{i.e.}, it is easier to improve the utility with the same amount of effort). 

\subsubsection{Conjunctive tasks}\label{subsec:r_in_conjunctive}
When $\rho < 1$ and $\rho \neq 0$, the task is conjunctive, and the teamwork outcome $G$ is a concave function in $g_i$ for all $i$ (\textit{i.e.} similar to the additive case,  equilibria can be computed by working through the standard first-order conditions of utility maximisation). Lemma \ref{lemma:existence_weak_link_group_productivity} states that in this scenario, the replacement correspondence is a function (with two possible domains depending on the value of $\rho$) satisfying the first-order conditions (FOCs) in Eq. (\ref{eq:FOC}). In an analysis we do not replicate here due to space constraints, \citeauthor{cornes2007weak} employ the share function (Definition \ref{def:share_function}) to demonstrate the existence and uniqueness of an equilibrium in general public good games with concave $G$. We can extend this finding to our context:

\begin{proposition}\label{prop:uniqueness_weak_link_group_productivity}
    Let $(\Gamma, \mathcal{T}, \sigma)$ be a teamwork game featuring a conjunctive task. Under the conditions of Lemma \ref{lemma:existence_weak_link_group_productivity}, the game has a unique equilibrium whenever the utility function $\widetilde{u_i}(x_i, \sigma(G))$ is such that the corresponding indifference map is asymptotic to the axes.
\end{proposition}

The interested reader can find a proof of Proposition \ref{prop:uniqueness_weak_link_group_productivity} in \ref{app:proof_conjunctive}. Applying the FOCs in Eq. (\ref{eq:FOC}) and using the utility function in Eq. (\ref{eq:modified-cobb-douglas-preferences}), best-responses can be computed with the following equation: 

\begin{equation}\label{eq:weak_link_FOC}
    \left.\frac{\partial \widetilde{u_i}}{\partial g_i}\right|_{\textbf{g}_{-i}} = \frac{\partial \widetilde{u_i}}{\partial x_i}\frac{\partial G}{\partial g_i}\left\{\frac{x_i}{\alpha}\frac{\sigma'(G)}{\sigma(G)} - \left(\frac{\partial G}{\partial g_i}\right)^{-1}\right\} \geq 0
\end{equation}  

with equality for \(g_i < p_i^\mathcal{T} \cdot \Delta_t\). Manipulating, we obtain the following expression for the replacement function in teamwork games with conjunctive tasks:

\begin{equation}\label{eq:replacement_function_weak_w_assessment}
     \frac{\sigma(G)}{\sigma^\prime(G)} \cdot G^{\rho -1} = \left(\Delta_t - \frac{r_i(G)}{p_i^{\mathcal{T}}}\right) \cdot \frac{\beta_i p_i^{\mathcal{T}}}{\alpha}\cdot r_i(G)^{\rho -1}
\end{equation}

\ref{app:comparative_statics_weaker} examines how \(r_i(G)\) changes as the parameters in this equation increase or decrease. To give a flavour of the resulting equilibria, let's study the case of strongly conjunctive tasks (\(\rho \rightarrow -\infty\)). For simplicity, let's consider a homogeneous team with \(p_i^{\mathcal{T}} = 1\) for all \(i\), and assume that all team members have equal importance: \(\beta_i = 1\) for all \(i\). In the limit \(\rho = -\infty\), the teamwork outcome is solely determined by the contribution of the team member with the lowest performance:

\[
    G = \min_{i\in\mathcal{I}} g_i
\]

By Proposition \ref{prop:uniqueness_weak_link_group_productivity}, the game has a unique equilibrium for $\rho \rightarrow -\infty$. Let \(\hat{G}(\rho)\) be the teamwork outcome at equilibrium. Since all players are identical, their contributions are the same due to symmetry and are given by 

\[
    \hat{g} = \left(\frac{1}{n}\right)^{1/\rho}\hat{G}(\rho)
\]

Setting the bracketed term to zero in Eq. (\ref{eq:weak_link_FOC}) and using \(\hat{g} = \left(\frac{1}{n}\right)^{1/\rho}\hat{G}(\rho)\), we obtain the equation that the best-responses must satisfy in the case of symmetric players:

\[
    \frac{\Delta_t - n^{-1/\rho}\hat{G}(\rho)}{\alpha} \cdot \frac{\sigma'(G)}{\sigma(G)} = n^{1-\frac{1}{\rho}}
\]

Taking the limit \(\rho \rightarrow -\infty\), we get the expression for \(\hat{G}(-\infty)\):

\begin{equation}\label{eq:weakest_limit}
    \hat{G}(-\infty) = \Delta_t - n\alpha \frac{\sigma(G)}{\sigma'(G)}
\end{equation}

Thus, in the case of tasks where the teamwork outcome is solely determined by the lowest individual contribution (and players are identical), all team members contribute the same, and the teamwork outcome becomes a function of the number of players. Specifically, \(\hat{G}(-\infty)\) decreases to zero as \(n \rightarrow \infty\). This implies that when a task is strongly conjunctive and there is a homogeneous team, increasing the team size results in a loss of productivity at the Nash equilibrium. This observation aligns with \citet{steiner1972group}'s assertion on the effect of team size on conjunctive tasks: ``when tasks are truly conjunctive (\textit{i.e.}, when success depends on the poorest member of the group), potential productivity decreases as the group is enlarged''.

\subsubsection{Disjunctive tasks}\label{subsec:r_in_disjunctive}

In this section, we will explore the dynamics of teamwork games involving disjunctive tasks, and we will see that such games can result in multiple equilibria. In these equilibria, some players contribute while others defect. The analysis required to reach this conclusion is more complex than the analyses from Sections \ref{subsec:r_in_additive} and \ref{subsec:r_in_conjunctive}. This complexity arises because disjunctive tasks operate within a parameter regime of $\rho > 1$, resulting in teamwork outcomes that are non-concave functions of the individual contributions $g_i$.

Due to the non-concavity of $G$ in $g_i$, equilibria cannot be determined using the standard first-order conditions for utility maximization (Eq. (\ref{eq:FOC})). Instead, a more in-depth analysis is necessary to identify these equilibria. However, it is still true that for any player $i$, a local maximum of their utility function is reached when $\frac{d\widetilde{u_i}}{dg_i} = 0$. Upon developing and simplifying this equation, we obtain the same expression as Eq. (\ref{eq:replacement_function_weak_w_assessment}) (but now for $\rho > 1$):

    \begin{equation}\label{eq:replacement_function_good_w_assessment}
        \frac{\sigma(G)}{\sigma^\prime(G)} \cdot G^{\rho -1} = \left(\Delta_t - \frac{g_i}{p_i^{\mathcal{T}}}\right) \cdot \frac{\beta_i p_i^{\mathcal{T}}}{\alpha}\cdot g_i^{\rho -1}
    \end{equation}

Or equivalently:
\begin{equation}\label{eq:G_local_max}
    G = \left(\frac{\beta_i p_i^{\mathcal{T}}}{\alpha} \cdot\left(\Delta_t - \frac{g_i}{p_i^{\mathcal{T}}}\right) \cdot \frac{\sigma^\prime(G)}{\sigma(G)}\right)^{\frac{1}{\rho -1}} \cdot g_i
\end{equation}

A teamwork contribution $g_i$ satisfying this local maximum equation will also be a global maximum if the payoff at it is greater than or equal to the payoff at $g_i = 0$:

    \begin{equation}\label{eq:global_max}
         \left(\Delta_t - \frac{g_i}{p_i^{\mathcal{T}}}\right)^{\alpha} \cdot \sigma(G) \geq \Delta_t^\alpha \cdot \sigma(G_{-i})
    \end{equation}

Notice that by definition, the values of $G_{-i}^*$ from Lemma \ref{lemma:best_response_correspondence_aggregative_games} and the corresponding critical point $g_i^*$ are obtained when Eqs. (\ref{eq:replacement_function_good_w_assessment}) and (\ref{eq:global_max}) are satisfied with equality. Because in this case, the set of best-responses can have more than one element according to lemma \ref{lemma:best_response_correspondence_aggregative_games}, a replacement \textbf{correspondence} has to be defined in turn:

\begin{equation}\label{eq:replacement_correspondence_good_shot}
    r_i(G) = \left\{g_i : g_i\in \mathcal{B}_i \left(\left[G^\rho - \beta_i g_i^\rho \right]^{\frac{1}{\rho}}\right)\right\}
\end{equation}

Hence, in a disjunctive task, player $i$ might choose from multiple possible strategies in an equilibrium resulting in total teamwork outcome $G$. From Lemma \ref{lemma:best_response_correspondence_aggregative_games}, we know that $g_i \in r_i(G)$ if and only if either:

\begin{equation}\label{ineq:best_response_correspondence}
    \begin{cases}
    g_i = 0 \text{ and } G_{-i} \geq G_{-i}^{*}, \text{ or} \\
    g_i > 0 \text{ and }
        \begin{cases}
            \frac{\sigma(G)}{\sigma^\prime(G)} \cdot G^{\rho -1} = \left(\Delta_t - \frac{g_i}{p_i^{\mathcal{T}}}\right) \cdot \frac{\beta_i p_i^{\mathcal{T}}}{\alpha}\cdot g_i^{\rho -1} \\
      
            \beta_i^{1/\rho} g_i \leq G \leq \left(\sigma^{-1}\left(\sigma(G)\left(1 - \frac{g_i}{p_i^{\mathcal{T}}\Delta_t}\right)^{\alpha}\right)^\rho + \beta_i g_i^\rho\right)^\frac{1}{\rho}
        \end{cases}
\end{cases}
\end{equation}

Where for the second inequality of the case $g_i > 0$, we have used the global maximum condition from Eq. (\ref{eq:global_max}). The expressions in (\ref{ineq:best_response_correspondence}) tell us that the replacement correspondence $r_i(G)$ in our case has two components: 
\begin{itemize}
    \item A component extending along the $X$- axis to the right of $G_{-i}^{*}$.
    \item A positive component, which is the reflection across the $45\degree$ line of the segment of the teamwork outcome $G$ that simultaneously fulfils the local and global maximum conditions (Eqs. (\ref{eq:G_local_max}) and (\ref{eq:global_max}) respectively).

\end{itemize}

To understand the geometry of the positive component of $r_i$, notice that the right-hand side of Equation (\ref{eq:replacement_function_good_w_assessment}) vanishes when $g_i$ is either equal to $0$ or to $p_i^{\mathcal{T}} \cdot \Delta_t$, and it has a peak at $g_i = \Delta_t p_i^{\mathcal{T}} \left(1-\frac{1}{\rho}\right)$. Additionally, it's worth noting that:

\begin{enumerate}
    \item The boundary line of the left-hand inequality in Eq. (\ref{ineq:best_response_correspondence}), crosses the graph of Eq. (\ref{eq:G_local_max}) at the point corresponding to the standalone value of player $i$ (the level of the teamwork outcome that player $i$ would contribute were they the sole contributor, $\overline{G}_i$):

    \[
        (\overline{g}_i, \overline{G}_i) = \left(\Delta_t\cdot p_i^{\mathcal{T}} - \frac{\sigma(\overline{G}_i)}{\sigma^\prime(\overline{G}_i)}\frac{\alpha}{\beta_i^{1/\rho}},\: \beta_i^{1/\rho}\cdot \Delta_t\cdot p_i^{\mathcal{T}} - \frac{\sigma(\overline{G}_i)}{\sigma^\prime(\overline{G}_i)}\alpha \right)
    \]
    \item By definition, the boundary line of the right-hand inequality in Eq. (\ref{ineq:best_response_correspondence}) crosses the graph of Eq. (\ref{eq:G_local_max}) to the left of $(\overline{g}_i, \overline{G}_i)$, at the critical points $(g_i^*, G_{i}^*)$, where     
        \[G_{i}^* = (G_{-i}^{*\rho} + \beta_i g_i^{*\rho})^{1/\rho}\]
\end{enumerate}

\begin{figure}[!h]
    \centering
    \includegraphics[width=0.5\linewidth]{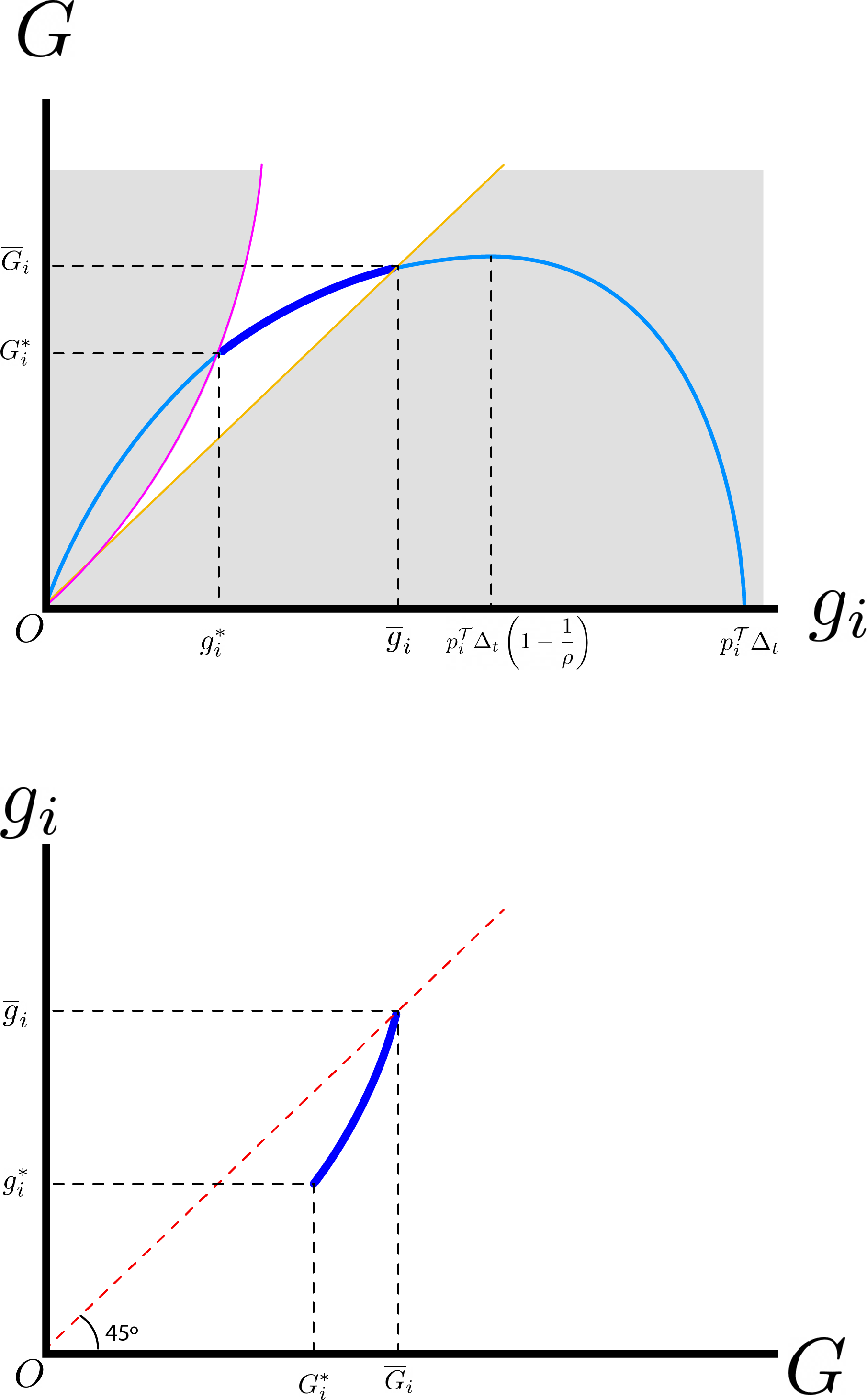}
    \caption{Top panel: The teamwork outcome \( G \), as given by Eq. (\ref{eq:G_local_max}), is depicted in two different shades of blue and has a bell shape. The boundary line of the left-hand inequality in Eq. (\ref{ineq:best_response_correspondence}), shown in fuchsia, intersects this graph at \((g_i^*, G_{i}^*)\). The boundary line of the right-hand inequality in Eq. (\ref{ineq:best_response_correspondence}), shown in light orange, is to the right of the fuchsia line and intersects the graph at \((\overline{g}_i, \overline{G}_i)\). The teamwork outcome \( G \), as given by Eq. (\ref{eq:G_local_max}), is depicted in two different shades of blue and has a bell shape. The boundary line of the left-hand inequality in Eq. (\ref{ineq:best_response_correspondence}), shown in fuchsia, intersects this graph at \((g_i^*, G_{i}^*)\). The boundary line of the right-hand inequality in Eq. (\ref{ineq:best_response_correspondence}), shown in light orange, is to the right of the fuchsia line and intersects the graph at \((\overline{g}_i, \overline{G}_i)\). The non-shaded area indicates the region of \(G\) that satisfies both the local and global maximum conditions simultaneously (bold blue segment). Bottom panel: Reflection of the teamwork outcome $G$ across the $45\degree$ line, showing only the region where $G$ meets the local and global maximum conditions. Figures adapted from \citep{cornes2007weak}.}

     \label{fig:positive_component}
\end{figure}

The top panel in Figure \ref{fig:positive_component} illustrates $G$ from equation (\ref{eq:G_local_max}) and how the different boundary lines mentioned intersect with its graph. The bottom panel represents the positive component of $r_i$ as the reflection across the $45\degree$ line of the segment of $G$ that fulfills the conditions for $g_i > 0$ in Eq. (\ref{ineq:best_response_correspondence}).

If both $(g_i^*, G_{i}^*)$ and $(\overline{g}_i, \overline{G}_i)$ are located to the left of (or at) the maximum of $G(g_i)$, both points will lie within the increasing portion of this function. Examining the $45\degree$ reflection of that function in the bottom panel of Figure \ref{fig:positive_component}, this indicates that the positive component of the replacement correspondence is single-valued on its domain: $\left[G_i^*, \overline{G}_i\right]$. To ensure this condition is met, it is easy to check that it must satisfy:

\[
    \frac{\rho}{\Delta_t p_i^{\mathcal{T}}}\geq \frac{\sigma^{\prime}(\overline{G}_i)}{\sigma (\overline{G}_i)}\cdot \frac{\beta_i^{1/\rho}}{\alpha}
\]

As a result, we get the following corollary:
\begin{corollary}\label{cor:replacemente_correspondence}
    In a teamwork game involving a disjunctive task, if $\frac{\rho}{\Delta_t p_i^{\mathcal{T}}}\geq \frac{\sigma^{\prime}(\overline{G}_i)}{\sigma (\overline{G}_i)}\cdot \frac{\beta_i^{1/\rho}}{\alpha}$, then the positive component of the replacement correspondence is single-valued on its domain: $\left[G_i^*, \overline{G}_i\right]$.
\end{corollary}

In essence, this corollary tells us that although a player in a teamwork game involving a disjunctive task does not have a replacement function but rather a replacement correspondence as per Eq. (\ref{eq:replacement_correspondence_good_shot}), when examining the two possible components of the replacement correspondence individually, they are single-valued. One component outputs zero, and the other is a function with a domain of $\left[G_i^*, \overline{G}_i\right]$. As a consequence, teamwork games involving disjunctive tasks may have multiple equilibria. In each of these equilibria, some players will contribute ($g_i>0$), while all others will defect ($g_i = 0$). Players who positively contribute to teamwork are called the ``active'' players in that equilibrium, and they form an \textbf{equilibrium set} $\mathcal{J} \subseteq \mathcal{I}$.

By Definition \ref{def:share_function}, whoever the active players are, their share correspondences will sum up to $1$ in equilibrium:
\begin{equation}\label{eq:share_in_eq}
    \sum_{j \in \mathcal{J}} s_j(\hat{G}) = 1
\end{equation}

Whereas $\hat{g}_i = 0 \: \forall i \notin \mathcal{J}$. If we can demonstrate that the share correspondences of the active players are indeed functions and that they are increasing, then the sum above would equal $1$ for each set of active players \(\mathcal{J}\) in a unique configuration of contributions. In other words, if we show that \(s_j(G)\) are increasing functions for the active players, then there will be a unique equilibrium for each set of active players. The following proposition guarantees that these conditions are met:

\begin{proposition}\label{prop:share_function}
    If $\frac{\rho}{\Delta_t p_i^{\mathcal{T}}}\geq \frac{\sigma^{\prime}(\overline{G}_i)}{\sigma (\overline{G}_i)}\frac{\beta_i^{1/\rho}}{\alpha}$, the graph of $s_i(G)$ in a teamwork game featuring a disjunctive task is the disjoint union of two sets: $\{(G,0): G\geq G_{-i}^* \}$ and $\{(G, s_i(G)): G_i^* \leq G \leq \overline{G}_i\}$. Plus, $s_i$ is continuous and strictly increasing, and satisfies 
    \[
    \begin{cases}
        s_i(G_i^*) = \beta_i \frac{g_i^{*\rho}}{G_i^{*\rho}} >0 \text{, and}\\
        s_i(\overline{G}_i) = 1
    \end{cases}
    \]
\end{proposition}

\begin{figure}[!h]
    \centering
    \includegraphics[width=.75\linewidth]{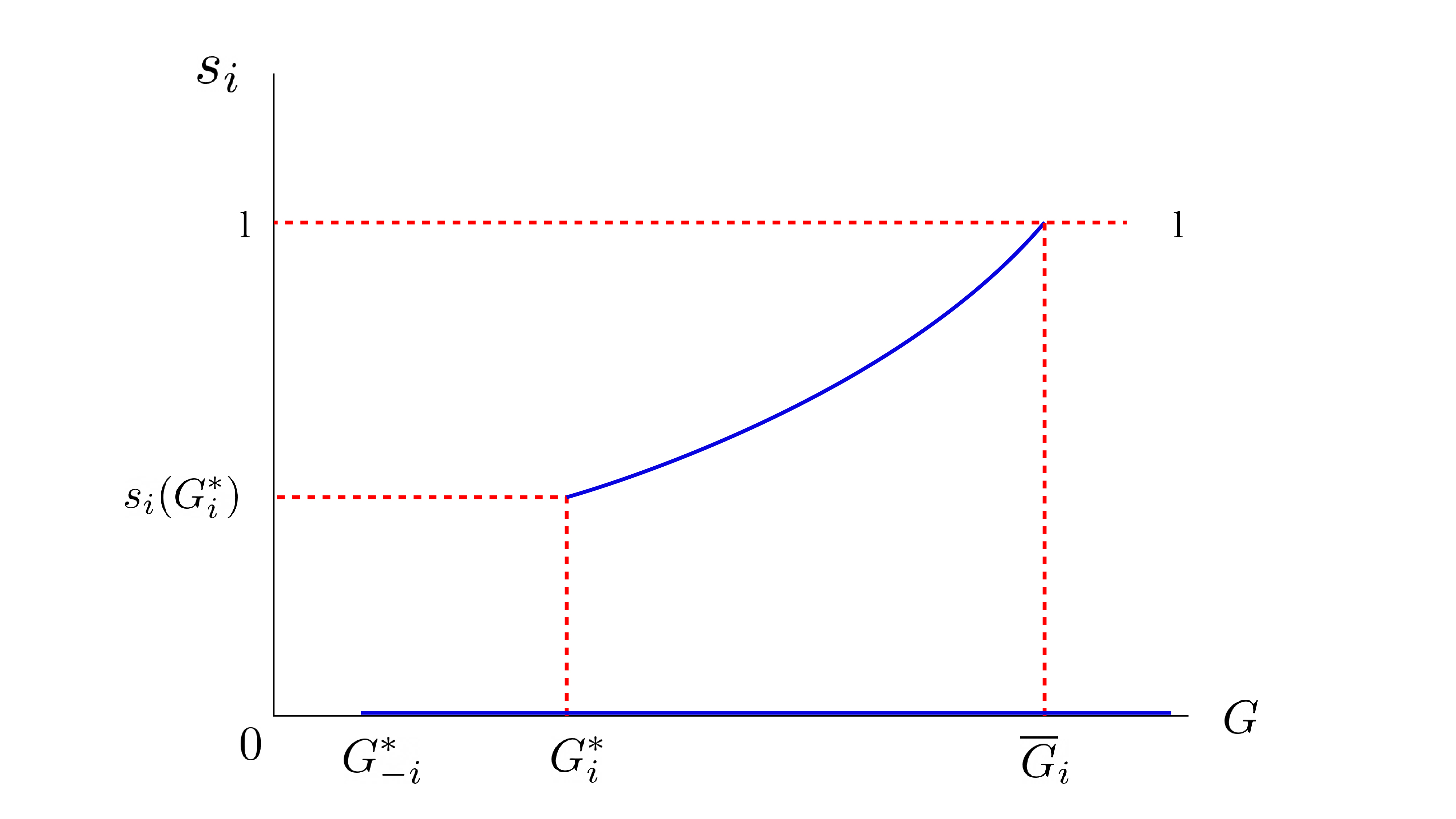}
    \caption{The two components of player $i$'s share correspondence when $\frac{\rho}{\Delta_t p_i^{\mathcal{T}}}\geq \frac{\sigma^{\prime}(\overline{G}_i)}{\sigma (\overline{G}_i)}\frac{\beta_i^{1/\rho}}{\alpha}$}
    \label{fig:share_correspondence_disjunctive}
\end{figure}

The interested reader can find proof of this Proposition in \ref{app:proof_prop_share_function}. Figure \ref{fig:share_correspondence_disjunctive} depicts this result.

We can now delineate equilibria with share functions. According to Proposition \ref{prop:share_function} detailed above, for $j \in \mathcal{J}$, $s_j(G)$ is an increasing function in $G$ and hence there is one equilibrium per set of active players. The level of that equilibrium is denoted by $\hat{G}(\mathcal{J})$, and the equilibrium is obtained through the following procedure:

\begin{enumerate}
    \item Take $s_i(\hat{G}(\mathcal{J})) = 0$ if $i \notin \mathcal{J}$.
    \item Compute $\left\{s_j(\hat{G}(\mathcal{J}))\right\}$ for $j \in \mathcal{J}$ such that Eq. (\ref{eq:share_in_eq}) is satisfied.
\end{enumerate}

Therefore, characterising equilibria is equivalent to characterising equilibrium sets, with necessary conditions for an equilibrium set following from Proposition \ref{prop:share_function}. That is, in an equilibrium set $\mathcal{J}$, the following conditions are fulfilled: 

\begin{align*}
    G_j^* \leq \hat{G}(\mathcal{J})\leq \overline{G}_i \: \forall j\in \mathcal{J}\\
    G_{-i}^* \leq \hat{G} (\mathcal{J}) \: \forall i \notin \mathcal{J}
\end{align*}

We can use these conditions to identify equilibrium sets systematically:
\begin{enumerate}
    \item Evaluate each possible subset $\mathcal{J} \subseteq \mathcal{I}$.
    \item Within each subset, we select the critical value $G_j^{*}$ of the player with the lowest tolerance for undersupplying the teamwork outcome. 
    \item Simultaneously, we choose the critical value $G_{-i}^{*}$ of the player outside the subset who also exhibits the least tolerance for the teamwork outcome being undersupplied by others. 
    \item  The maximum of these two values yields $G^{*}(\mathcal{J})$.
    \item For $\mathcal{J}$ to qualify as an active set, it must consist of players whose individual standalone values $\overline{G}_j$ all exceed or equal this critical value $G^{*}(\mathcal{J})$, while also satisfying 
    \[\sum \limits_{j\in \mathcal{J}}s_j(G^{*}(\mathcal{J}))\leq 1\]
\end{enumerate}

More formally, let $G^{*}(\mathcal{J})$ be the maximum among 1) the highest critical value $G_j^{*}$ of the active players and 2) the highest critical value of the inactive players, $G_{-i}^{*}$:

    \[G^{*}(\mathcal{J}) = \max\{\max\limits_{j\in\mathcal{J}}G_j^{*},\max\limits_{i\notin\mathcal{J}} G_{-i}^{*} \}\]
    
With this definition, we can adopt \citeauthor{cornes2007weak}'s characterisation of equilibrium sets in non-concave public goods\footnote{See \citep{cornes2007weak} for further clarification.} for teamwork games with disjunctive tasks:

\begin{proposition}\label{prop:equilibria_good_shot}
Suppose that $\frac{\rho}{\Delta_t p_i^{\mathcal{T}}}\geq \frac{\sigma^{\prime}(\overline{G}_i)}{\sigma (\overline{G}_i)} \frac{\beta_i^{1/\rho}}{\alpha}$ $\forall i \in \mathcal{I}$ in a teamwork game featuring a disjunctive task. A non-empty set $\mathcal{J}\subseteq \mathcal{I}$ is an equilibrium set if and only if it satisfies: 
\[
\begin{cases}
    & G^{*}(\mathcal{J}) \leq \min \limits_{j\in\mathcal{J}}{\overline{G}_j} \\
    & \sum \limits_{j\in \mathcal{J}}s_j(G^{*}(\mathcal{J})) \leq 1
\end{cases} 
\]    
\end{proposition}

It logically follows from this proposition that non-empty subsets of equilibrium sets themselves form equilibrium sets. Finally, we can borrow a corollary in \citep{cornes2007weak} and adapt it regarding the teamwork outcomes at possible equilibria in teamwork games involving disjunctive tasks:

\begin{corollary}\label{cor:eq_subsets}
    If $\frac{\rho}{\Delta_t p_i^{\mathcal{T}}}\geq \frac{\sigma^{\prime}(\overline{G}_i)}{\sigma (\overline{G}_i)}\cdot \frac{\beta_i^{1/\rho}}{\alpha}$ $\forall i \in \mathcal{I}$ and $\mathcal{J}$ is an equilibrium set, any $\mathcal{K}\subset \mathcal{J}$ is also an equilibrium set and $\hat{G}(\mathcal{K}) > \hat{G}(\mathcal{J})$.
\end{corollary}

In other words, if we take a subset \(\mathcal{K} \subset \mathcal{J}\) of active players, the teamwork outcome produced by the players in \(\mathcal{K}\) will be strictly greater than the outcome produced by the bigger set of players in \(\mathcal{J}\). As cautioned by \citet{cornes2007weak}, this result indicates that free-riders will consistently prefer a smaller set of active players, enabling them to exploit a larger share of the teamwork outcome. In our case, when a task is disjunctive —meaning higher individual contributions within the team lead to a greater impact on the team outcome— rational players who abstain from contributing will prefer that the workload is shouldered by as few teammates as possible. The proof to this corollary is the same as in \citep{cornes2007weak}.

\subsection{Equilibria in teamwork games vs equilibria in general public good games}
Before delving into the details of the multi-agent model we propose for learning equilibrium in teamwork games, let's briefly compare our results with those of \citet{cornes2007weak}. Table \ref{tab:model_comparison} summarizes the replacement functions (or correspondences) obtained in Subsection \ref{subsec:NE} for teamwork games, alongside the corresponding functions for general public good (GPG) games as studied by \citet{cornes2007weak}.
\begin{table}[!h]
\centering
\caption{Replacement functions/correspondences in teamwork and general public good games}
\label{tab:model_comparison}
\begin{subtable}{\linewidth}
\centering
\caption{Replacement functions/correspondence for the three different task types in teamwork games.}
\label{tab:replacement_functions_teamwork}
\setlength{\tabcolsep}{10pt}
\renewcommand{\arraystretch}{1.5} 
\begin{tabular}{|p{0.6\textwidth}|p{0.72\textwidth}|}
\cline{2-2}
\multicolumn{1}{c|}{}                  & \multicolumn{1}{c|}{\cellcolor[HTML]{E3E3E3} Teamwork Games} \\ \hline
\multicolumn{1}{|l|}{\cellcolor[HTML]{E3E3E3} Additive $\mathcal{T}$($\rho = 1$)}    & $r_i(G) = \max \left\{0, p_i^{\mathcal{T}} \Delta_t-\frac{\alpha}{\beta_i}\cdot \frac{\sigma(G)}{\sigma'(G)} \right\}$ \\ \hline
\multicolumn{1}{|l|}{\cellcolor[HTML]{E3E3E3} Conjunctive $\mathcal{T}$($\rho < 1$)}& $\frac{\sigma(G)}{\sigma^\prime(G)} \cdot G^{\rho -1} = \left(\Delta_t - \frac{r_i(G)}{p_i^{\mathcal{T}}}\right) \cdot \frac{\beta_i p_i^{\mathcal{T}}}{\alpha}\cdot r_i(G)^{\rho -1}$ \\ \hline
\multicolumn{1}{|l|}{\cellcolor[HTML]{E3E3E3} Disjunctive $\mathcal{T}$($\rho >1$)} & 
\parbox[c]{\hsize}{\[
    \begin{cases}
    r_i(G) = 0 \text{ and } G_{-i} \geq G_{-i}^{*}, \text{ or} \\
    r_i(G) > 0 \text{ and }
        \begin{cases}
            \frac{\sigma(G)}{\sigma^\prime(G)} \cdot G^{\rho -1} = \left(\Delta_t - \frac{r_i(G)}{p_i^{\mathcal{T}}}\right) \cdot \frac{\beta_i p_i^{\mathcal{T}}}{\alpha}\cdot r_i(G)^{\rho -1} \\
      
            \beta_i^{\frac{1}{\rho}} r_i(G) \leq G \leq \left(\sigma^{-1}\left(\sigma(G)\left(1 - \frac{r_i(G)}{p_i^{\mathcal{T}}\Delta_t}\right)^{\alpha}\right)^\rho + \beta_i r_i(G)^\rho\right)^\frac{1}{\rho}
        \end{cases}
    \end{cases}
\] }
\\ \hline
\end{tabular}
\end{subtable}

\begin{subtable}{\linewidth}
\caption{Replacement functions/correspondence for canonical, concave and convex public goods}
\label{tab:replacement_functions_public_good}
\centering
\setlength{\tabcolsep}{10pt}
\renewcommand{\arraystretch}{1.5} 
\begin{tabular}{|p{0.6\textwidth}|p{0.72\textwidth}|}
\cline{2-2}
\multicolumn{1}{c|}{}                  & \multicolumn{1}{c|}{\cellcolor[HTML]{E3E3E3} General Public Good Games} \\ \hline
\multicolumn{1}{|l|}{\cellcolor[HTML]{E3E3E3} Canonical GPG ($\rho = 1$)}    & $r_i(G) =  \max \left\{ 0, w_i - \frac{\alpha}{\beta_i} G\right\}$ \\ \hline
\multicolumn{1}{|l|}{\cellcolor[HTML]{E3E3E3} Concave GPG ($\rho < 1$)}& $r_i(G)^{\rho-1}(w_i - r_i(G)) = \frac{\alpha}{\beta_i} \cdot G^\rho$ \\ \hline
\multicolumn{1}{|l|}{\cellcolor[HTML]{E3E3E3} Convex GPG ($\rho > 1$)} & 
\parbox[c]{\hsize}{\[
    \begin{cases}
    r_i(G) = 0 \text{ and } G_{-i} \geq G_{-i}^{*}, \text{ or} \\
    r_i(G) > 0 \text{ and }
        \begin{cases}
            r_i(G)^{\rho-1}(w_i - r_i(G)) = \frac{\alpha}{\beta_i} \cdot G^\rho  \\
      
            \beta_i^{\frac{1}{\rho}} r_i(G) \leq G \leq 
            \beta_i^{\frac{1}{\rho}} r_i(G)w_i^{\alpha}\left[w_i^{\alpha\cdot \rho} - (w_i -r_i(G)^{\alpha \cdot \rho}\right]^{\frac{-1}{\rho}}
        \end{cases}
    \end{cases}
\] }
\\ \hline
\end{tabular}
\end{subtable}
\end{table}

In a GPG game, \( w_i \) represents player \( i \)'s exogenous income (as opposed to \( p_i^\mathcal{T}\Delta_t \) in teamwork games). Additionally, \citeauthor{cornes2007aggregative} do not consider heterogeneity among players (\textit{i.e.}, players with different levels of expertise). In fact, expertise is not explicitly modelled in their game, so we assume \( p_i^{\mathcal{T}} = 1 \: \forall i \) in GPG games.

Examining the value of the substitution parameter \( \rho \), we can see that GPG games involving canonical public goods correspond to teamwork games involving additive tasks in our model. Similarly, there is a mapping between GPG games with concave public goods and teamwork games involving conjunctive tasks, and between GPG games with convex public goods and teamwork games involving disjunctive tasks.

The results in Table \ref{tab:model_comparison} demonstrate that our findings extend those of \citeauthor{cornes2007weak}. By setting \(\sigma(G) = G\) and \(p_i^{\mathcal{T}} = 1\), we can reproduce the expressions for \(r_i(G)\) found in table \ref{tab:replacement_functions_public_good}. Notice how incorporating a non-trivial evaluation function in teamwork games influences the strategy \(g_i\) selected by player \(i\) through the denominator \(\frac{\sigma(G)}{\sigma'(G)}\). As detailed in Section \ref{subsub:conditions_r_in_additive}, this ratio represents the relationship between the evaluation function \(\sigma(G)\) and the rate at which the assessment grows as \(G\) increases. Since the rate of growth \(\sigma'(G)\) is positive by definition, \(\frac{\sigma(G)}{\sigma'(G)}\) indicates how valuable it is for players to invest their time in improving their assessment rather than enjoying their leisure time. In other words, it provides information about the \textit{opportunity cost} of the evaluation in terms of enjoyed leisure. Consequently, holding other parameters constant, an optimal contribution \(g_i\) will increase as this fraction decreases, as it becomes easier to improve the assessment score with the same amount of effort.
Table \ref{tab:running_example_NE_all} presents the Nash equilibria when two agents with expertise \( p^{\mathcal{T}} = 1 \) play three teamwork games: one involving an additive task, another involving a conjunctive task, and a third involving a disjunctive task\footnote{The agents in this example have expertise \( p^{\mathcal{T}} = 1 \) to ensure the results are comparable with the equilibria of a general public good game}. In all these games, we take the evaluation function (Def. \ref{def:assessment_function}) to be a logistic function:
\[
    \sigma(G) = \frac{d}{1+ \mathrm{e}^{-\gamma(G-b)}}
\] 

and choose the passing threshold \( b=5 \). Furthermore, we model the players' preferences with
\[
    \widetilde{u_i}(x_i, \sigma(G)) = x_i^\alpha \cdot \sigma(G)
\]

Similar to Table \ref{tab:running_example_NE_teamwork_games}, let's assume that enjoying leisure is quadratically more important to the agents than receiving a good evaluation, with $\alpha = 2$ for all players. For comparison, we include the corresponding Nash equilibria of the corresponding general public good (GPG) games.

\begin{table}[!h]
\centering
\caption{Nash equilibria for two agents with expertise \( p^{\mathcal{T}} = 1 \) playing three types of teamwork games and a general public good (GPG) game. The required teamwork outcome for passing is \( G=5 \).}
\label{tab:running_example_NE_all}

\begin{subtable}{\textwidth}
\centering
\begin{tabular}{|c|c|l|c|c|}
\cline{1-2} \cline{4-5}
\cellcolor[HTML]{E3E3E3}\textbf{Additive $\mathcal{T}$} & \cellcolor[HTML]{E3E3E3}\textbf{Canonical PG} & & \cellcolor[HTML]{E3E3E3}\textbf{Conjunctive $\mathcal{T}$} & \cellcolor[HTML]{E3E3E3}\textbf{Concave PG} \\ \cline{1-2} \cline{4-5} 
(30\%, 30\%)                                            & (20\%, 20\%)                             & \multicolumn{1}{c|}{} & (52\%, 52\%)                                     & (20\%, 20\%)                           \\ \cline{1-2} \cline{4-5} 
\end{tabular}
\end{subtable}

\vspace{1em}

\begin{subtable}{\textwidth}
\centering
\begin{tabular}{|c|c|}
\hline
\cellcolor[HTML]{E3E3E3}\textbf{Disjunctive $\mathcal{T}$} & \cellcolor[HTML]{E3E3E3}\textbf{Convex PG} \\ \hline
\begin{tabular}[c]{@{}c@{}}(56\%, 0\%)\\ (0\%, 56\%)\end{tabular} & \begin{tabular}[c]{@{}c@{}}(33\%, 0\%)\\ (0\%, 33\%)\end{tabular} \\ \hline
\end{tabular}
\end{subtable}

\end{table}

Just as in Table \ref{tab:running_example_NE_teamwork_games}, a Nash equilibrium is represented by a vector of strategies \(\mathbf{a}\), indicating the percentage of time each player dedicates to the teamwork outcome in equilibrium. Notice that in the cases shown, the introduction of the evaluation function shifts the equilibria, resulting in greater time contributions compared to public good games. Furthermore, for our chosen parameters, the equilibrium in GPG games with canonical and concave public goods is identical: in both cases, both players allocate \(20\%\) of their turn to the public good\footnote{Note that this does not imply that the value of the public good in equilibrium, $\hat{G}$, is the same. In fact, in the GPG games shown here, \(\hat{G}_{\text{canonical}} = 4\), while \(\hat{G}_{\text{concave}} = 2\).}. In a teamwork game, this symmetry breaks, and players differentiate between additive and conjunctive tasks. In all the displayed equilibria in teamwork games, the team passes (\(G \geq 5\)). Furthermore, note that both in the teamwork game involving a disjunctive task and the GPG game involving a convex public good, there are two equilibria. In each of them, one player is the sole contributor, while the other free-rides.

\section{Multiagent multi-armed bandit learning}\label{sec:MA-MAB_learning} 
In previous sections, we presented a formal model of a teamwork game and characterised the replacement correspondences for different types of tasks that teams may face. These replacement correspondences allow for the characterisation of Nash Equilibria (NE) through the application of Proposition \ref{prop:eq_aggregate}. In this section, we propose a multiagent multi-armed bandit (MA-MAB) framework where agents' learning converges to the approximate NE of teamwork games.

The rationale behind this choice of learning framework is twofold. Firstly, this approach effectively captures the exploration-exploitation dilemma faced by decision-makers with multiple choices. Secondly, as discussed by \citet{leslie2005individual}, action selection methods based on smooth best responses, such as the soft-max method \citep{sutton2018reinforcement}, can guide a multiagent MAB system towards strategies approximating Nash equilibria. The use of smooth best responses means that Nash equilibria (NE) are no longer fixed points in the strategy space. Instead, the algorithm converges to Nash distributions—joint strategies where each agent plays smooth best responses to rewards from interactions with others. Under certain conditions, the learning algorithm can be applied blindly to any game and, if convergence occurs, a Nash distribution will have been reached \citep{govindan2003short, leslie2005individual}.
\subsection{Overview}

Consider an agent participating in a team who must decide how much effort to contribute. As we have seen, this decision is influenced by various factors, including the team's composition, the nature of the task, and team interaction dynamics. Conceptualising this decision-making as a teamwork game, each team member's decision-making process resembles that of a multi-armed bandit problem. Here, the ``arms'' represent discrete values indicating the percentage of a decision maker's total available time allocated to the task. An independent learning agent, denoted by $i$, only learns from its own observations, actions, and rewards, without taking into account experiences from other agents \citep{marl-book}. This means that each learning agent only considers the effects of other agents' actions as part of the environment dynamics. The following sections present the elements of a multiagent multi-armed bandit learning framework where the agents play an underlying teamwork game.

Section \ref{subsec:environment} presents the elements of our multiagent framework related to the environment with which the agents interact, including the conditions under which a task is considered complete and the reward signal that agents receive for their contributions to the teamwork outcome. Section \ref{subsec:agents} specifies the elements that define our agents, including their action selection mechanism and the learning algorithm employed.

\subsection{Teamwork environment}\label{subsec:environment}

Our learning environment uses a teamwork game $(\Gamma, \mathcal{T}, \sigma)$ as the underlying game. The agents have one turn of common length $\Delta_t$ to complete a task $\mathcal{T} = (G^{\mathcal{T}}, \left[\Delta_t, N=1\right], \Theta)$ of typology specified by the parameters in $\Theta = \{\rho, \beta_1, \ldots, \beta_n\}$. Because the task is one-shot, the fulfilment conditions in Eq. (\ref{eq:end_conditions}):

\[
    \left\{\begin{matrix}
        G\geq G^{\mathcal{T}}  \\
        t \leq N \\
    \end{matrix}\right.
\]

are simplified, and the task is considered successfully finished if the units of work output by the team reach the minimum $G^{\mathcal{T}}$:

    \[
        G  \geq G^{\mathcal{T}}
    \]

agent $i$ is given a fixed weight $\beta_i$ at the beginning of the game, so their contribution $g_i$ is aggregated with the rest as:
\[
    G =  \left(\sum \limits_{i\in \mathcal{I}} \beta_i \cdot g_i^\rho\right)^{\frac{1}{\rho}}
\]
After this only turn, agent $i$ receives a reward $R_i$ that depends on the number of leisure units that $i$ has been able to enjoy, $x_i$, and the assessment $\sigma(G)$, which is common to all agents. This reward is given by
 the utility function of the underlying teamwork game:

\[
    R_i = x_i^\alpha \cdot \sigma(G) \in \mathbb{R}
\]

\subsection{Learning agents}\label{subsec:agents}
Our set of agents, $\mathcal{I} = \{1, \ldots, n\}$, consists of $n$ independent $k$-armed bandits. Each agent $i$ has an intrinsic, fixed expertise level $p_i^{\mathcal{T}}$. Following the convention established in Section \ref{sec:model}, we assume all agents have a leisure capacity $p_i^L = 1$ (although, in general, this parameter could be different from $1$). Given a task $\mathcal{T}$, agent $i$ must decide what percentage $a_i \in \mathcal{A}_i$ of time $\Delta_t$ to allocate to $\mathcal{T}$. The action vector $\textbf{a} = (a_1, \ldots, a_n)$ represents the strategy profile of all players. This strategy profile maps to a vector of contributions $\textbf{g} = (g_1, \ldots, g_n)$, where
\[
    g_i = \Delta_t \cdot p_i^{\mathcal{T}} \cdot a_i \text{ for all } i.
\]
If $i$ chooses action $a_i$, their private contribution representing leisure units allocated to private activities will be
\[
x_i = \Delta_t - \frac{g_i}{p_i^{\mathcal{T}}}
\]

Notice that this equation is correct as long as the convention $p_i^L = 1 \: \forall i$ is followed. We manage $n$ independent $k$-armed bandits, each selecting actions from a Boltzmann distribution with a temperature parameter $\tau$:
\[
    \mathbb{P}(a) = \frac{\mathrm{e}^{Q_t(a)/(Q_t^\text{max}\cdot \tau)}}{\sum \limits_a \mathrm{e}^{Q_t(a)/(Q_t^\text{max}\cdot \tau)}}
\]
Each agent can distinguish actions (time percentages) with a granularity of $2$ decimals. That is to say: agents can choose among the $10^2 + 1$ actions in $\mathcal{A}_i = \{0, 10^{-2}, \ldots, 1 - 10^{-2}, 1\}$. At episode $t+1$, the value estimates of agent $i$ are computed incrementally:

\[
    Q^{t+1}_{i} = Q^t_{i} + l_r^{t} \cdot (R^t_{i} - Q^t_{i})
\]

Here, $Q_t^{i}$ represents agent $i$'s value estimate at episode $t$, while $R_t^{i}$ denotes the reward that $i$ receives in episode $t$. The learning rate parameter, denoted by $l_r^{t} \in \left(0,1\right]$, is the same for all agents and decays with the inverse of time:

\begin{equation}\label{eq:lr}
    l_r^t = \left(\frac{k}{k + t}\right)^{-1}
\end{equation}

As justified in Section \ref{sec:preliminaries}, this algorithm implements smooth best responses: if $\tau$ is sufficiently low and the agents interact for a reasonably long period, then convergence will only occur at a Nash distribution \citep{leslie2005individual}.

\section{Results and discussion}\label{sec:experiments}

The primary aim of this section is to address whether a MA-MAB framework, as described in Section \ref{sec:MA-MAB_learning}, can converge to approximations of equilibria of teamwork games after learning. If convergence is possible, we will examine the nature of these empirical approximations and how they are impacted by changes in the various components of our model. We focus on teams consisting of two agents, leaving the study of larger teams for future work. More specifically, we conduct the following analyses:
\begin{enumerate}
    \item We examine the consistency of the agents' learned strategies with the game-theoretic predictions of the underlying game's Nash Equilibria.
    \item We examine the repercussions on agent team productivity stemming from the interplay between team composition and task types.
    \item We explore how the difficulty of assessment influences agent team productivity.
    \item We delve into the individual strategies adopted by agents and how they are impacted by different task types, team compositions, and assessment difficulties.
    \item Finally, we expand our analysis beyond analytically
    solvable games and empirically study the policies of our agents after learning in a binary pass/fail scenario.
\end{enumerate}

\subsection{Experimental setup}\label{subsec:setup}
Our experimental setup involves teams consisting of two multi-armed bandits, each equipped with 101 arms, with each arm representing a discrete value indicating a percentage of the decision maker’s total available time allocated to the task. We train each agent during $5\cdot 10^4$ episodes of $1$ step each each. Throughout the experiments, we consider that all of the agents' contributions are equally valuable ($\beta_1 = \beta_2 = 1 $) and assume a standard leisure time parameterization, $p_1^{L} = p_2^{L} = 1$. The learning rate parameter, denoted by $l_r^{t} \in \left(0,1\right]$, is the same for all agents and decays with the inverse of time as per Eq. (\ref{eq:lr}), where $k>0$ is determined using an automatic hyperparameter optimization framework \citep{optuna_2019}. By default, we conduct one-shot games with a turn duration of $\Delta_t = 10$ time units, and we set the common parameter $\alpha$ in the utility function to $2$ (Eq. (\ref{eq:modified-cobb-douglas-preferences})). This value of $\alpha$ models agents for whom their free time is squaredly more important than the teamwork outcome. Both $\Delta_t$ and $\alpha$ are convenient choices we have made to ensure clarity in presenting the results, but in general, any $\Delta_t, \alpha > 0$ could have been chosen. The reward function guiding agent $i$'s policy is derived from the underlying utility in the respective teamwork game:

\[
R_i \equiv \widetilde{u_i}(g_i, \sigma(G)) = x_i^\alpha \cdot \sigma(G) = \Delta_t^\alpha\cdot(1 - a_i)^{\alpha} \cdot \sigma(G)
\]

\begin{figure}[!h]
    \centering
    \includegraphics[width=.65\linewidth]{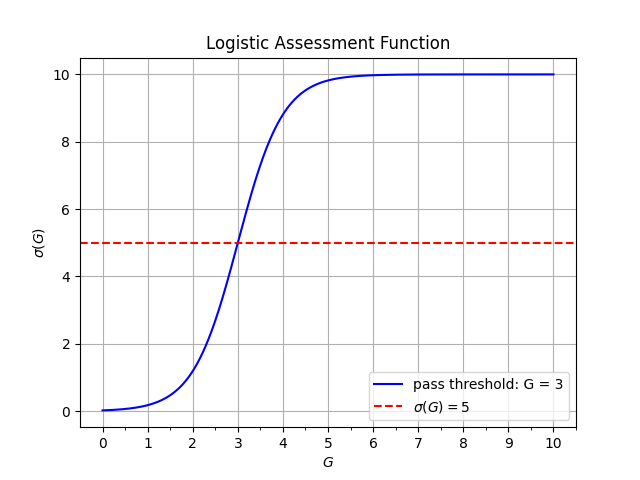}
    \caption{Logistic evaluation function with passing threshold \(b=3\), steepness parameter \(\gamma = 2\), and right-hand asymptote \(d=10\).}
    \label{fig:logistic_assessment}
\end{figure}

Notice that \(\sigma(G)\) is the same for both agents, and the second equality follows from Eq. (\ref{eq:private_good}). As for the evaluation, we seek a function that can represent continuous assessment. Ideally, it should be readily applicable to provide a continuous representation of evaluations based on varying levels of merit. As discussed below, the logistic function (Fig. \ref{fig:logistic_assessment}) meets these requirements:

\begin{equation}\label{eq:logistic_function}
    \sigma(G) = \frac{d}{1+ \mathrm{e}^{-\gamma(G-b)}}
\end{equation}

The logistic function takes the teamwork outcome \(G\) as input and returns an evaluation within the interval \(\left[0,d\right] \subset \mathbb{R}\). The parameter \(\gamma\) indicates how many units of the teamwork outcome are necessary to raise the assessment by one point. When \(G=b\), the teamwork outcome receives an evaluation of \(\frac{d}{2}\), making \(b\) the passing threshold. Notably, the logistic function allows us to create a continuous approximation of qualitative evaluations, similar to a gradable pass/fail system. Furthermore, by stacking logistic functions, we could achieve a continuous representation of evaluations based on varying levels of merit. Some of the key properties of the logistic function include:

\begin{itemize}
\item Non-negative assessment: The function approaches the right horizontal asymptote, $d$, representing the maximum attainable reward. The left asymptote is $\sigma(G) = 0$, although, in our model, $G$ remains non-negative, preventing it from being reached.
\item Evaluation steepness: The curve steepness, controlled by parameter $\gamma$, determines the rate of evaluation change concerning $G$.
\item Passing threshold: The parameter $b$ of the curve represents the midpoint of the evaluation criteria, and it determines the passing threshold. 
\end{itemize}

As discussed in Section \ref{subsec:r_in_additive}, the ratio \(\frac{\sigma(G)}{\sigma'(G)}\) indicates how valuable it is for players to invest their time in improving their assessment rather than enjoying their leisure time. In the case of the logistic function, the parameters \(b\) and \(\gamma\) jointly provide this information. If \(b\) decreases, then the passing threshold lowers, making it easier to pass (and making it less worthwhile to work on the task). If \(\gamma\) decreases, then the slope of the curve decreases and more teamwork outcome units are needed to increase the evaluation by one point (and also inhibiting contributions). Therefore, \(b\) and \(\gamma\) provide information about the \textit{opportunity cost} of the evaluation in terms of enjoyed leisure. In this paper, we keep \(\gamma\) constant in all experiments for the sake of parsimony. Consequently, it will be the change in \(b\) from one experiment to another that marks differences in the fraction $\frac{\sigma(G)}{\sigma'(G)}$, and we will use the term \textit{evaluation hardness} to refer to the effect of \(b\) in the following discussion.

The logistic function in Eq.(\ref{eq:logistic_function}) fulfils all the criteria to serve as an evaluation function (Def. \ref{def:assessment_function}). The resulting reward, expressed as a function of agent $i$'s contribution $g_i$ and the provision of the teamwork outcome $G$, $\widetilde{u_i}(g_i, \sigma(G))$, remains non-negative across all values.

\begin{figure}[!h]
    \centering
    \includegraphics[width=1\linewidth]{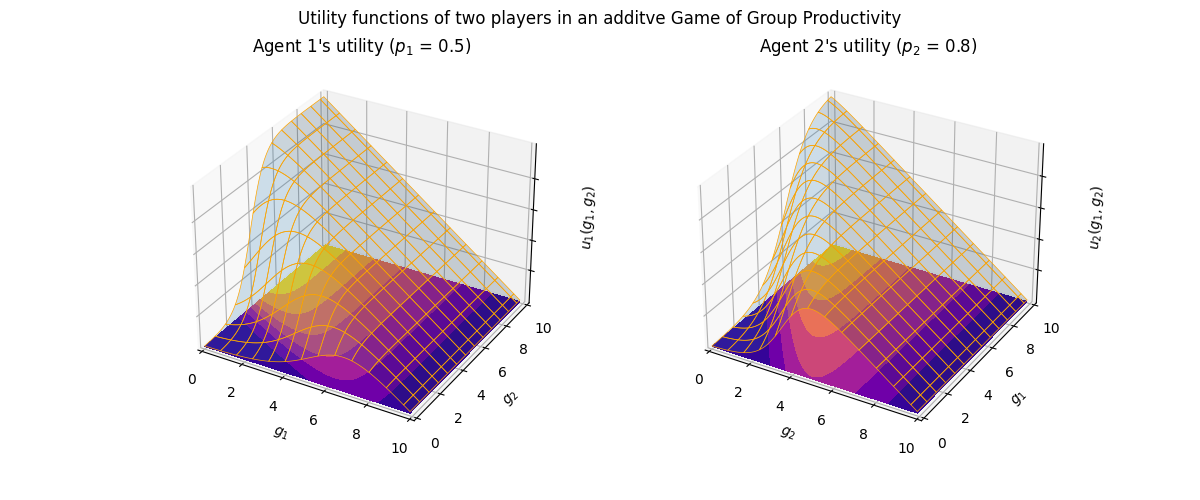}
    \caption{Utilities $\widetilde{u_i}$ of two agents in a team, shown as functions of the agent's work units contribution $g_i$ and the rest of the team's contribution $G_{-i}$. Left: Utility for an agent with expertise $p_1^\mathcal{T} = 0.5$. Right: Utility for their teammate with expertise $p_2^\mathcal{T} = 0.8$.}
    \label{fig:utility_function}
\end{figure}

Figure \ref{fig:utility_function} depicts the utilities for each agent in a two-agent team: Agent 1 with $p_1^\mathcal{T} = 0.5$ and Agent 2 with $p_2^\mathcal{T} = 0.8$. The logistic evaluation function is employed. The task is additive (\(\rho=1\)) with a passing threshold \(b=5\) and turn duration \(\Delta_t = 10\). It's worth noting that the most favourable scenario for Agent 1 occurs when their teammate contributes maximally ($g_{2} = 10$) while they refrain from any action ($g_1 = 0$). When neither agent contributes ($g_1 = g_{2} = 0 \: \forall i$), both agents perceive a utility close to zero due to the evaluation function. However, neither agent desires to contribute maximally, as it would mean sacrificing leisure time and receiving a null utility.
In this case, the individual contributions (in work units) in equilibrium are $\left(g_1 = 1.25, g_2 = 4.24\right)$ (Eq. (\ref{eq:individual_contribution})).

\subsection{Consistency with game-theoretic predictions}\label{subsec:consistency}
First, we scrutinise the alignment of our agents' learned strategies with the NE of the underlying teamwork game. A total of 140 experiments are conducted, involving pairs of agents with diverse expertise combinations drawn from a set ranging from \(p_i^\mathcal{T} = 0.1\) to \(p_i^\mathcal{T} = 0.9\).
Additionally, we explore different permutations of task types, ranging from strongly conjunctive tasks (\(\rho = -100\)) to strongly disjunctive tasks (\(\rho = 100\)) (Table \ref{tab:task_types}).

\begin{table}[]
\centering
\caption{Task types used in the experiments, characterised by the value of $\rho$}
\label{tab:task_types}
\begin{tabular}{|c|c|l|c|c|}
\hline
\rowcolor[HTML]{E3E3E3} 
\begin{tabular}[c]{@{}c@{}}Strongly \\ Conjunctive\end{tabular} & \begin{tabular}[c]{@{}c@{}}Quasi-\\ Conjunctive\end{tabular} & Additive                        & \begin{tabular}[c]{@{}c@{}}Quasi-\\ Disjunctive\end{tabular} & \begin{tabular}[c]{@{}c@{}}Strongly\\ Disjunctive\end{tabular} \\ \hline
$\rho = -100$                                                   & $\rho \in \left\{-10, -3, 0.5\right\}$                       & \multicolumn{1}{c|}{$\rho = 1$} & $\rho \in \left\{3, 10\right\}$                              & $\rho = 100$                                                   \\ \hline
\end{tabular}
\end{table}

We use three evaluation conditions represented by thresholds $b \in \{3,5,7\}$ referred to as \textbf{soft}, \textbf{medium}, and \textbf{hard} assessment treatments, respectively (Table \ref{tab:evaluation_conditions}).

\begin{table}[!h]
\centering
\caption{Evaluation conditions used in the experiments, represented by passing thresholds \(b \in \{3,5,7\}\)}
\label{tab:evaluation_conditions}
\begin{tabular}{|c|c|c|}
\hline
\rowcolor[HTML]{E3E3E3} 
Soft    & Medium  & Hard  \\ \hline
$b = 3$ & $b = 5$ & $b=7$ \\ \hline
\end{tabular}
\end{table}

\begin{figure}[!h]
    \centering
    \begin{subfigure}{0.5\textwidth}
        \centering
        \includegraphics[width=\linewidth]{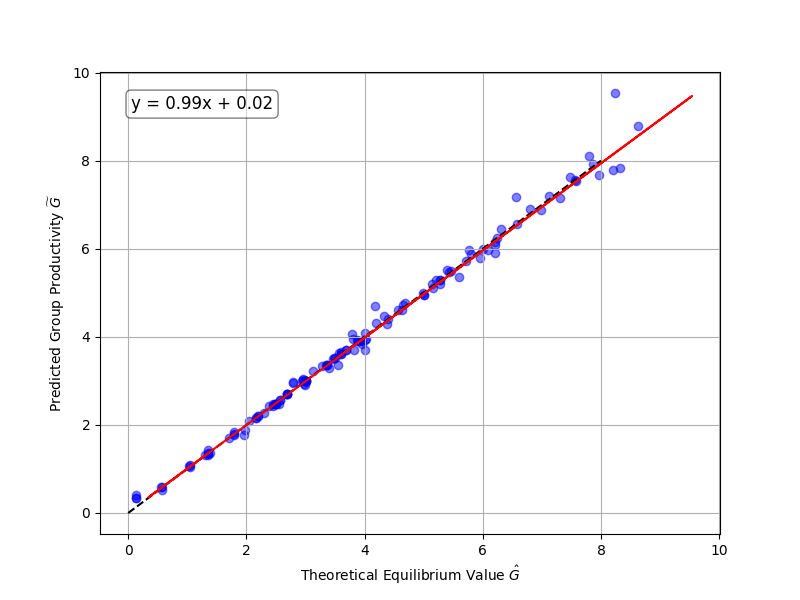}
        \caption{Scatter plot}
        \label{fig:scatter_plot}
    \end{subfigure}
    \begin{subfigure}{0.5\textwidth}
        \centering
        \includegraphics[width=\linewidth]{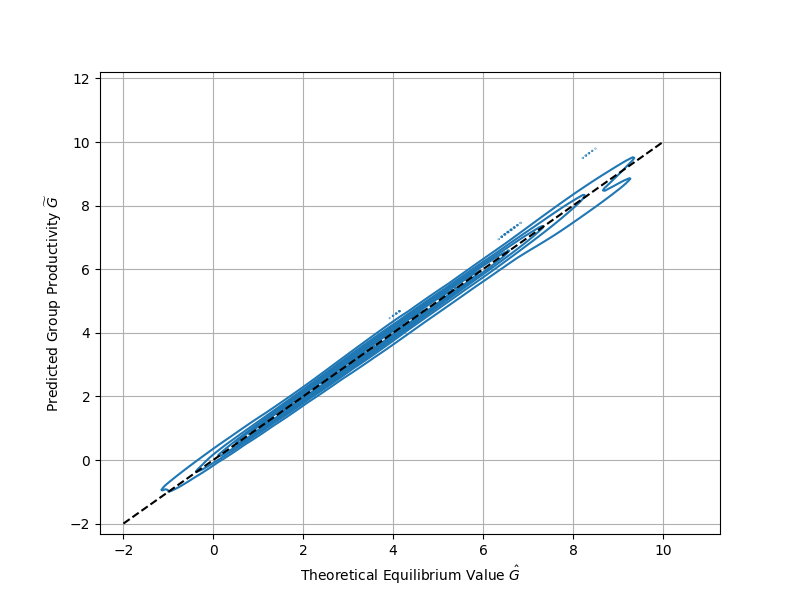}
        \caption{Density estimation contour}
        \label{fig:density_plot}
    \end{subfigure}
    \caption{Correspondence between predicted team productivity (horizontal axis) and theoretical NE values of team productivity (vertical axis).}
    \label{fig:scatter_and_density}
\end{figure}

For each experimental configuration \([(p_1, p_2), \rho, b]\), we train the pair of agents. Each training consists of playing a teamwork game \(5 \times 10^4\) times. The learned policies for each experimental configuration, \(\widetilde{\textbf{a}} = (a_1, a_2)\), result in an experimental teamwork outcome \(\widetilde{G}\). This experimental result is compared with the theoretical value of the teamwork outcome at the Nash equilibrium, \(\hat{G}\), computed according to Sections (\ref{subsec:r_in_additive}, \ref{subsec:r_in_conjunctive}, and \ref{subsec:r_in_disjunctive}) respectively. Figure \ref{fig:scatter_and_density} shows the scatter plot and the density estimation contour for these pairs \((\widetilde{G}, \hat{G})\).

Overall, these results underscore the robustness of our multiagent system, as evidenced by the remarkable agreement between our agents' learned strategies and the corresponding NE strategies. This alignment validates the efficacy of our learning framework to approximate the Nash equilibria of teamwork games. Furthermore, the near-perfect fit observed between the theoretical and empirical values suggests that our system is well-designed and theoretically grounded, paving the way for its application in diverse real-world scenarios beyond the scope of traditional game theory assumptions.
On the left side of Figure \ref{fig:scatter_and_density}, we present a scatter plot depicting the relationship between the theoretical equilibrium value $\hat{G}$ and the empirical team productivity $\widetilde{G}$ derived from the MA-MAB setting. The horizontal axis represents the theoretical equilibrium value $\hat{G}$, while the vertical axis represents the empirical or predicted team productivity. Additionally, the plot includes the identity line $y=x$ (dashed black line) and a regression line (shown in red). The parameters of the linear regression are $(\alpha= 0.99, \beta=0.02)$, with a goodness of fit of $\xi^2 = 0.992$. 

On the other hand, the density estimation contours displayed on the right side of the figure highlight the regions where the majority of the probability mass is concentrated. These contours provide evidence of the consistency between our agents' learned strategies and the NE predictions.

\subsection{Interaction between team composition and task type}\label{subsec:composition_and_type}

\begin{figure}[h]
    \centering
    \begin{subfigure}{0.32\textwidth}
        \centering
        \includegraphics[width=\linewidth]{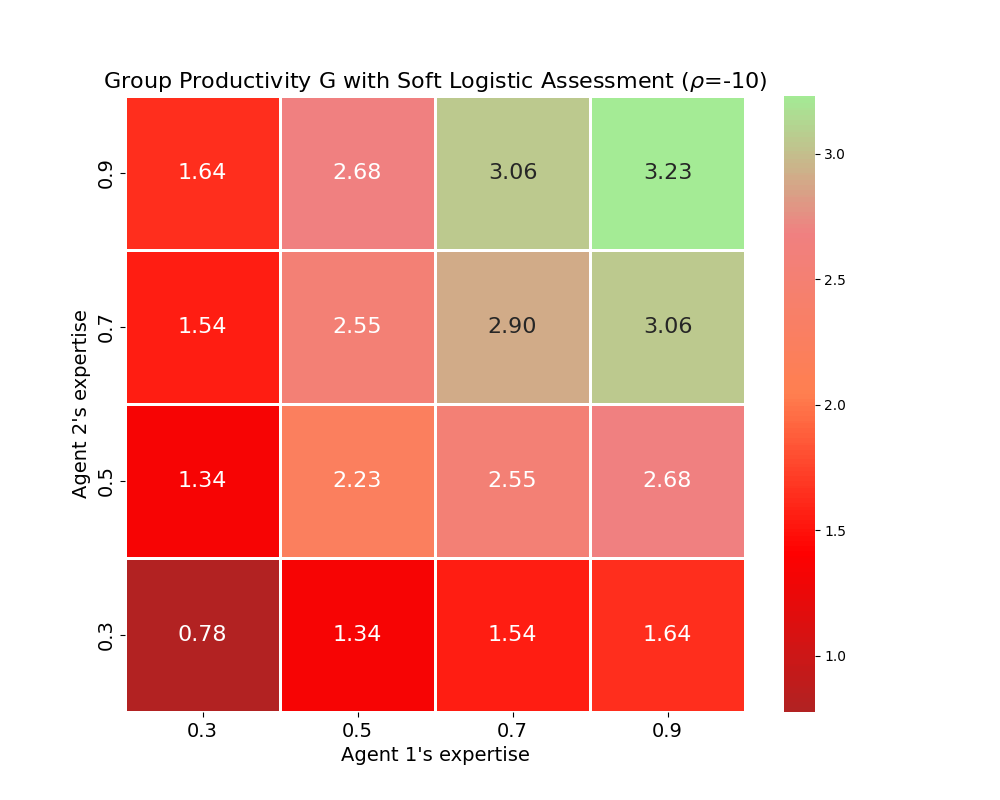}
        \label{subfig:weak_soft}
    \end{subfigure}
    \begin{subfigure}{0.32\textwidth}
        \centering
        \includegraphics[width=\linewidth]{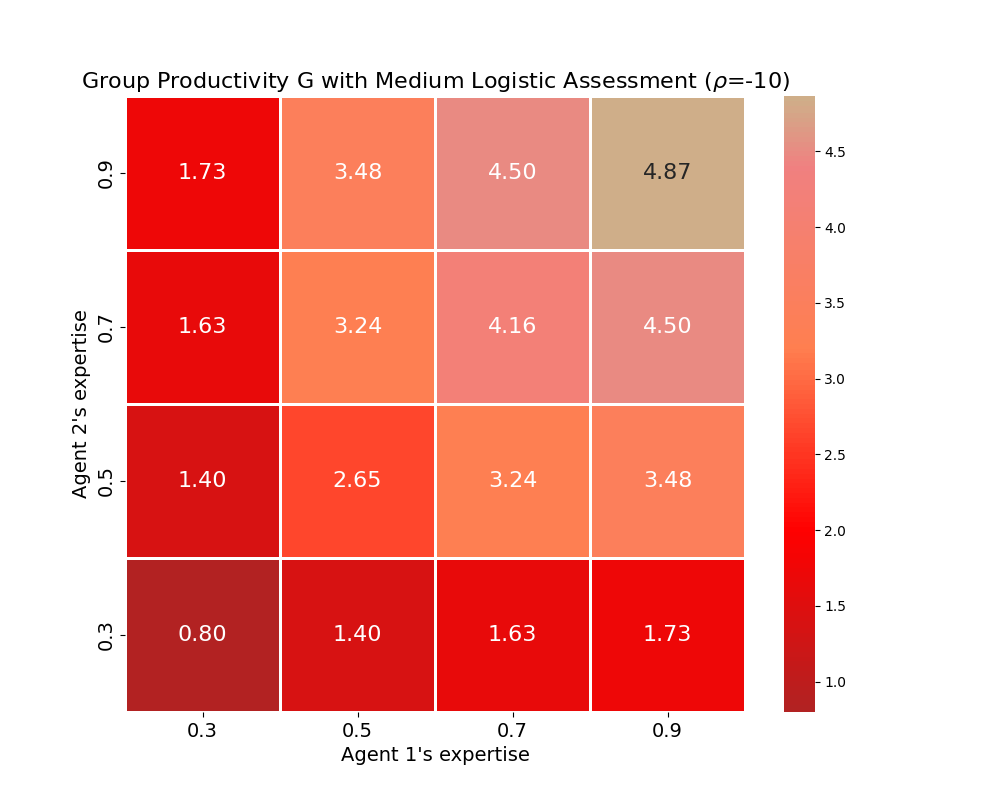}
        \label{subfig:weak_medium}
    \end{subfigure}
    \begin{subfigure}{0.32\textwidth}
        \centering
        \includegraphics[width=\linewidth]{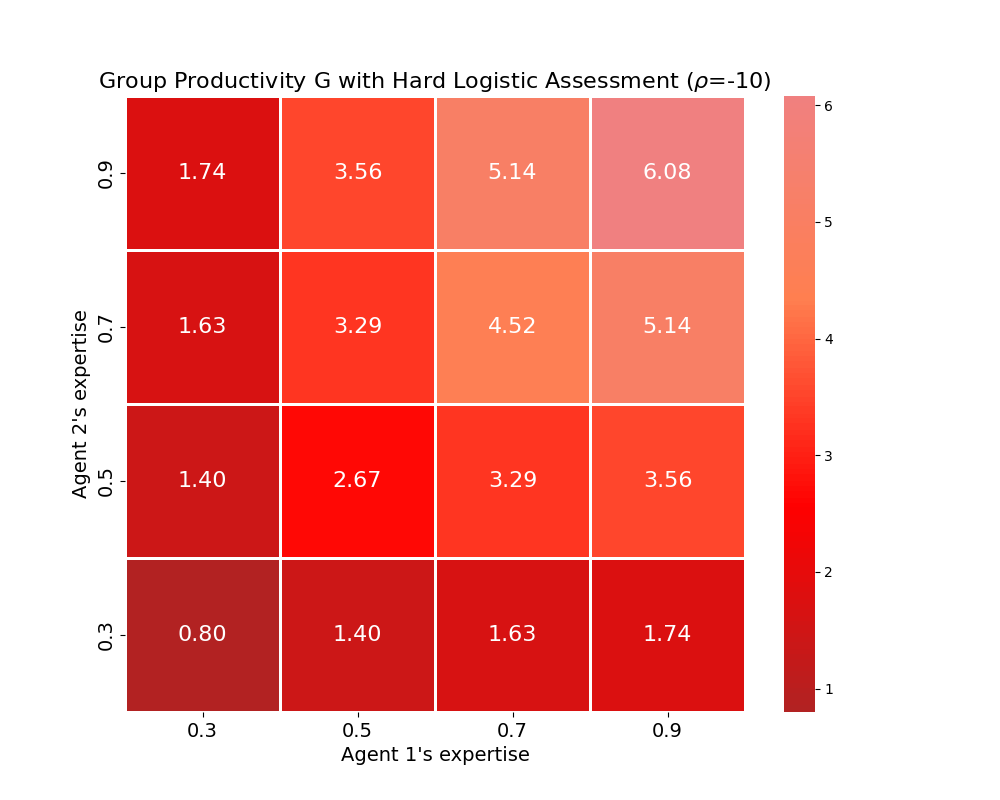}
        \label{subfig:weak_hard}
    \end{subfigure}

    \begin{subfigure}{0.32\textwidth}
        \centering
        \includegraphics[width=\linewidth]{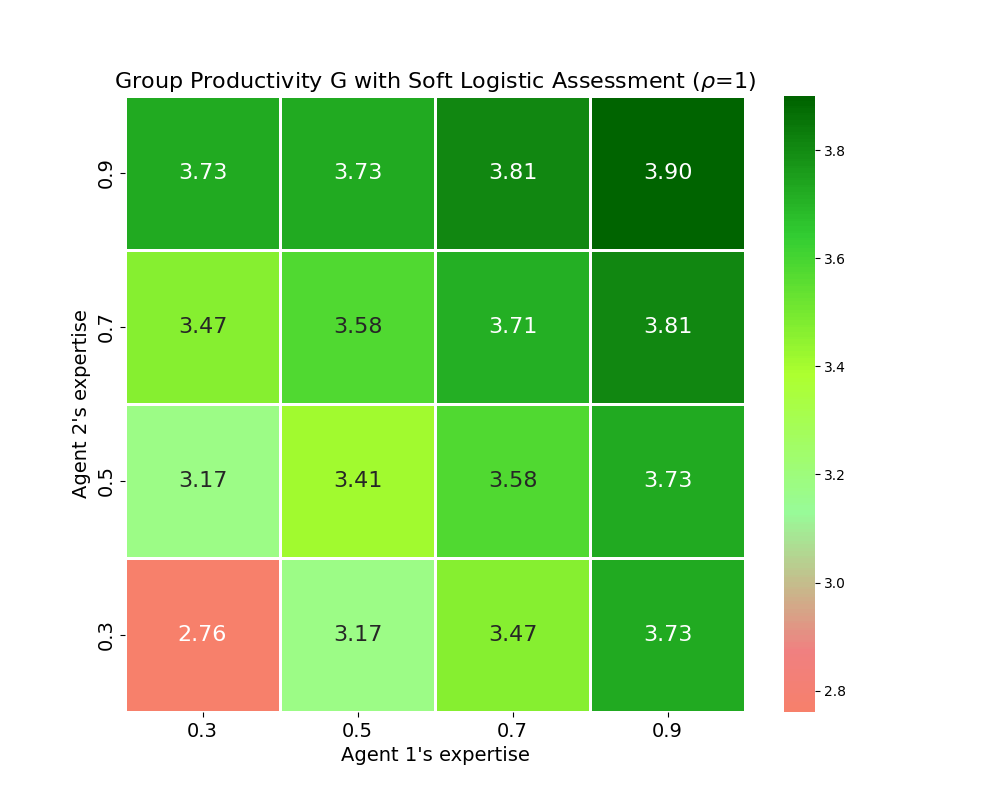}
        \label{subfig:BBV_soft}
    \end{subfigure}
    \begin{subfigure}{0.32\textwidth}
        \centering
        \includegraphics[width=\linewidth]{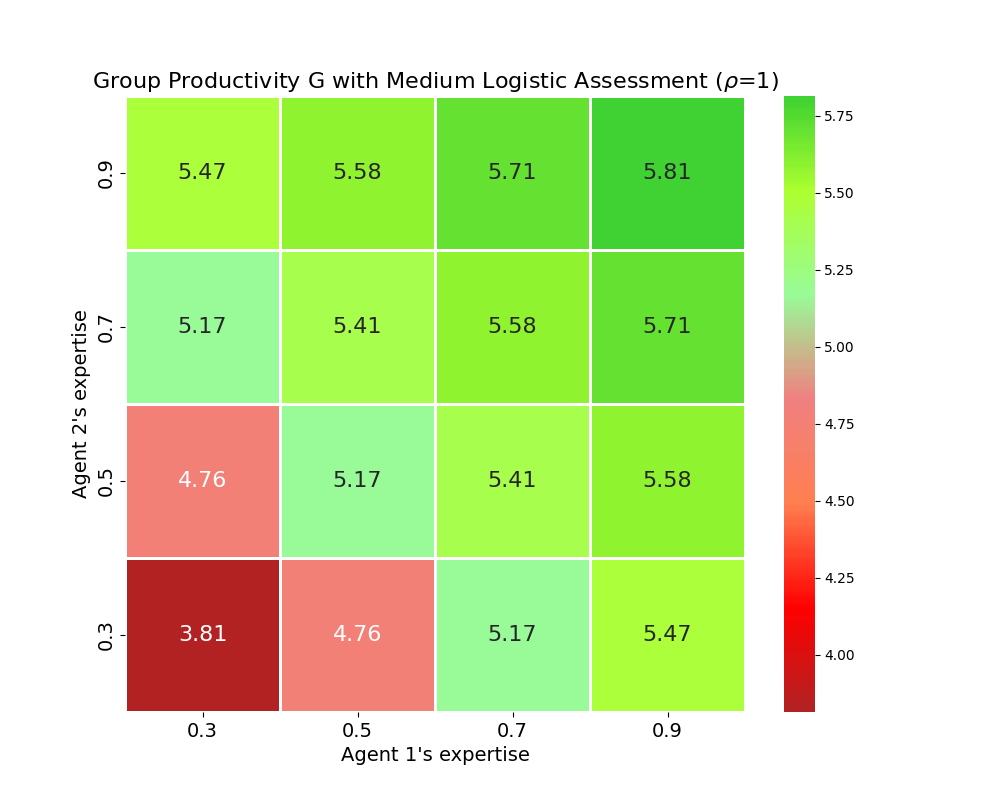}
        \label{subfig:BBV_medium}
    \end{subfigure}
    \begin{subfigure}{0.32\textwidth}
        \centering
        \includegraphics[width=\linewidth]{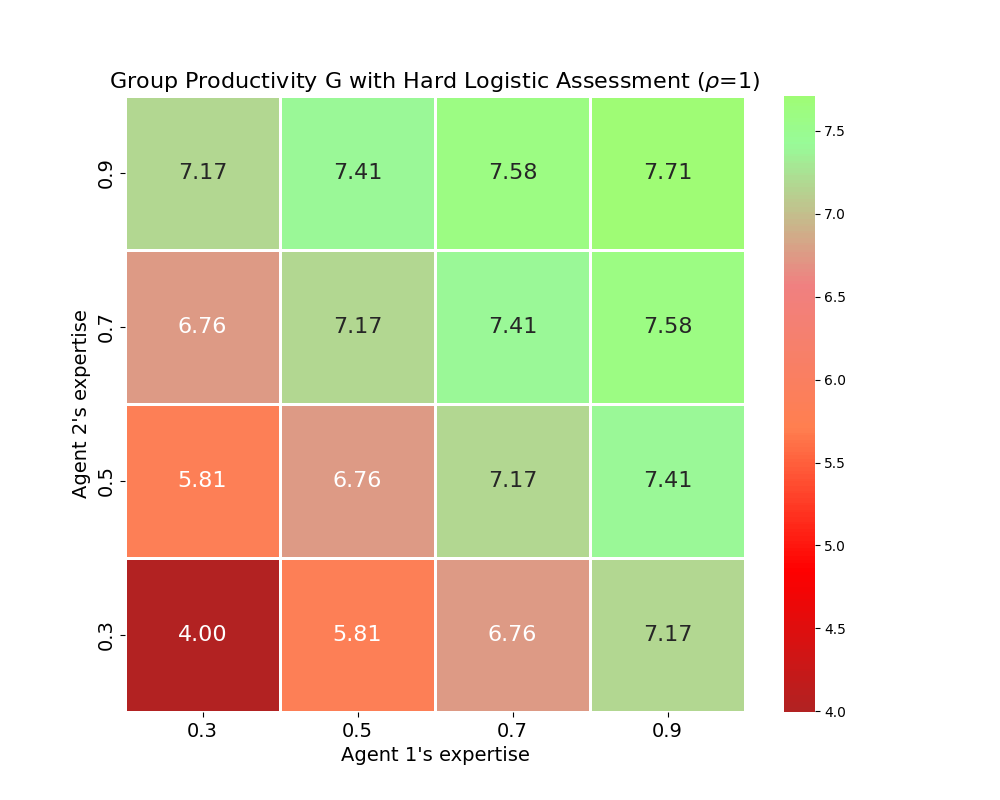}
        \label{subfig:BBV_hard}
    \end{subfigure}

    \begin{subfigure}{0.32\textwidth}
        \centering
        \includegraphics[width=\linewidth]{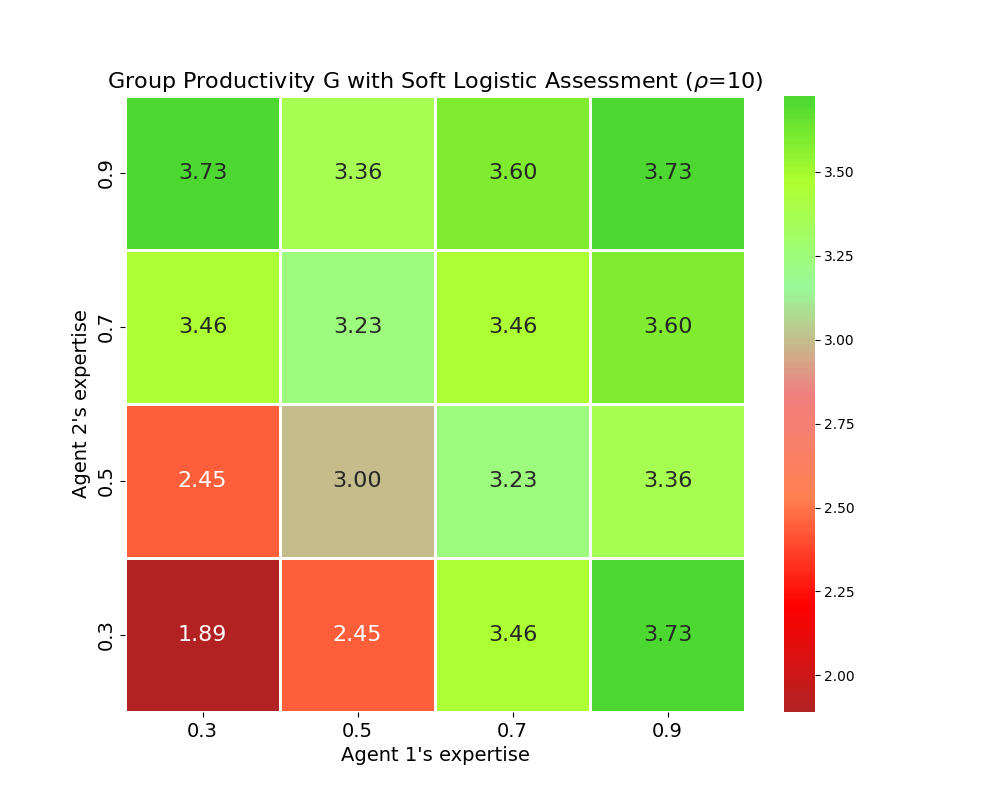}
        \label{subfig:good_soft}
    \end{subfigure}
    \begin{subfigure}{0.32\textwidth}
        \centering
        \includegraphics[width=\linewidth]{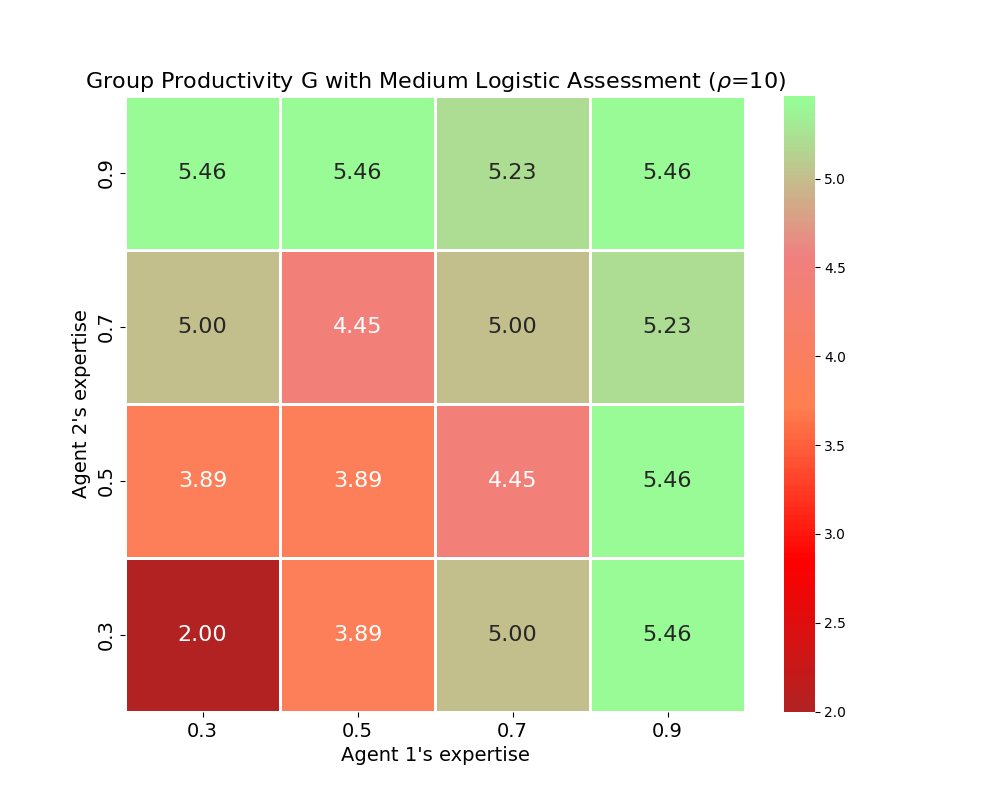}
        \label{subfig:good_medium}
    \end{subfigure}
    \begin{subfigure}{0.32\textwidth}
        \centering
        \includegraphics[width=\linewidth]{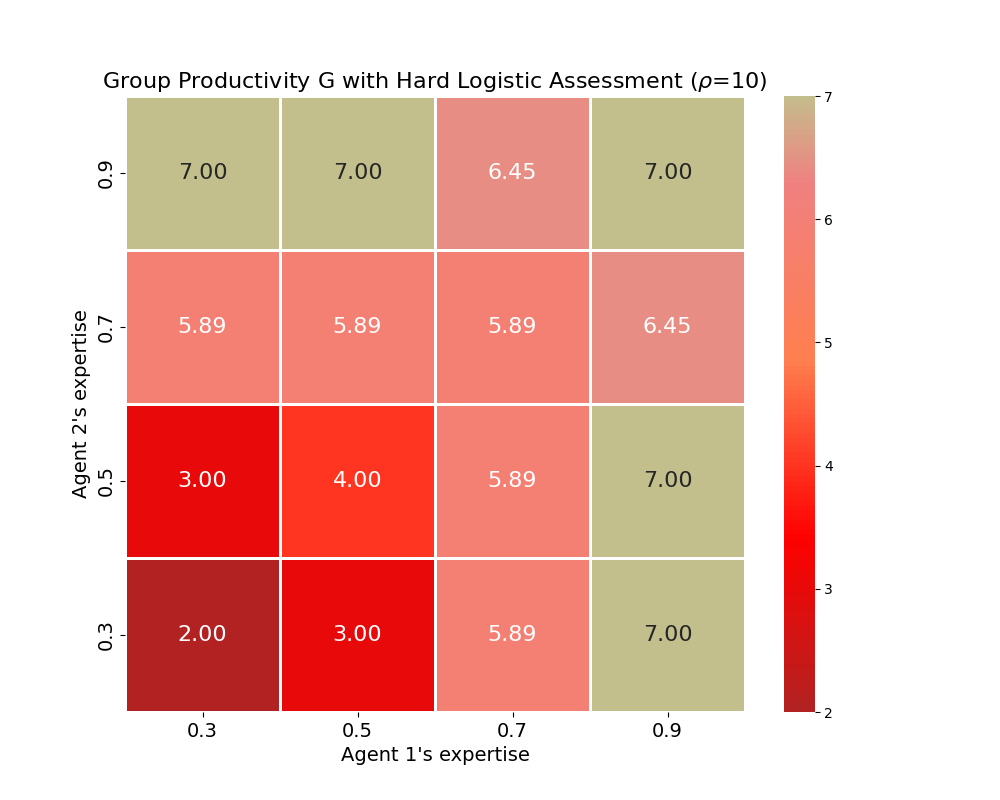}
        \label{subfig:good_hard}
    \end{subfigure}
    
   \caption{Variation in team productivity across different teams for 1) conjunctive tasks (upper row, $\rho = -10$), 2) additive tasks (middle row, $\rho = 1$), and 3) disjunctive tasks (lower row, $\rho = 10$), dependent on the passing threshold $b$ (left column: $b=3$, middle column: $b=5$, right column: $b=7$). Colour legend: red indicates low productivity, while green indicates high productivity, with the passing threshold determining the colour transition.}
    \label{fig:combined}
\end{figure}

In this section, we aim to highlight our main observations regarding how task type affects team productivity. We will use Figure \ref{fig:combined} to illustrate our findings. 

This figure utilises a colour scale to represent the levels of team productivity predicted by our model for different teams. We consistently compare the same teams, formed by all possible combinations of pairs of agents with expertise levels extracted from \(p_i^{\mathcal{T}} \in \{0.3, 0.5, 0.7, 0.9\}\), and create a plot for quasi-conjunctive, additive, and disjunctive tasks (\(\rho \in \{-10,1,10\}\)). We consider passing thresholds from soft to high: \(b \in \{3,5,7\}\).
The current discussion emphasises the changes in team productivity as the task type varies. Thus, we will be describing how team productivity changes as one moves \textbf{across rows} in the figure.

Focusing on conjunctive tasks (first row of Fig. \ref{fig:combined}), we observe that as one moves perpendicularly away from the diagonal of homogeneous teams, performance drops. For example, there is a decrease of $38\%$ (from $G = 2.65$ to $G = 1.63$) in team productivity between the homogeneous pair $\textbf{p} = (0.5, 0.5)$ and the heterogeneous pair $\textbf{p} = (0.3, 0.7)$ in the conjunctive task with medium assessment treatment. We interpret this fact as a confirmation of Steiner's assertion that \textbf{conjunctive tasks are potentially better carried out by homogeneous teams} (with similar expertise levels) than heterogeneous ones \citep{steiner1972group}, at least in the case of rational agents operating under the utility functions considered here.

Regarding additive tasks (middle row), we observe that they consistently yield the highest outcomes no matter what column (assessment treatment) we look at. This aligns with findings in social psychology, where additive tasks typically result in higher team potential than the best team member’s potential. Visually, team productivity remains constant or nearly constant for a given passing threshold along the direction perpendicular to the identity diagonal of the represented matrices. For example, pairs with expertise levels ($p_i^{\mathcal{T}} = 0.5$, $p_j^{\mathcal{T}} = 0.7$) and ($p_i^{\mathcal{T}} = 0.3$, $p_j^{\mathcal{T}} = 0.9$), exhibit very similar performances among themselves. Again, we consider this a confirmation of Steiner's hypothesis that \textbf{team productivity is unaffected by the heterogeneity of team members in additive tasks} \citep{steiner1972group}.

Finally, the productivity observed in the disjunctive task falls somewhere between that observed in the conjunctive and additive tasks across columns. A visual examination of the bottom row of Fig. \ref{fig:combined} reveals that, unlike the conjunctive scenario, team performance $G$ declines as one moves perpendicularly to the diagonal of homogeneous teams. For instance, consider the pairs with expertise levels $\textbf{p} = (0.5, 0.5)$ and $\textbf{p}=(0.3, 0.7)$ in the disjunctive task with medium assessment treatment. The first team produces $G = 3.89$ units of teamwork outcome, whereas the second team produces $G = 5.00$. This results in an increase in productivity of $29\%$. This observation reinforces Steiner's proposition that \textbf{disjunctive tasks may be more effectively tackled by teams with diverse expertise levels} rather than by those with uniform capabilities \citep{steiner1972group}.

\subsection{Interaction between team composition and assessment treatment}\label{subsec:assessment_impact}
Based on our theoretical analyses in Section \ref{subsec:NE}, the difficulty of the assessment should influence productivity. In this section, we investigate whether this impact is observed in the empirical approximations of equilibrium teamwork outcomes achieved by our agents. Given a task, we can study how productivity varies from a soft assessment treatment to a hard assessment treatment as we adjust the parameter $b$ in the logistic function (Eq. \ref{eq:logistic_function}).  To make these comparisons, we look again at Fig. \ref{fig:combined} and turn our attention to changes in $G$ \textbf{column-wise}. Take for instance the second row (additive task) in Fig \ref{fig:combined}. In the soft assessment condition (left column), all teams except one, comprising two agents with low expertise ($p = 0.3$), pass. However, as the pass threshold is raised, more teams fail. In the hard assessment condition, only teams with an aggregated expertise of at least 60\% of the total possible expertise pass:

\[\frac{\sum \limits_{j = 1}^n p_j^{\mathcal{T}}}{n} \geq 60\%\] 

Given a fixed team, productivity generally increases as the pass threshold is raised. However, these increments are consistently smaller when resulting productivity falls short of the new passing threshold. A consistent behavioural pattern emerges across different task types: when the task is easy enough, productivity increases as $b$ is raised, and it increases even more when passing becomes possible at the next level of evaluative difficulty. This trend ceases once the task requirements exceed the team's capacity

While it's evident that teams typically exhibit poorer performance on conjunctive tasks (top row), this does not imply an absence of potential improvement. Indeed, our plots reveal that when agents have the opportunity to enhance their performance (\textit{i.e.}, when their expertise levels allow for increased contributions without significant sacrifices in their free time), more rigorous evaluation criteria result in higher team productivity values during conjunctive tasks. However, when agents have sufficiently low expertise levels, this incentive mechanism becomes futile, as the cost of increasing contributions outweighs the potential benefits in terms of their free time. The scenario depicted on the upper row of Fig. \ref{fig:combined} illustrates an intrinsically challenging task, where under the soft assessment treatment, only the top two teams ($\textbf{p} = (0.7, 0.9)$ and $\textbf{p} = (0.9, 0.9)$) manage to succeed. To enhance performance, it may be necessary to provide agents with additional time $\Delta_t$, or to soften the assessment by reducing the evaluation threshold $b$.

As for disjunctive tasks (bottom row), we once again notice a trend towards increased team productivity as the evaluation criteria become more stringent. The nature of the disjunctive task alters the threshold that defines what constitutes ``low'' team expertise compared to the conjunctive scenario. In a conjunctive task, hardening the assessment with an increase of $b$ from $3$ to $5$ barely elicited a response in productivity levels from pairs of agents whose maximum expertise was less than or equal to $0.5$. For example, the pair \(\textbf{p} = (0.3, 0.3)\) only raised their teamwork outcome from $0.78$ to $0.8$. With a further increase of $b$ from \(5\) to \(7\), the agents became even more insensitive to the evaluation threshold, with only teams boasting minimum expertise of \(p_i^{\mathcal{T}} \geq 0.7\) showing an increase in contributions by at least one-tenth. For example, the pairs with \(\textbf{p} = (0.7, 0.9)\) raised their teamwork outcome from \(4.5\) to \(5.14\), a \(14\%\). However, the pairs \(\textbf{p} = (0.5, 0.7)\) only increased their outcome by \(1\%\).
In the disjunctive setting, all teams react to the initial increase in assessment hardness, and those with at least one member with high expertise $p_i^{\mathcal{T}} \geq 0.7$ respond to the subsequent difficulty escalation. Notably, we observe a decrease in team productivity during the transition from ``medium'' to ``hard'' for the pairs with low expertise levels ($p_{1,2}^{\mathcal{T}} \leq 5$). Thus, in this case, an increase in the demand threshold has a detrimental effect on team productivity, \textbf{a phenomenon akin to a loss of motivation in human teams \citep{steiner1972group} that can be explained as a direct consequence of our agents' reward-maximising behaviour}.

Finally, we observe that as the task becomes more challenging, the amount by which productivity improves, compared to the previous difficulty level, becomes smaller. In other words, while productivity still tends to increase with increasing difficulty, the rate of this increase diminishes. Specifically, for the studied additive, conjunctive and disjunctive tasks, we consistently observe a reduction in the rate at which the output productivity $G$ increases as the assessment becomes more stringent (\textit{i.e.}, with higher $b$).

\subsection{Individual strategies}\label{subsec:assessment_impact_individual}

The previous sections have focused on the team outcome $G$ and how it changes as the assessment criteria become more demanding or the nature of the task is altered. Now we study how the assessment difficulty and the task type affect the \textbf{individual} strategies followed by the agents. As a reminder, by \textit{strategies} we refer to the percentage $a_i$ of their turn that the agents choose to allocate to the task. An agent with expertise $p_i^{\mathcal{T}}$ who takes strategy $a_i$ makes a teamwork contribution $g_i$ (Eq. (\ref{eq:individual_contribution})). Table \ref{tab:strategies_additive} presents the learned strategies teams of two agents for an additive task ($\rho = 1$), where $\Delta_t = 10$ and $\alpha = 2$ for all agents. The table showcases strategies for the pairs who are assigned an additive task, categorised under three assessment treatments: soft (lighter-shaded), medium (medium-shaded), and hard (darker-shaded). For simplicity, we fill only the upper diagonal of the matrices in the table, which are symmetric.

\begin{table}[htbp]
\caption{Individual team contribution across different teams for an additive task under three different assessment treatments. For simplicity, we fill only the upper diagonal of the matrices, which are symmetric.}
\label{tab:strategies_additive}
    \begin{subtable}{\linewidth}
        \caption{Soft Assessment Treatment ($b = 3$)}
        \label{tab:strategies_additive_3_matrix}
        \centering
        \begin{tabular}{r|rrrr}
            \rowcolor[HTML]{EEE0FF}
            \toprule
            \diagbox{$p_2^\mathcal{T}$}{$p_1^\mathcal{T}$}& 0.30 & 0.50 & 0.70 & 0.90 \\
            \midrule
            \cellcolor[HTML]{EEE0FF}0.30 & (46\%, 46\%) & (20\%, 52\%) & (0\%, 50\%) & (0\%, 41\%) \\
            \cellcolor[HTML]{EEE0FF}0.50 &  &  (34\%, 34\%) & (16\%, 40\%) &(0\%, 41\%) \\
            \cellcolor[HTML]{EEE0FF}0.70 & &  & (27\%, 27\%)&  (23\%, 22\%) \\
            \cellcolor[HTML]{EEE0FF}0.90 &   &  &   & (22\%, 22\%) \\
            \bottomrule
        \end{tabular}

    \end{subtable}

    \begin{subtable}{\linewidth}
        \centering
        \caption{Medium Assessment Treatment ($b = 5$)}
        \label{tab:strategies_additive_5_matrix}
        \begin{tabular}{r|rrrr}
            \rowcolor[HTML]{CFC3FF}
            \toprule
            \diagbox{$p_2^\mathcal{T}$}{$p_1^\mathcal{T}$}& 0.30 & 0.50 & 0.70 & 0.90 \\
            \midrule
            \cellcolor[HTML]{CFC3FF}0.30 & (64\%, 64\%) & (46\%, 68\%) & (20\%, 66\%) & (0\%, 61\%) \\
            \cellcolor[HTML]{CFC3FF}0.50 & &(52\%, 52\%) & (34\%, 53\%) & (16\%, 53\%) \\
            \cellcolor[HTML]{CFC3FF}0.70 & & & (40\%, 40\%) & (27\%, 43\%) \\
            \cellcolor[HTML]{CFC3FF}0.90 & & & &  (32\%, 32\%)  \\
            \bottomrule
        \end{tabular}
 
    \end{subtable}

    \begin{subtable}{\linewidth}
        \centering
        \caption{Hard Assessment Treatment ($b = 7$)}
        \label{tab:strategies_additive_7_matrix}
        \begin{tabular}{r|rrrr}
            \rowcolor[HTML]{AB96FF}
            \toprule
           \diagbox{$p_2^\mathcal{T}$}{$p_1^\mathcal{T}$} & 0.30 & 0.50 & 0.70 & 0.90 \\
            \midrule
            \cellcolor[HTML]{AB96FF} 0.30 & (67\%, 67\%) & (64\%, 78\%) & (46\%, 77\%) & (20\%, 73\%) \\
            \cellcolor[HTML]{AB96FF}0.50 & & (68\%, 68\%) & (52\%, 66\%) & (34\%, 63\%) \\
            \cellcolor[HTML]{AB96FF}0.70 & & & (53\%, 53\%) & (40\%, 53\%)  \\
            \cellcolor[HTML]{AB96FF}0.90 & & & & (43\%, 43\%)  \\
            \bottomrule
        \end{tabular}

    \end{subtable}
\end{table}
We consistently observe in this table that the agent with the highest expertise level contributes more time than their counterpart in all pairs. Remarkably, elevating the evaluation level does not diminish individual contributions in any observed case, albeit there are instances of no discernible effect. For instance, the less proficient agent in the $(p_1^{\mathcal{T}} = 0.3, p_2^{\mathcal{T}} = 0.9)$ pair contributes nothing under soft or medium evaluation (free-riding), only engaging with a 20\% commitment when the passing threshold reaches $b=7$.

Still in the additive case, the average contribution of an agent with expertise $p_i^{\mathcal{T}}$ increases across all teams as the passing threshold rises. However, transitioning from a soft to a medium evaluation system yields a more pronounced effect than transitioning from medium to hard. In other words, agents consistently increase their task engagement as evaluation demands intensify, with a more significant augmentation observed when transitioning from a soft to medium evaluation compared to a medium to hard evaluation.

For further understanding of individual strategies in equilibrium for conjunctive and disjunctive tasks, readers can refer to similar tables like Table \ref{tab:strategies_additive} in \ref{app:individual_effort_tables_dis_conj}. In the conjunctive scenario, we notice that the weakest member of all possible pairs in all assessment treatments contributes more effort than their counterpart. This phenomenon is akin to the \textbf{Köhler effect} observed in human teams, where an inferior team member performs a difficult task better in a team or cooperative situation than expected based on their individual performance \citep{feltz2011buddy}. Furthermore, increasing the evaluation level did not decrease individual contributions in any observed case, although there were some instances where there was no discernible effect. 

On the other hand, in the disjunctive case, elevating the evaluation level causes a drop in individual contributions when the expertise of the strongest member is sufficiently high ($> 0.7$). For instance, if we look at the corresponding table in \ref{app:individual_effort_tables_dis_conj}, we can see that the less proficient agent in the $(p_1 = 0.5, p_2 = 0.7)$ pair contributes in one of the equilibria under soft evaluation, but free-rides as soon as the passing threshold is heightened. This phenomenon mirrors the \textbf{social loafing} effect observed in human teams, where some members exploit the efforts of others while free-riding.

Table \ref{tab:avg_individual_increments_all} displays the percentage increase in the average team contribution of each agent to the additive, conjunctive and disjunctive task as the passing threshold was raised from soft to medium and from medium to hard.

\begin{table}[htbp]
    \caption{Percentage increase in average dedication ($\Delta_{\%}(\overline{a}_i)$) of each agent to different tasks as the passing threshold rises}
    \label{tab:avg_individual_increments_all}
    \centering
    \begin{subtable}{\linewidth}
        \centering
        \caption{Additive Task ($\rho = 1$)}
        \begin{tabular}{rrr}
            \toprule
            \rowcolor[HTML]{E3E3E3} & Soft ($b=3$) to Medium ($b=5$) & Medium ($b=5$) to Hard ($b=7$) \\
            \rowcolor[HTML]{E3E3E3} $p_i^{\mathcal{T}}$ & $\Delta_{\%}(\overline{a}_i)$ & $\Delta_{\%}(\overline{a}_i)$ \\             
            \midrule
            0.3  & $97\%$            & $52\%$           \\ 
            0.5  & $104\%$           & $50\%$           \\ 
            0.7  & $68\%$            & $39\%$           \\ 
            0.9  & $45\%$            & $35\%$           \\ 
            \bottomrule
        \end{tabular}

    \end{subtable}

    \begin{subtable}{\linewidth}
        \centering
        \caption{Conjunctive Task ($\rho = -10$)}
        \begin{tabular}{rrr}
            \toprule
            \rowcolor[HTML]{E3E3E3} & Soft ($b=3$) to Medium ($b=5$) & Medium ($b=5$) to Hard ($b=7$) \\
            \rowcolor[HTML]{E3E3E3} $p_i^{\mathcal{T}}$ & $\Delta_{\%}(\overline{a}_i)$ & $\Delta_{\%}(\overline{a}_i)$ \\    
            \midrule
            0.3  & 5\%            & 0\%            \\
            0.5  & 27\%           & 1\%            \\
            0.7  & 46\%           & 11\%           \\
            0.9  & 53\%           & 24\%           \\ 
            \bottomrule
        \end{tabular}

    \end{subtable}

    \begin{subtable}{\linewidth}
        \caption{Disjunctive Task ($\rho = 10$)}
        \centering
        \begin{tabular}{rrr}
            \toprule
            \rowcolor[HTML]{E3E3E3} & Soft ($b=3$) to Medium ($b=5$) & Medium ($b=5$) to Hard ($b=7$) \\
            \rowcolor[HTML]{E3E3E3} $p_i^{\mathcal{T}}$ & $\Delta_{\%}(\overline{a}_i)$ & $\Delta_{\%}(\overline{a}_i)$ \\    
            \midrule
            0.3  & 6\%           & 0\%          \\
            0.5  & 4\%           & -36\%         \\
            0.7  & 42\%           & 18\%         \\
            0.9  & 49\%            & 28\%          \\
            \bottomrule
        \end{tabular}

    \end{subtable}

\end{table}

In all cases but one ($\rho = 10$, $p_i^{\mathcal{T}} = 0.5$), the percentage increase as the evaluation hardens is non-negative. That is, the more demanding the evaluation, the more committed the agents. In the mentioned exception, the weakest agent learns not to contribute as soon as the task becomes maximally difficult (hard treatment) and their counterpart possesses greater expertise, hence the observed negative difference in the table. These increments tend to be higher overall when agents face an additive task compared to other cases, although this is not always the case (the agent with expertise $p_i^{\mathcal{T}} = 0.9$ always learns to lead regardless of the task type, exhibiting a steady pattern and consistently increasing its contributions by a similar amount for a given assessment treatment).

Observing Table \ref{tab:avg_individual_increments_all}, we discern that agents with lower expertise levels demonstrate a heightened sensitivity to the type of task at hand. For example, in the case of $p_1^\mathcal{T} = 0.3$ in an additive task, the agent's effort nearly doubles when the assessment is hardened and the passing threshold shifts from $b=3$ to $b=5$. Examining Table \ref{tab:strategies_additive_3_matrix}, this notable escalation stems from the agent's non-contribution under the soft assessment treatment ($b=3$) when their counterpart exhibits higher expertise ($p_2^\mathcal{T} = 0.7$). Likewise, transitioning from medium to hard assessment treatment prompts this agent to contribute even when their counterpart possesses substantial expertise ($p_2^\mathcal{T} = 0.9$). In conjunctive tasks ($\rho = -10$), this agent consistently contributes, leaving minimal scope for performance improvement. Finally, in disjunctive tasks ($\rho = 10$), the agent refrains from contributing altogether in any equilibrium where their counterparts surpass them significantly in expertise (\textit{i.e.}, $p_2^\mathcal{T} > 0.5$). This equilibrium remains resilient to variations in assessment treatment, indicating that \textbf{hardening the assessment does not induce changes in this agent's policy}. In general, agents with high expertise ($p_i^{\mathcal{T}} \geq 0.7$) exhibit more steadfast policies concerning task types. Regardless of the task, they consistently enhance their contributions as the passing threshold rises, with a comparable degree of increase across all tasks. Hence, these agents are more influenced by shifts in evaluation criteria than by the inherent nature of the task.

\subsection{A binary pass/fail assessment scenario}\label{subsec:discontinous_rewards}

To explore the versatility of our MA-MAB simulation platform beyond game-theoretical contexts, we explore a scenario featuring a discontinuous reward function, mirroring a binary pass/fail evaluation scenario. Our reward function is now defined as:
\begin{equation}\label{eq:discontinuous_reward}
R_i  = \Delta_t^\alpha\cdot(1 - a_i)^{\alpha} \cdot H(G)
\end{equation}

where $H(G)$ is a modified version of the Heaviside function:

\begin{figure}[h]
    \centering
    \includegraphics[width=0.5\linewidth]{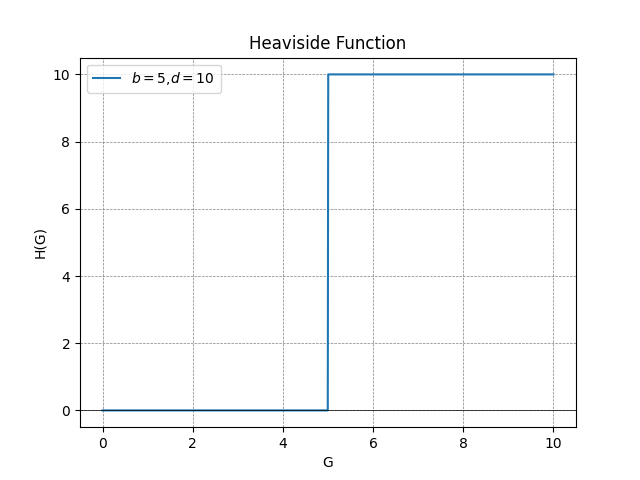}
    \caption{A Heaviside function with parameters $b = 3$ and $d = 10$}
    \label{fig:enter-label}
\end{figure}

\begin{figure}[!h]
    \centering
    \includegraphics[width=0.55\linewidth]{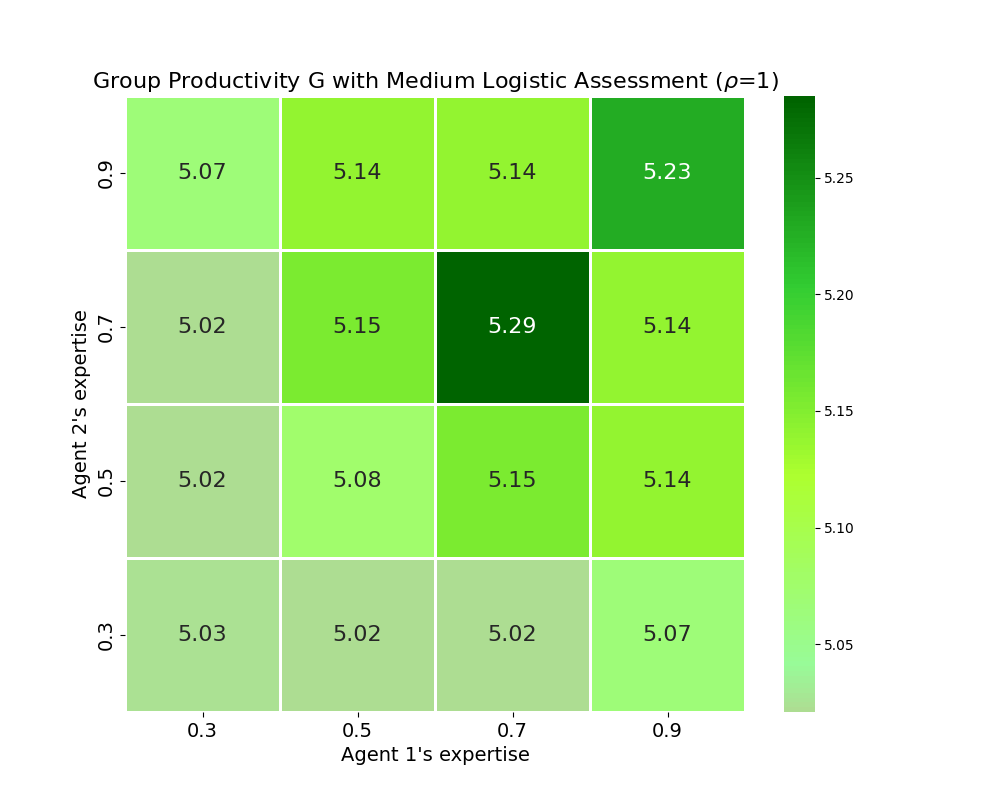}
    \caption{team productivity for an additive task under the medium assessment ($b=5$) treatment and using a Heaviside evaluation function. Colour Legend: Red would denote low productivity under the passing threshold, while Green signifies high productivity over the passing threshold.}
    \label{fig:group_productivity_discrete_additive}
\end{figure}

The function returns 0 for \(G < b\) and $d$ for \(G \geq b\). In other words, it acts as a step function that ``switches on'' at \(G = b\):

\[
 H(G)=\left\{\begin{matrix}
        0 & \text{if }G < b  \\
        d & \text{if }G \geq b \\
    \end{matrix}\right.
\]
Note how $d$ here assumes the role of the right horizontal asymptote in Eq. (\ref{eq:logistic_function}), while $b$ represents the passing evaluation point. To maintain consistency with previous sections, we set the maximum evaluation possible $d=10$. We consistently compare the same teams as in Section \ref{subsec:composition_and_type}, formed by all possible combinations of pairs of agents with expertise levels extracted from $p_i^{\mathcal{T}} \in \{0.3, 0.5, 0.7, 0.9\}$. We concentrate on additive tasks under the medium assessment treatment ($\rho = 1$, $b=5$). For each experimental configuration $[(p_1^{\mathcal{T}}, p_2^{\mathcal{T}}), \rho=1, b=5]$, we train the pair of agents, with each training consisting of playing a teamwork game with a Heavise evaluation function $5 \times 10^4$ times. The learned policies for each experimental configuration, $\widetilde{\textbf{a}} = (a_1, a_2)$, result in an experimental teamwork outcome $\widetilde{G}$. Similar to Section \ref{subsec:composition_and_type}, we employ a colour scale to depict teamwork outcome under this discontinuous reward regime (Figure \ref{fig:group_productivity_discrete_additive}). Red signifies lower productivity levels, while green denotes higher productivity levels. To compute the displayed quantities in the figure, each game was learned three times, resulting in three values of the teamwork outcome for each team $\textbf{p}$, $\{\widetilde{G}_1^{\textbf{p}}, \widetilde{G}_2^{\textbf{p}}, \widetilde{G}_3^{\textbf{p}}\}$. From each sample, we calculate the mean $\overline{G}$. Figure \ref{fig:group_productivity_discrete_additive} shows the resulting $\overline{G}$ for all teams under this binary pass/fail assessment scenario. To ensure the accuracy of average value $\overline{G}$, we verified that the percentage dispersion of $\widetilde{G}$ in each sample was less than $2\%$, where the percentage dispersion is computed as:

\[ \text{Percentage Dispersion of } G = \frac{\max\{\widetilde{G}_1^{\textbf{p}}, \widetilde{G}_2^{\textbf{p}}, \widetilde{G}_3^{\textbf{p}}\} - \min{\widetilde{G}_1^{\textbf{p}}, \widetilde{G}_2^{\textbf{p}}, \widetilde{G}_3^{\textbf{p}}\}} }{\overline{G}} \times 100\% \]

Table \ref{tab:dispersion} presents the percentage dispersion of the samples $\{\widetilde{G}_1^{\textbf{p}}, \widetilde{G}_2^{\textbf{p}}, \widetilde{G}_3^{\textbf{p}}\}$ for each team. For simplicity, we fill only the upper diagonal of the table, which is symmetric.

\begin{table}[!h]
    \caption{Percentage dispersion of the teamwork outcome samples $\{\widetilde{G}_1^{\textbf{p}}, \widetilde{G}_2^{\textbf{p}}, \widetilde{G}_3^{\textbf{p}}\}$ for each team. In all cases, the percentage dispersion is lower than $2\%$}
    \label{tab:dispersion}
    \centering
        \begin{tabular}{r|rrrr}
            \rowcolor[HTML]{E3E3E3}
            \toprule
            \diagbox{$p_2^\mathcal{T}$}{$p_1^\mathcal{T}$}& 0.30 & 0.50 & 0.70 & 0.90 \\
            \midrule
            \cellcolor[HTML]{E3E3E3}0.30 & $0\%$ & $0.4\%$ & $0.6\%$ & $1.19\%$ \\
            \cellcolor[HTML]{E3E3E3}0.50 &  &  1\% & 1.78\% & 0.2\% \\
            \cellcolor[HTML]{E3E3E3}0.70 & &  & 0.2\% &  0.18\% \\
            \cellcolor[HTML]{E3E3E3}0.90 &   &  &   & 0\% \\
            \bottomrule
        \end{tabular}
\end{table}

We observe that the strategies agents converge to when working on an additive task under the medium assessment treatment show some variability. However, the final team productivity $G$ consistently remains the same across all experiments for the pair, with very low dispersion. A notable detail we have observed is that when pairs are highly heterogeneous (\textit{i.e.} agents show large differences in expertise), particularly in cases such as {($p_i^{\mathcal{T}}=0.3, p_j^{\mathcal{T}}=0.7$), ($p_i^{\mathcal{T}}=0.3, p_j^{\mathcal{T}}=0.9$), ($p_i^{\mathcal{T}}=0.5, p_j^{\mathcal{T}}=0.9$)}, the less skilled agent refrains from contributing. Interestingly, in this case, their behaviour is consistent, with variance again falling below $2\%$. This contrasts with the continuous assessment scenario depicted in Table \ref{tab:strategies_additive_5_matrix}, where such behaviour was observed only in the cases of the two maximally heterogeneous pairs ($p_i^{\mathcal{T}}=0.3, p_j^{\mathcal{T}}=0.9$). Therefore, the inclusion of a binary pass/fail evaluation function causes agents to polarise in their behaviour compared to continuous evaluation scenarios. In situations where there are significant differences between them (such as when the expertise gap is 0.4 or greater), the less skilled agent chooses to withdraw, thereby leaving their counterpart to shoulder the entire workload. Nonetheless, these adapted strategies ensure the team consistently meets the assignment criteria, as indicated by the absence of red hues in Figure \ref{fig:group_productivity_discrete_additive}.

\section{Conclusions, limitations, and future work}\label{sec:conclusions}

Game theory and computer science have extensively studied the concept of teamwork. In both fields, a team is perceived as a set of players or agents that need to cooperate with each other. Game theory assumes that all players in a team share the same utility function and examines teamwork using cooperative game theory tools. On the other hand, computer science, particularly Multiagent Systems (MAS), views teams as groups of self-interested agents who require cooperation to achieve individual success, rendering intentionally defective strategies unviable. Although both approaches have proven useful for studying environments where cooperation is in the interest of all players or agents, they are inadequate for modelling human teamwork.

Humans are capable of collaboration and recognise its benefits. However, collaboration may not be in the interest of all members of a human team, and establishing binding collaboration agreements is often unattainable in this context. Consequently, defection occurs.

This paper positions itself at the intersection of game theory and multiagent systems, aiming to address the question of how to learn theoretically grounded predictions of team performance in settings where cooperation is not enforced.

To answer this, we have dedicated the first part of this paper to formalising the concept of an aggregative teamwork game: a non-cooperative game that incorporates the elements identified by \citeauthor{steiner1972group} in social psychology as most informative for explaining the productivity of a human team. After presenting this model, we characterised the Nash equilibria (NE) of teamwork games. Notably, these equilibria extend beyond those seen in general public good games, with variations arising from the novel elements introduced in our model.

The second part of the article proposes a multiagent multi-armed bandit (MA-MAB) framework in which agents learn strategic behaviour that approximates the NE of the game. We validated the proposed MA-MAB system by empirically proving the convergence of the learned strategies towards the approximated NE of the game. After this validation, we analysed the impact of isolated variables on the teamwork outcome.

When we studied team productivity across different task types, our findings aligned with \citeauthor{steiner1972group}'s hypotheses on team composition and productivity. Specifically, conjunctive tasks where the weakest members have the most influence are better performed by homogeneous teams (similar expertise levels). In additive tasks (where contributions are summed), team composition does not affect productivity. Conversely, when the strongest members have the most influence (disjunctive tasks), heterogeneous teams (varying expertise levels) are more effective than homogeneous ones.

In analysing the effects of increasing the difficulty of the evaluation on teamwork outcome, we find that if the task is easy enough, team productivity increases. However, when the evaluation becomes so difficult that the task requirements exceed the capacity of the team, we observe that what \citeauthor{steiner1972group} would call a loss of motivation is actually an NE in the teamwork game, to which our reward-maximising agents' strategies converge.

When examining how individual strategies were impacted by task type and assessment difficulty, we found that agents with lower expertise adapted their strategies based on task type, but their contributions varied less with assessment difficulty. In conjunctive tasks, these agents contributed more compared to the other types of tasks, which mirrors the Köhler effect in human teams. Conversely, in additive and disjunctive tasks, the least skilled agent contributed less, sometimes completely free-riding. This phenomenon mirrors the social loafing effect in human teams. Meanwhile, agents with high expertise levels were more responsive to changes in evaluation criteria, contributing more as assessment difficulty increased. With equal assessment difficulty, high-expertise agents behaved nearly the same in all tasks. 

In our last experiment, we expanded our analysis beyond analytically solvable games and empirically studied the policies of our agents after learning in a binary pass/fail scenario. Our findings showed that when pairs of individuals had high levels of heterogeneity, less skilled individuals tended to refrain from contributing, leaving their partner to handle the entire workload.

Our approach has some limitations. For example, we have only considered teamwork as a single interaction, while real-world teamwork typically involves multiple interactions over time. Furthermore, we have assumed that the parameter $\alpha$, which measures the importance of agents' free time, is the same for all agents. In the future, it would be interesting to explore how agents' policies change when $\alpha$ is randomly distributed across the population. Similarly, the learning rate is currently the same for all agents and decays at the same pace, which may not be optimal \citep{christianos2021scaling}. Future studies will explore scenarios where parameter sharing is implemented judiciously rather than indiscriminately. We also plan to expand our simulations to include more agents, study the effects of team size on productivity, and incorporate discretionary tasks that have different weights for each agent's contribution.

In conclusion, our work has paved the way for a new research direction in cooperative AI \citep{dafoe2020open}, particularly in multiagent systems. This approach aims to achieve a better understanding of the social dynamics that drive collective behaviour in teams where cooperation is not enforced. We believe that incorporating factors that enhance the ecological validity of our model will enable us to provide a valuable tool for designing teamwork settings that favour cooperation in teams and societies in general.

\clearpage
\appendix

\section{Proof of lemma \ref{lemma:assessment_guarantees} }\label{app:assessment_guarantees}
In a teamwork game, player $i$'s utility is determined by their private contribution $x_i$ and the assessment that the team outcome $G$ receives:

\begin{equation}
 \widetilde{u_i}(x_i, \sigma(G))
\end{equation}

Let $\widetilde{u_i}(x_i, G)$ be the utility function of some general public good game. The properties outlined in Definition \ref{def:assessment_function} ensure that the new utility function $\widetilde{u_i}(x_i, \sigma(G))$ is both differentiable and quasi-concave: 

\begin{itemize}
     \item Differentiability: Let $\sigma(G)$ and $\widetilde{u_i}(x, G)$ be both differentiable. For $f(x,G) = \widetilde{u_i}(x_i, \sigma(G))$ to be differentiable, we need to ensure that its partial derivatives are well-defined.

    \begin{itemize}
        \item Since \(\sigma\) does not depend on \(x\), the partial derivative of \(\widetilde{u_i}(x, \sigma(G))\) with respect to \(x\) is the partial derivative of \(\widetilde{u_i}(x, G)\) evaluated at \((x, \sigma(G))\):
        \[
        \frac{\partial \widetilde{u_i}(x, \sigma(G))}{\partial x} = \frac{\partial \widetilde{u_i}(x, G)}{\partial x} \bigg|_{(x, \sigma(G))}.
        \]
        
        \item Applying the chain rule and considering that \(\sigma(\cdot)\) is differentiable, we have:
        \[
        \frac{\partial \widetilde{u_i}(x, \sigma(G))}{\partial G} = \frac{\partial \widetilde{u_i}(x, G)}{\partial G} \bigg|_{(x, \sigma(G))} \cdot \sigma'(G).
        \]
    \end{itemize}

    Since both partial derivatives exist and are continuous, \(f(x, G) = \widetilde{u_i}(x, \sigma(G))\) is differentiable.
    
    \item Quasi-Concavity: Let \(\widetilde{u_i}(x, G)\) be a quasi-concave function and \(\sigma(G)\) be an increasing function. We want to show that \(f(x, G) = \widetilde{u_i}(x, \sigma(G))\) is quasi-concave. Consider two arbitrary points \((x_1, G_1)\) and \((x_2, G_2)\) and a \(\lambda \in [0, 1]\). The convex combination of these points is
        \[
            (\lambda x_1 + (1-\lambda) x_2, \lambda G_1 + (1-\lambda) G_2).
        \]

We evaluate the function \(f\) at this point:
\[
f(\lambda x_1 + (1-\lambda) x_2, \lambda G_1 + (1-\lambda) G_2) = \widetilde{u_i}(\lambda x_1 + (1-\lambda) x_2, \sigma(\lambda G_1 + (1-\lambda) G_2)).
\]
Without loss of generality, assume \( G_1 \leq G_2 \). Since \(\sigma\) is an increasing function, and the convex combination of \(G_1\) and \(G_2\) lies within the interval \([G_1, G_2]\), it follows that
\[
\sigma(\lambda G_1 + (1-\lambda) G_2) \in [\sigma(G_1), \sigma(G_2)].
\]

Now, using the quasi-concavity of \(\widetilde{u_i}\):
\[
\widetilde{u_i}(\lambda x_1 + (1-\lambda) x_2, \sigma(\lambda G_1 + (1-\lambda) G_2)) \geq \min \{ \widetilde{u_i}(x_1, \sigma(G_1)), \widetilde{u_i}(x_2, \sigma(G_2)) \}.
\]

This is precisely
\[
f(\lambda x_1 + (1-\lambda) x_2, \lambda G_1 + (1-\lambda) G_2) \geq \min \{ f(x_1, G_1), f(x_2, G_2) \}.
\]

Thus, we have shown that \(f(x, G) = \widetilde{u_i}(x, \sigma(G))\) is quasi-concave.
\end{itemize}

To determine whether goods $x$ and $G$ remain normal goods under the preferences of the composite utility function $f(x, G) = \widetilde{u_i}(x, \sigma(G))$, we need to analyse how the demands for $x$ and $G$ change in response to a change in income:

Since $\widetilde{u_i}$ is a utility function in which $x$ and $G$ are normal goods, we know that $\left.\frac{\partial x}{\partial I}\right|_{u_i} > 0$ and $\left.\frac{\partial G}{\partial I}\right|_{u_i} > 0$, where ``$\left.\right|_{u_i}$'' denotes \textit{under the preferences of $u_i$} and $I$ is income. 

\begin{itemize}
    \item Given that \(G\) is a normal good under the preferences of \(\widetilde{u_i}\), we know that \(\left.\frac{\partial G}{\partial I}\right|_{u_i} > 0\), indicating that an increase in income leads to an increase in the demand for \(G\) under \(\widetilde{u_i}\).

    \item Now, consider the composite function \(f(x, G) = \widetilde{u_i}(x, \sigma(G))\). To demonstrate that \(G\) remains a normal good under the preferences of \(f\), we need to show that \(\left.\frac{\partial G}{\partial I}\right|_f > 0\) also under these preferences.

    Since \(G\) is an argument of \(\widetilde{u_i}\), we can use the chain rule to derive \(G\) with respect to \(I\) under the preferences of \(f\):

        \[
            \left.\frac{\partial G}{\partial I}\right|_f = \frac{\partial \widetilde{u_i}}{\partial G}\cdot \frac{d\sigma}{dG} \cdot \left.\frac{\partial G}{\partial I}\right|_{u_i}
        \]

\end{itemize}

Given that \(\frac{\partial \widetilde{u_i}}{\partial G}\) is positive (as \(G\) is a normal good under the preferences of \(\widetilde{u_i}\) ), and \(\frac{d\sigma}{dG}\) is positive (as \(\sigma\) is an increasing function), we can see that $\left.\frac{\partial G}{\partial I}\right|_f$  remains positive under the preferences of \(f\). Hence, both goods $x$ and $G$ remain normal under the new utility function $f(x, G) = \widetilde{u_i}(x, \sigma(G))$ QED.

\section{Proof of Lemma \ref{lemma:existence_weak_link_group_productivity} and Proposition \ref{prop:uniqueness_weak_link_group_productivity}}\label{app:proof_conjunctive}

In a central lemma in \citep{cornes2007weak}, the authors go on to show the circumstances under which given a general public good game with a concave CES aggregator, a well-defined replacement function exists:

\begin{lemma}\label{lemma:existence_weak_link}
    Let $\Gamma =\left(\mathcal{A}_i, \widetilde{u_i}\right)_{i\in \mathcal{I}}$ be a generalised aggregative game featuring a concave CES social composition function. Under increasing, normal preferences, there is a unique $g_i \in \left(0,w_i\right]$ for each player satisfying the first-order condition: 
    \[
         \left.\frac{\partial \widetilde{u_i}}{\partial g_i}\right|_{\textbf{g}_{-i}} \geq 0
    \]
    with equality if $g_i < w_i$. Furthermore, the feasibility condition $G^\rho \geq \beta_i g_i^\rho$ is satisfied for $G\geq \overline{G}_i$ if $0 < \rho < 1$ and for $G\leq \overline{G}_i$ if $\rho < 0$.
\end{lemma}

This lemma implies that in general public good games involving a concave aggregator, the replacement correspondence $r_i(G)$ is a well-defined function satisfying the first order conditions and has domain $\left[\overline{G}_i, \infty\right)$ if $0 < \rho < 1$ and $\left[0, \overline{G}_i\right]$ if $\rho < 0$.
A later proposition in that same paper clarifies the existence and uniqueness of equilibria in this model: 
\begin{proposition}
    Under the conditions of the Lemma above, the game has a unique equilibrium whenever the utility function $\widetilde{u_i}(x_i, G)$ is such that the corresponding indifference map is asymptotic to the axes.
\end{proposition}

In a teamwork game, player $i$'s utility is a function of $x_i$ and of the assessment $\sigma(G)$:

\[
    \widetilde{u_i}(x_i, \sigma(G)) 
\]
The properties of $\sigma(G)$ in Definition \ref{def:assessment_function} ensure that if $i$'s preferences represented by the utility function $\widetilde{u_i}(x_i, G)$ were increasing and normal, then they would remain increasing and normal under $\widetilde{u_i}(x_i, \sigma(G)) $ as justified in Lemma \ref{lemma:assessment_guarantees}. Hence, the Lemma in this appendix still applies in a teamwork game $(\Gamma, \mathcal{T}, \sigma)$ whenever the aggregator (\textit{i.e.} the teamwork outcome) is concave. This is the case of teamwork games involving a conjunctive task. In our case, $g_i < w_i$ is replaced by $g_i < p_i^{\mathcal{L}}\cdot \Delta_t$. Likewise, we can adapt the above proposition by using the teamwork game utility function $\widetilde{u_i}(x_i, \sigma(G))$.

\section{Analysis of a player's replacement function in teamwork games with conjunctive tasks}\label{app:comparative_statics_weaker}
Let us analyse how the replacement function in a conjunctive task ($\rho < 1$) is impacted by changes in each of the elements of a teamwork game:
\[
        r_i(G)^{\rho-1}\left(\Delta_t - \frac{r_i(G)}{p_i^{\mathcal{T}}}\right) = \frac{\alpha}{\beta_i p_i^{\mathcal{T}}} \cdot G^{\rho-1}\cdot\frac{\sigma(G)}{\sigma^\prime(G)}
\]

\begin{itemize}
    \item Deriving both sides of the equation \textit{w.r.t.} $G$ shows that the derivative of the left-hand side must be negative:

    \[
        \frac{dr_i}{dG} \cdot r_i^{\rho -2} \cdot \left[\rho \left(\Delta_t - \frac{r_i}{p_i^{\mathcal{T}}}\right) - \Delta_t\right] = \frac{\alpha}{\beta_i p_i^{\mathcal{T}}}G^{\rho -2}\left[(\rho -1)\frac{\sigma(G)}{\sigma^{\prime}(G)} + G \cdot \frac{\sigma^{\prime}(G)^2 - \sigma^{\prime\prime}(G)\sigma(G)}{\sigma^{\prime}(G)^{2}}\right]
    \]
    Indeed, using that $\frac{r_i(G)}{p_i^{\mathcal{T}}} \in \left[0, \Delta_t \right]$ and $(\rho -1) < 0$, we see that the term in brackets in the left-hand side of the equation above must be negative. Thus, there are two behavioural regimes for $r_i(G)$ as $G$ increases
    \begin{itemize}
        \item If the term in brackets of the right-hand side of the equation is positive, then $\frac{dr_i(G)}{dG} <0$. This is true whenever $\frac{|\rho -1|}{G} < \frac{\sigma^{\prime}(G)^2 - \sigma^{\prime\prime}(G)\sigma(G)}{\sigma^{\prime}(G)\sigma(G)}$.
        \item If the term in brackets of the right-hand side of the equation is negative, $\frac{dr_i(G)}{dG} >0$, which happens whenever $\frac{|\rho -1|}{G} > \frac{\sigma^{\prime}(G)^2 - \sigma^{\prime\prime}(G)\sigma(G)}{\sigma^{\prime}(G)\sigma(G)}$.
    \end{itemize}
    
    This resonates with the qualitative regimes that \citet{cornes2007weak} distinguished (see Section \ref{subsub:conditions_r_in_conjunctive}): if $\rho$ is sufficiently negative as in the second case, a substitution effect occurs and encourages player $i$ to contribute more since a given increase in $G$ generated by their contribution now has a lower opportunity cost in terms of private consumption. Otherwise, an income effect dominates and the preferred contributions by $i$ will decrease as $G$ increases.

    \item When we derive \textit{w.r.t} $\alpha$, we obtain the equation:

    \[
        \frac{dr_i}{d\alpha} r_i^{\rho -2} \left((\rho -1)\Delta_t - \frac{\rho}{p_i^{\mathcal{T}}}r_i\right) = \frac{G^{\rho -1}}{\beta_i p_i^{\mathcal{T}}}\cdot \frac{\sigma(G)}{\sigma^{\prime} (G)}
    \]
    The right-hand side of this equation is positive because $\sigma'(G) > 0$, and the term in parenthesis on the left is negative by the same argument as above. Hence, we conclude that $\frac{dr_i(G)}{d\alpha} <0 $. This result is in line with what was observed in Section \ref{subsec:r_in_additive}: as the importance that a player gives to their leisure time increases, their optimal contributions decrease.

    \item Deriving \textit{w.r.t.} $\beta$:

    \[
        \frac{dr_i}{d\beta_i} r_i^{\rho -2} \left((\rho -1)\Delta_t - \frac{\rho}{p_i^{\mathcal{T}}}r_i\right) = -\frac{G^{\rho -1} \alpha}{\beta_i^{2} p_i^{\mathcal{T}}}\cdot\frac{\sigma(G)}{\sigma^{\prime}(G)}
    \]

    we see that the right-hand term is negative. Again, the factor in parenthesis on the left is negative. Hence, the derivative is positive to maintain the sign: $\frac{dr_i(G)}{d\beta_i} >0$. The higher the weight or importance that a player's contribution has, the higher the optimal contribution will be. 
    
    \item Regarding the term $\frac{\sigma(G)}{\sigma^{\prime}(G)}$, we had already established that this coefficient represents the ratio between the incentive $\sigma(G)$ and the slope of the tangent curve to the incentive $\sigma(G)$, and thus is a measure of \textit{opportunity cost} in terms of leisure. Deriving with respect to it yields:

    \[
        \frac{dr_i}{d (\sigma/\sigma^{\prime})} r_i^{\rho -2}\left((\rho -1)\Delta_t - \frac{\rho}{p_i^{\mathcal{T}}}r_i\right) = \frac{\alpha}{\beta_i p_i^{\mathcal{T}}}G^{\rho -1}
    \]

    Using similar arguments to the above, it is straightforward to see that $\frac{dr_i(G)}{d (\sigma/\sigma^{\prime})} < 0$. Just as in Section \ref{subsec:r_in_additive}, higher values of this ratio (\textit{i.e.} higher opportunity cost in terms of enjoyed leisure) induce lower contributions.

    \item When deriving with respect to the turn duration, $\Delta_t$, one obtains:

    \[
        \frac{dr_i}{d\Delta_t} = \frac{p_i^{\mathcal{T}} \cdot r_i}{\rho\cdot r_i + p_i^{\mathcal{T}}(1-\rho)\Delta_t}
    \]
    Using $\rho -1 <0$, we see that the fraction's numerator and denominator are positive. Thus, we conclude that $\frac{dr_i(G)}{d\Delta_t} >0$, meaning that the more time a player has to perform a team task, the more they will contribute.

    \item Finally, the derivative \textit{w.r.t.} the expertise level $p_i^{\mathcal{T}}$ yields, after some re-arrangement of the terms:

\[
    \frac{dr_i}{dp_i^{\mathcal{T}}} \cdot \left[r_i^{\rho -2}\cdot(\rho - 1) \cdot \left(\Delta_t - \frac{r_i}{p_i^{\mathcal{T}}}\right) - \frac{r_i^{\rho -1}}{p_i^{\mathcal{T}}} \right] =  - \frac{\alpha}{\beta_i \left(p_i^{\mathcal{T}}\right)^{2}} G^{\rho -1} \frac{\sigma(G)}{\sigma^{\prime}(G)} - \frac{r_i^\rho}{\left(p_i^{\mathcal{T}}\right)^{2}}
\]
    Using that $\rho -1 < 0$ and $r_i(G) \in \left[0, p_i^{\mathcal{T}}\Delta_t\right]$, we can see that $\frac{dr_i(G)}{dp_i^{\mathcal{T}}} >0$. Again, just as in the canonical case, higher expertise levels induce higher contributions to player $i$.
\end{itemize}

\section{Proof of lemma \ref{lemma:best_response_correspondence_aggregative_games}}\label{app:proof_lemma}
Given a \textbf{general public good game} with Cobb-Douglas preferences (Equation (\ref{eq:cobb-douglas_preferences})) and $\rho >1$, \citet{cornes2007weak} demonstrate in a lemma the existence of three potential behavioural regimes within the set of best responses:

\begin{lemma}
    There exists a threshold value $G_{-i}^{*}>0$ and a positive real-valued function $b_i$ on $\left[0, G_{-i}^{*}\right]$ such that:
    \begin{equation}
        \mathcal{B}_i(G_{-i}) = 
            \begin{cases}
                \left\{0\right\} & \text{if } G_{-i} > G_{-i}^{*} \\
                \left\{0, b_i(G_{-i})\right\} & \text{if } G_{-i} = G_{-i}^* \\
                \left\{b_i(G_{-i})\right\} & \text{if } G_{-i} < G_{-i}^* \\
            \end{cases}
    \end{equation}
\end{lemma}

Their proof revolves around the observation that the stationary points of the Cobb-Douglas utility function satisfy an expression that can be decomposed into a term dependent on $g_i$, denoted $\phi(g_i)$, and another dependent on $G_{-i}$:

\begin{equation}\label{eq:proof_lemma}
    \frac{d\widetilde{u_i}}{dg_i} = \phi(g_i) - \alpha \cdot G_{-i}^{\rho} = 0
\end{equation}

The authors establish that $\phi(g_i)$ attains a maximum $\phi_{\max}$, implying that Equation (\ref{eq:proof_lemma}) holds as long as $\alpha \cdot G_{-i}^{\rho}$ is less than or equal to $\phi_{\max}$. Therefore, $\phi_{\max} \geq \alpha \cdot G_{-i}^{\rho}$ serves as a necessary and sufficient condition for the stationary points of the utility function. Completing the proof involves: 1) demonstrating the well-defined nature of $G_{-i}^{*}$, 2) assigning the curve in the upper panel of Figure \ref{fig:utility_good_shot} to the case $G_{-i} > G_{-i}^{*}$, 3) associating the middle panel with $G_{-i} = G_{-i}^{*}$, and 4) assigning the lower panel to the case $G_{-i} < G_{-i}^{*}$.

The approach that we will follow regarding teamwork games mirrors that of \citeauthor{cornes2007weak}: we will take our utility function to be 

\[
\widetilde{u}_i = x_i^{\alpha}\cdot \sigma(G)
\]

and we will break down $\frac{\partial \widetilde{u_i}}{\partial g_i}$ into two components: one dependent on $(g_i, \sigma(G))$, and another related to $G_{-i}$; and we will prove that $\phi(g_i, \sigma(G))$ attains a maximum. After that, we will show that $G_{-i}^{*}$ is well-defined. Then, we will reason how each regime of $G_{-i}$ relates to the panels in Figure \ref{fig:utility_good_shot}.

The stationary points of utility function satisfy:

\[
    \frac{\partial \widetilde{u_i}}{\partial g_i} = \phi(g_i, \sigma(G)) - f(G_{-i})
\]
By definition, the threshold points $(g_i^{*}, G_i^{*})$ would satisfy $\frac{\partial \widetilde{u_i}}{\partial g_i} = 0$. If we demonstrate that $\phi(g_i, \sigma(G))$ has a maximum $\phi_\text{max} \: \forall G$, then there will exist critical points $(g_i^{*}, G_i^{*})$ as long as $f(G_{-i}) \leq \phi_\text{max}$.

We have established that the condition $\frac{\partial \widetilde{u_i}}{\partial g_i}=0$ is equivalent to:
\[
    \frac{\sigma(G)}{\sigma^\prime(G)} \cdot G^{\rho -1} = \left(\Delta_t - \frac{g_i}{p_i^{\mathcal{T}}}\right) \cdot \frac{\beta_i p_i^{\mathcal{T}}}{\alpha}\cdot g_i^{\rho -1}
\]

Manipulating this equation, we obtain:

\[
    \alpha^{\frac{\rho}{\rho -1}}\left(\beta_i g_i^\rho + G_{-i}^\rho\right) = \left[\left(\Delta_t - \frac{g_i}{p_i^{\mathcal{T}}}\right)\cdot \left(\frac{\sigma^\prime(G)}{\sigma(G)}\right)\cdot \beta_i p_i^{\mathcal{T}}\right]^{\frac{\rho}{\rho -1}}{g_i}^{\rho}
\]

So we can write:

\[
    \phi(g_i, \sigma(G)) = \left[\left(\Delta_t - \frac{g_i}{p_i^{\mathcal{T}}}\right)\cdot \left(\frac{\sigma^\prime(G)}{\sigma(G)}\right)\cdot\beta_i p_i^{\mathcal{T}}\right]^{\frac{\rho}{\rho -1}}{g_i}^{\rho} - \alpha^{\frac{\rho}{\rho -1}}\beta_i g_i^\rho
\]

Thus, $\phi(0, \sigma) = 0$ and $\phi(p_i^{\mathcal{T}}\cdot\Delta_t, \sigma) = - \alpha^{\frac{\rho}{\rho -1}}\beta_i (p_i^{\mathcal{T}}\cdot \Delta_t)^\rho$. Fixing $G$, it's straightforward to verify that $\phi(g_i, \sigma)$ has another root, that we denote $c$, at

\[
    c = p_i^{\mathcal{T}} \Delta_t - \frac{\alpha}{\beta_i^{1/\rho}}\cdot \frac{\sigma(G)}{\sigma^{\prime}(G)}
\]

By Rolle's Theorem, since $\phi$ is continuous for any  $g_i \in \left[0, p_i^{\mathcal{T}}\Delta_t\right]$ and differentiable for any $g_i \in \left(0, p_i^{\mathcal{T}}\Delta_t\right)$, $\phi$ will have at least one critical point for a fixed value of $G$. By the Weierstrass' Theorem, we know that the absolute maximum and minimum of $\phi$ are attained on $\left[0, p_i^{\mathcal{T}}\Delta_t\right]$. If we can demonstrate that at some critical point $g_i^{\prime}$, $\phi(g_i^{\prime}, \sigma(G))>0 \: \forall G$, then there will be an absolute maximum in $\left(0, p_i^{\mathcal{T}}\Delta_t\right)$.

Manipulating, we get the following equation for the critical points of $\phi$ by imposing $\frac{\partial \phi}{\partial g_i} = 0$:

\[
    \left(\Delta_t - \frac{g_i}{p_i^{\mathcal{T}}}\right)^{k}\cdot \left(\frac{\sigma^\prime}{\sigma}\right)^{k} \cdot \left(\beta_i\cdot p_i^{\mathcal{T}}\right)^{k} g_i^{\rho} \cdot \left(\frac{-k/p_i^{\mathcal{T}}}{\Delta_t - \frac{g_i}{p_i^{\mathcal{T}}}} + k\frac{\sigma(G)}{\sigma^{\prime}(G)}\beta_i \left(\frac{g_i}{G}\right)^{\rho -1}\cdot \frac{\sigma\sigma^{\prime \prime} - (\sigma^{\prime})^{2}}{(\sigma^{\prime})^{2}} + \frac{\rho}{g_i}\right) - \alpha^{k}\beta_i \rho g_i^{\rho-1} = 0
\]

where $k = \frac{\rho}{\rho -1}$. We can extract the common factor $\left(\Delta_t - \frac{g_i}{p_i^{\mathcal{T}}}\right)^{k}\cdot \left(\frac{\sigma^\prime(G)}{\sigma(G)}\right)^{k} \cdot \left(\beta_i\cdot p_i^{\mathcal{T}}\right)^{k} g_i^{\rho}$ from this equation and substitute it into the expression of $\phi(g_i, \sigma)$ to obtain the sign of $g_i(c, \sigma)$. Doing so, we verify that $g_i(c, \sigma(G)) >0 \: \forall G$ if and only if:

\[
    \frac{-k/p_i^{\mathcal{T}}}{\Delta_t - \frac{g_i}{p_i^{\mathcal{T}}}}  + k \frac{\sigma\sigma^{\prime \prime} - (\sigma^{\prime})^{2}}{\sigma \cdot \sigma^{\prime}} \beta_i \cdot \left(\frac{g_i}{G}\right)^{\rho-1} < 0
\]

Definition \ref{def:assessment_function} of the evaluation function guarantees that this inequality holds for any value of $G$. Hence, the function $\phi$ has a maximum $\phi_{\max} \: \forall G$, Q.E.D.

The correspondence of each panel in Fig. (\ref{fig:utility_good_shot}) to the different regimes of \(G_i\) follows the same reasoning as in \citep{cornes2007weak}. Due to space constraints, we do not repeat that reasoning here.

We can solve for $G_{i}^\rho$ in the global maximum condition (Eq. (\ref{eq:global_max})) and substitute in the local maximum condition (\ref{eq:G_local_max}) to check if $G_{-i}^{*}$ is well-defined:
\[
      \left(\Delta_t - \frac{g_i^{*}}{p_i^{\mathcal{T}}}\right)^{\alpha}\cdot \sigma\left(\left(\frac{\sigma^{'}(G_i^{*})}{\sigma(G_i^{*})}\frac{\beta_i}{\alpha}p_i^{\mathcal{T}}\right)^{\frac{1}{\rho -1}} \cdot \left(\Delta_t - \frac{g_i^{*}}{p_i^{\mathcal{T}}}\right)^{\frac{1}{\rho -1}}g_i^{*}\right) = \Delta_t^{\alpha} \cdot \sigma\left(\left(\left(\frac{\sigma^{'}(G_i^{*})}{\sigma(G_i^{*})}\frac{\beta_i}{\alpha}p_i^{\mathcal{T}}\right)^{\frac{\rho}{\rho -1}} \cdot \left(\Delta_t - \frac{g_i^{*}}{p_i^{\mathcal{T}}}\right)^{\frac{\rho}{\rho -1}} -\beta_i\right)^{\frac{1}{\rho}}g_i^{*}\right)
\]

For $\beta_i >0$, the argument of $\sigma$ in this equation's right-hand side (RHS) is smaller than that of the left-hand side (LHS). Given that $\sigma$ is monotonically increasing, we have that the quotient $\frac{\sigma_{\text{RHS}}}{\sigma_{\text{LHS}}} < 1$. Thus, we can write 

\begin{equation}
    g_i^{*} = p_i^{\mathcal{T}}\Delta_t \left(1 - \left(\frac{\sigma_{\text{RHS}}}{\sigma_{\text{LHS}}}\right)^{1/\alpha}\right) > 0
\end{equation}

By differentiating both sides of the global maximum condition (\ref{eq:global_max}) and substituting in the local maximum equation (\ref{eq:G_local_max}), we derive an expression of $G_{-i}^{*}$ in terms of $g_i^{*}$:

\[
\frac{\sigma(G_{-i}^{*})}{\sigma^{'}(G_{-i}^{*})}G_{-i}^{*(\rho -1)} = \left(\Delta_t - \frac{g_i^{*}}{p_i^{\mathcal{T}}}\right)\frac{\beta_i p_i^{\mathcal{T}}}{\alpha}g_i^{*(\rho -1)}
\]

which proves that $G_{-i}^{*}>0$, confirming that  $G_{-i}^{*}$ is well-defined.

\section{Proof of Proposition \ref{prop:share_function}}\label{app:proof_prop_share_function}
By Def. \ref{def:share_function}, the share correspondence is a function whenever the replacement correspondence is a function. In particular, given a teamwork game involving a disjunctive task and under the conditions of the Proposition, we know by Corollary \ref{cor:replacemente_correspondence} that both the null component and the positive component of the replacement correspondence are single-valued, and thus $s_i(G)$ can also be decoupled in two components: a null one and a positive one defined over the same domains as $r_i(G)$. By definition of the share function, of $G_i^*$ and of $\overline{G}_i$, we obtain the conditions that $s_i(G)$ must satisfy in the proposition.

To see that the positive component of $s_i$ is an increasing function, we use Eq. (\ref{eq:G_local_max}) and write it in terms of the share function (Definition \ref{def:share_function}):

    \begin{equation}\label{eq:G_as_func_of_s}
        G = \Delta_t\frac{p_i^{\mathcal{T}} \cdot \beta_i^{1/\rho}}{s_i(G)^{1/\rho}} - \frac{\sigma(G)}{\sigma^{\prime}(G)}\frac{\alpha}{s_i(G)}
    \end{equation}

Which yields:

\[
    \frac{dG}{ds_i(G)} = \left(1 + \frac{\sigma^{\prime}(G)^{2}- \sigma(G)\cdot \sigma^{\prime \prime}(G)}{\sigma^{\prime}(G)^{2}}\cdot \frac{\alpha}{s_i(G)}\right)^{-1}\cdot \frac{1}{s_i(G)}\left(\frac{\sigma(G)}{\sigma^{\prime}(G)}\frac{\alpha}{s_i(G)} - \frac{p_i^{\mathcal{T}}}{\rho}\frac{\beta_i^{1/\rho}\Delta_t}{s_i(G)^{1/\rho}}\right)
\]

Given the properties of the evaluation function, it is straightforward to see that the first factor in this equation is positive for all $G$. Using the condition $\frac{\rho}{\Delta_t p_i^{\mathcal{T}}}\geq \frac{\sigma^{\prime}(\overline{G}_i)}{\sigma (\overline{G}_i)}\cdot \frac{\beta_i^{1/\rho}}{\alpha}$, and the fact that for disjunctive tasks $s_i(G)^{-1/\rho} < s_i(G)^{-1}$, we can check that the second factor is positive too. Hence, $s_i(G)$ is a strictly increasing function of $G$ in the domain $\left[G_i^*, \overline{G}_i\right]$. Furthermore, $s_i(G)$ is the inverse of a continuous function (Eq. (\ref{eq:G_as_func_of_s})) on a compact domain and therefore itself continuous. These observations allow us to adapt Proposition 6.1 in \citep{cornes2007weak} into Proposition \ref{prop:share_function} QED.

\section{Individual efforts in disjunctive and conjunctive tasks: tables}\label{app:individual_effort_tables_dis_conj}
\begin{table}[htbp]
\caption{Individual effort across different teams for a \textbf{conjunctive} task ($\rho = -10$) under three different assessment treatments. For simplicity, we fill only the upper diagonal of the matrices, which are symmetric.}
\label{tab:strategies_conjunctive}
    \begin{subtable}{\linewidth}
        \caption{Soft Assessment Treatment ($b = 3$)}
        \label{tab:strategies_conjunctive_3_matrix}
        \centering
        \small
        \begin{tabular}{r|rrrr}
            \rowcolor[HTML]{EEE0FF}
            \toprule
            \diagbox{$p_2$}{$p_1$}& 0.30 & 0.50 & 0.70 & 0.90 \\
            \midrule
            \cellcolor[HTML]{EEE0FF}0.30 & (28\%, 28\%) & (46\%, 30\%) & (53\%, 26\%) & (56\%, 22\%) \\
            \cellcolor[HTML]{EEE0FF}0.50 &  &  (48\%, 48\%) & (53\%, 40\%) &(55\%, 34\%) \\
            \cellcolor[HTML]{EEE0FF}0.70 & &  & (44\%, 44\%)&  (46\%, 37\%) \\
            \cellcolor[HTML]{EEE0FF}0.90 &   &  &   & (38\%, 38\%) \\
            \bottomrule
        \end{tabular}

    \end{subtable}

    \begin{subtable}{\linewidth}
    \caption{Medium Assessment Treatment ($b = 5$)}
        \label{tab:strategies_conjunctive_5_matrix}
        \centering
        \small
        \begin{tabular}{r|rrrr}
            \rowcolor[HTML]{CFC3FF}
            \toprule
            \diagbox{$p_2^\mathcal{T}$}{$p_1^\mathcal{T}$}& 0.30 & 0.50 & 0.70 & 0.90 \\
            \midrule
            \cellcolor[HTML]{CFC3FF}0.30 & (29\%, 29\%) & (49\%, 31\%) & (56\%, 27\%) & (59\%, 23\%) \\
            \cellcolor[HTML]{CFC3FF}0.50 & &(57\%, 57\%) & (68\%, 52\%) & (72\%, 45\%) \\
            \cellcolor[HTML]{CFC3FF}0.70 & & & (64\%, 64\%) & (67\%, 55\%) \\
            \cellcolor[HTML]{CFC3FF}0.90 & & & &  (58\%, 58\%)  \\
            \bottomrule
        \end{tabular}

    \end{subtable}

    \begin{subtable}{\linewidth}
        \caption{Hard Assessment Treatment ($b = 7$)}
        \label{tab:strategies_conjunctive_7_matrix}
        \centering
        \small
        \begin{tabular}{r|rrrr}
            \rowcolor[HTML]{AB96FF}
            \toprule
           \diagbox{$p_2^\mathcal{T}$}{$p_1^\mathcal{T}$} & 0.30 & 0.50 & 0.70 & 0.90 \\
            \midrule
            \cellcolor[HTML]{AB96FF} 0.30 & (29\%, 29\%) & (49\%, 31\%) & (56\%, 27\%) & (59\%, 23\%) \\
            \cellcolor[HTML]{AB96FF}0.50 & & (57\%, 57\%) & (69\%, 52\%) & (73\%, 46\%) \\
            \cellcolor[HTML]{AB96FF}0.70 & & & (69\%, 69\%) & (77\%, 64\%)  \\
            \cellcolor[HTML]{AB96FF}0.90 & & & & (72\%, 72\%)  \\
            \bottomrule
        \end{tabular}
    \end{subtable}
\end{table}

\begin{table}[htbp]
    \caption{Individual effort across different teams for a \textbf{disjunctive} task ($\rho =10$) under three different assessment treatments. For simplicity, we fill only the upper diagonal of the matrices, which are symmetric.}
    \label{tab:strategies_disjunctive}
    \begin{subtable}{\linewidth}
    \caption{Soft Assessment Treatment ($b = 3$)}
    \label{tab:strategies_disjunctive_3_matrix}
    \centering
    \footnotesize
        \begin{tabular}{r|rrrr}
            \rowcolor[HTML]{EEE0FF}
            \toprule
            \diagbox{$p_2^\mathcal{T}$}{$p_1^\mathcal{T}$}& 0.30 & 0.50 & 0.70 & 0.90 \\
            \midrule
            \cellcolor[HTML]{EEE0FF}0.30 & [(63\%, 0\%), (0\%, 63\%)] & [(63\%, 0\%), (0\%, 60\%)] & (0\%, 50\%) & (0\%, 41\%) \\
            \cellcolor[HTML]{EEE0FF}0.50 &  &  [(60\%, 0\%), (0\%, 60\%)] & [(60\%, 0\%), (0\%, 50\%)] & [(60\%, 0\%), (0\%, 41\%)] \\
            \cellcolor[HTML]{EEE0FF}0.70 & &  & [(50\%, 0\%), (0\%, 50\%)]&  [(50\%, 0\%), (0\%, 41\%)] \\
            \cellcolor[HTML]{EEE0FF}0.90 &   &  &   & [(41\%, 0\%), (0\%, 41\%)]\\
            \bottomrule
        \end{tabular}

    \end{subtable}

    \begin{subtable}{\linewidth}
        \centering
        \caption{Medium Assessment Treatment ($b = 5$)}
        \label{tab:strategies_disjunctive_5_matrix}
        \footnotesize
        \begin{tabular}{r|rrrr}
            \rowcolor[HTML]{CFC3FF}
            \toprule           
            \diagbox{$p_2^\mathcal{T}$}{$p_1^\mathcal{T}$}& 0.30 & 0.50 & 0.70 & 0.90 \\
            \midrule
            \cellcolor[HTML]{CFC3FF}0.30 & [(67\%, 0\%), (0\%, 67\%)] & [(67\%, 0\%), (0\%, 78\%)] & (0\%, 71\%) & (0\%, 61\%) \\
            \cellcolor[HTML]{CFC3FF}0.50 & &[(78\%, 0\%), (0\%, 78\%)] & (0\%, 71\%) & (0\%, 61\%) \\
            \cellcolor[HTML]{CFC3FF}0.70 & & & [(71\%, 0\%), (0\%, 71\%)] & [(71\%, 0\%), (0\%, 61\%)] \\
            \cellcolor[HTML]{CFC3FF}0.90 & & & &  [(61\%, 0\%), (0\%, 61\%)]  \\
            \bottomrule
        \end{tabular}
    \end{subtable}

    \begin{subtable}{\linewidth}
        \centering
        \caption{Hard Assessment Treatment ($b = 7$)}
        \label{tab:strategies_disjunctive_7_matrix}
        \footnotesize
        \begin{tabular}{r|rrrr}
            \rowcolor[HTML]{AB96FF}
            \toprule
           \diagbox{$p_2^\mathcal{T}$}{$p_1^\mathcal{T}$} & 0.30 & 0.50 & 0.70 & 0.90 \\
            \midrule
            \cellcolor[HTML]{AB96FF} 0.30 & [(67\%, 0\%), (0\%, 67\%)] & [(67\%, 0\%), (0\%, 80\%)]& (0\%, 84\%) & (0\%, 78\%) \\
            \cellcolor[HTML]{AB96FF}0.50 & & [(80\%, 0\%), (0\%, 80\%)] & (0\%, 84\%) & (0\%, 78\%) \\
            \cellcolor[HTML]{AB96FF}0.70 & & & [(84\%, 0\%), (0\%, 84\%)] & [(84\%, 0\%), (0\%, 78\%)]   \\
            \cellcolor[HTML]{AB96FF}0.90 & & & & [(78\%, 0\%), (0\%, 78\%)] \\
            \bottomrule
        \end{tabular}
    \end{subtable}

\end{table}

\clearpage
\bibliographystyle{elsarticle-num-names} 
\bibliography{cas-refs.bib}

\end{document}